\pdfoutput=1
\documentclass[a4paper,fleqn,usenatbib]{mnras}

\usepackage{newtxtext,newtxmath}

\usepackage[T1]{fontenc}
\usepackage{ae,aecompl}

\usepackage{wasysym}
\usepackage{booktabs}
\usepackage{subfig}
\usepackage{pdflscape}
\usepackage{color}

\usepackage{graphicx}	
\usepackage{amsmath}	
\usepackage{amssymb}	
\usepackage{cite}
\usepackage{natbib}
\usepackage{hyperref}

\newcommand{\chandra}{{\it Chandra}}
\newcommand{\planck}{{\it Planck}}
\newcommand{\hubble}{{\it Hubble}}
\newcommand{\rosat}{{\it ROSAT}}
\newcommand{\xmm}{{\it XMM-Newton}}
\newcommand{\suzaku}{{\it Suzaku}}

\title[X-ray/SZ Galaxy Cluster Deprojections]{Thermodynamic Profiles of Galaxy Clusters from a Joint X-ray/SZ Analysis}

\author[J. A. Shitanishi et al.]{
Jennifer A. Shitanishi,$^{1}$\thanks{E-mail: shitanis@usc.edu}
Elena Pierpaoli,$^{1}$
Jack Sayers,$^{2}$
Sunil R. Golwala,$^{2}$
\newauthor
Silvia Ameglio,$^{1}$
Adam B. Mantz,$^{3}$ 
Tony K. Mroczkowski,$^{4}$
\newauthor 
Elena Rasia$^{5}$
and Seth Siegel$^{2,6}$
\\
$^{1}$ Department of Physics and Astronomy, University of Southern California, Los Angeles, CA 90089-0484, USA  \\
$^{2}$ Division of Physics, Math, and Astronomy, California Institute of Technology, Pasadena, CA, 91125, USA \\ 
$^{3}$ Kavli Institute for Particle Astrophysics and Cosmology, Stanford University, Stanford, CA 94305-4085, USA \\ 
$^{4}$ ESO - European Organization for Astronomical Research in the Southern hemisphere, Karl-Schwarzschild-Str. 2, D-85748 Garching b. Munchen, Germany \\
$^{5}$ INAF, Osservatorio Astronomico di Trieste, via Tiepolo 11, I-34131, Trieste, Italy \\
$^{6}$ Department of Physics, McGill University, 3600 Rue University, Montreal, QC H3A 2T8, Canada 
}

\date{Accepted XXX. Received YYY; in original form ZZZ}

\pubyear{2017}

\begin{document}
\label{firstpage}
\pagerange{\pageref{firstpage}--\pageref{lastpage}}
\maketitle

\begin{abstract}

We jointly analyze Bolocam Sunyaev-Zeldovich (SZ) effect and \chandra\ X-ray data for a set of 45 clusters to derive gas density and temperature profiles without using spectroscopic information. The sample spans the mass and redshift range $3 \times 10^{14} M_{\astrosun} \le M_{500} \le 25 \times 10^{14} M_{\astrosun}$ and $0.15\le z \le 0.89$. We define cool--core (CC) and non--cool core (NCC) subsamples based on the central X-ray luminosity, and 17/45 clusters are classified as CC. In general, the profiles derived from our analysis are found to be in good agreement with previous analyses, and profile constraints beyond $r_{500}$ are obtained for 34/45 clusters. In approximately 30\% of the CC clusters our analysis shows a central temperature drop with a statistical significance of $>3\sigma$; this modest detection fraction is due mainly to a combination of coarse angular resolution and modest S/N in the SZ data. Most clusters are consistent with an isothermal profile at the largest radii near $r_{500}$, although 9/45 show a significant temperature decrease with increasing radius. The sample mean density profile is in good agreement with previous studies, and shows a minimum intrinsic scatter of approximately 10\% near $0.5 \times r_{500}$. The  sample mean temperature profile is consistent with isothermal, and has an intrinsic scatter of approximately 50\% independent of radius. This scatter is significantly higher compared to earlier X-ray-only studies, which find intrinsic scatters near 10\%, likely due to a combination of unaccounted for non-idealities in the SZ noise, projection effects, and sample selection. 

\end{abstract}




\section{Introduction}

Characterization of the gaseous intracluster medium (ICM) is important to the study of both cosmology and galaxy cluster astrophysics. For instance, measurements of the redshift-dependent halo mass function at cluster scales have produced tight constraints on a range of cosmological parameters, and ICM observables have played a central role in nearly all such surveys to date \citep[e.g.,][]{2009ApJ...692.1060V, 2010MNRAS.406.1759M, 2016A&A...594A..24P, 2016ApJ...832...95D}. These measurements rely on an accurate and well understood connection between ICM properties and underlying halo mass. For example, at fixed mass, what is the average shape and intrinsic scatter of the ICM thermodynamic profiles as a function of radius? The answer to this question is influenced by a range of complex physical processes, such as the radiative cooling and feedback from active galactic nuclei that tend to be important in the cluster core and the active accretion that occurs in the outer regions of the cluster \citep{2007ApJ...668....1N,2009ApJ...705.1129L}. The relative contributions of these physical processes to clusters' thermodynamic states is currently not well known, but can be studied with measurements of ICM density, pressure, and temperature profiles.

The ICM gas in clusters has foremost been studied through X-ray observations.  Imaging and spectroscopy provide density and temperature profiles, respectively \mbox{\citep{1988xrec.book.....S}}.  As a supplement to X-rays, recent improvements in instrumentation have enabled Sunyaev-Zel'dovich (SZ) effect observations to provide meaningful constraints on the ICM. The signal is proportional to the integrated pressure along the line of sight and pressure profiles can therefore be determined directly from SZ effect observations \citep{2010ApJ...716.1118P,2013A&A...550A.131P,2013ApJ...768..177S}. Due to the different dependences of X-ray and SZ signals on gas density and temperature, combining X-ray surface brightness maps with SZ maps allows the simultaneous  recovery  of both physical quantities \citep{2007MNRAS.382..397A}. Moreover, the weaker dependence of the SZ signal on density results in an observed brightness profile that falls more slowly with radius, which can often facilitate studies of a given cluster's outskirts.

There is a growing body of work based on joint X-ray and SZ fitting. Combining these multiwavelength data has yielded estimates of the cosmic distance scale \citep{2006ApJ...647...25B} and smooth thermodynamic profiles \citep{2006ApJ...652..917L}.  More recently, \citet{2013A&A...551A..22E} combined \rosat\ gas density and \planck\ pressure profiles to find that entropy profiles continue to rise beyond $r_{200} \approx 1.5 \times r_{500}$. \footnote {Throughout this paper, $r_{\Delta}$ denotes the distance at which the average density within is $\Delta$ times the critical density of the universe. } As another example, "Joint Analysis of Cluster Observations" (JACO) was developed by \citet{2013ApJ...767..116M} to combine X-ray, SZ, and weak lensing data to find cluster masses, and was most recently used on Cluster Lensing and Supernova survey with \hubble\ (CLASH) clusters to measure their mass and gas profiles \citep{2016arXiv161205377S}. In addition, X-ray and SZ surface brightness data have been combined to obtain precise thermodynamic profiles independent of X-ray spectroscopy in single clusters \citep{2009ApJ...694.1034M,2009A&A...506..623N,2010A&A...519A..29B,2017A&A...597A.110R}, in general finding good agreement with the spectroscopically-derived results.  Most recently, the \xmm\ cluster outskirts project (X-COP) completed a joint analysis on the cluster Abell 2319 using Planck SZ maps and \xmm\ X-ray surface brightness and spectroscopic measurements \citep{2017arXiv170802954G}. 

In addition, numerous efforts, mainly using X-ray data, have been made to characterize the properties of ensemble--average thermodynamic profiles. Such analyses require clusters of different masses and redshifts to be scaled to a common reference, which is generally done based on the self-similar relations derived from the simplifying scenario of clusters forming from a purely gravitational spherical collapse \citep{1986MNRAS.222..323K}. These ensemble studies of many clusters can then show trends and/or scatter away from perfect self-similarity, thus revealing the degree to which non-gravitational physics occurs (e.g., departures from hydrostatic equilibrium). 

For example, \citet{2006ApJ...640..691V} used \chandra\ exposures of 13 relaxed clusters at $z < 0.23$ to derive average density and temperature profiles.  The density profile was found to be consistent with self-similarity with a scatter of approximately 15\%, while the temperature profile decreases beyond $0.2 \times r_{500}$ and has even lower scatter.  \citet{2008A&A...486..359L}, based on \xmm\ exposures of a sample comprised of both CC and NCC clusters at $z < 0.3$ with no evidence of recent merger activity, found an average temperature profile with a clear drop beyond $0.2 \times r_{180} \approx 0.3 \times r_{500}$ and an intrinsic scatter of 6\%. \citet{2012A&A...541A..57E} used \rosat\ to study 31 clusters at $0.04 < z < 0.2$: half CC clusters and the other half NCC.  They found a 10--20\% scatter in the average density profile at intermediate radii, increasing to 30\% at $r_{200} \approx 1.5r_{500}$.  \citet{2014ApJ...794...67M}, using \chandra\ observations of 80 clusters spanning a broad redshift range, compared results for cluster samples centered at $z=0.46$ and $z=0.82$. They found the higher redshift sample to have 30\% lower average temperatures near the core and a steeper drop in temperature at large radii. \citet{2015MNRAS.450.2261M} stacked \chandra\ emission measure profiles of 320 clusters spanning a redshift range of $0.056 < z <1.24$ to derive an average density and intrinsic scatter; they found a scatter of $20\%$ at $r_{500}$ which increases to $30\%$ at $r_{200} \approx 1.5r_{500}$).  \citet{2016MNRAS.456.4020M} constrained the mean density and temperature profiles and intrinsic scatters of a sample of 40 relaxed, CC clusters using \chandra\ data, finding a temperature scatter of approximately 10\% at all radii. Furthermore, \citet{2017ApJ...843...72B} simultaneously fit cluster pressure profiles to X-ray spectroscopic data and Planck SZ measurements, finding a 10\% intrinsic scatter within $r_{500}$ and an increase at larger radii.

In sum, the small intrinsic scatters found in these and other observational studies indicate that cluster thermodynamic profiles are reasonably well approximated by a universal shape, at least for relatively relaxed clusters outside of the central core region. This conclusion is also supported by a range of numerical simulations \citep[e.g.,][]{2007ApJ...668....1N,2012ApJ...758...75B,2015ApJ...813L..17R,2015ApJ...806...68L,2017MNRAS.467.3827P,2017MNRAS.468..531B}. While this agreement is encouraging, simulations, while greatly improved recently, still cannot reproduce all of the small-scale physical phenomena which take place in the cluster core. In addition, the bulk of the observational constraints are focused on the inner regions of the cluster within $r_{500}$, although this situation is quickly changing (see, e.g., \citealt{2017ApJ...843...72B,2017arXiv170802954G}). Furthermore, at least one of the two primary X-ray telescopes used to study clusters (\xmm\ and \chandra) has a significant spectral calibration bias \citep{2015A&A...575A..30S}. Looking forward, joint SZ/X-ray analyses offer a promising tool to address the two latter issues. Specifically, they in general allow for studies to larger radii compared to analyses based solely on X-ray data, and they also allow for full thermodynamic constraints without the use of spectroscopic X-ray information.

In this work, we combine SZ images from Bolocam, a mm-wave bolometric imager that operated from the Caltech Submillimeter Observatory (CSO), with surface brightness maps from \chandra, NASA's flagship X-ray observatory, to recover density and temperature profiles for the BOXSZ sample of 45 clusters defined in \citet{2015ApJ...806...18C}.  Markov Chain Monte Carlo (MCMC) is used to fit smooth parametric models and a model consisting of concentric shells with uniform properties of the ICM density and temperature profiles of each cluster. In addition, sample mean profiles and the intrinsic scatter about these mean profiles are determined for both the full cluster sample and various sub-samples. The standard flat $\Lambda$CDM model is used, with $\Omega_{m}=0.3$, $\Omega_{\Lambda}=0.7$, and the dimensionless Hubble parameter $h=0.7$.  The outline of the paper is as follows: Sec.~\ref{sec:ClusterSample} describes the cluster sample, Sec.~\ref{sec:DataReduction} details the data reduction, and Sec.~\ref{sec:Method} reviews the modeling and fitting methods. Sec.~\ref{sec:AnalysisofMockClusterObservations} describes detailed consistency tests based on mock observations of smooth cluster models, Sec.~\ref{sec:Results} reviews the individual and joint cluster results, and Sec.~\ref{sec:Conclusions} presents the overall conclusions.

\section{Cluster Sample}
\label{sec:ClusterSample}

\begin{figure}
	\includegraphics[width=\columnwidth]{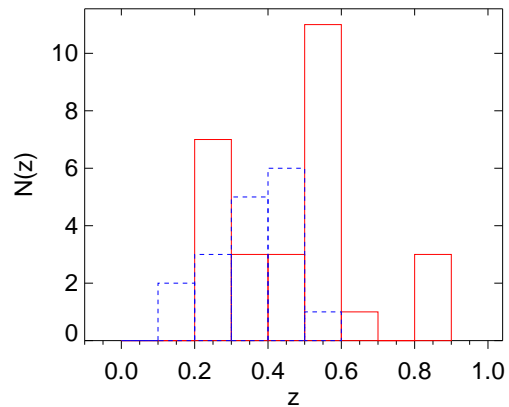}
	\caption{Cluster redshift distribution of the sample.  The red solid line indicates NCC clusters while the blue dotted line indicates CC clusters.}
	\label{fig:clusterz}
\end{figure}

The Bolocam X-ray/SZ (BOXSZ) sample consists of 45 clusters observed by both Bolocam and \chandra, and Tab.~\ref{tab:clusterobsinfo} lists some important characteristics of the clusters. The images from both instruments are approximately $14'$ in size. Given the relatively high median redshift of the sample, $z=0.44$ (see Fig.~\ref{fig:clusterz}), these images in general contain information beyond $r_{500}$ and therefore allow for studies of the clusters' outskirts. The clusters' masses, as estimated from X-rays according to the procedures described in \citet{2010MNRAS.406.1773M}, span the range  $M_{500} \simeq 3-25 \times 10^{14} M_{\astrosun}$.

As in \citet{2013ApJ...768..177S}, CC and NCC clusters are differentiated using an X-ray luminosity ratio cut. If the luminosity within $0.05r_{500}$ is greater than 0.17 times the total luminosity within $r_{500}$, then the cluster is classified as CC. Within the BOXSZ sample, the CC clusters have a lower median redshift compared to the NCC clusters.  In addition, 15 of the 16 highest-redshift clusters are NCC.  Although this trend matches what is expected based on cluster formation models, the trend within the BOXSZ sample is most likely due to selection effects. Again following the convention of \citet{2013ApJ...768..177S}, clusters are classified as disturbed using the X-ray centroid shift parameter, which quantifies the fractional variations in X-ray centroids computed within circular apertures increasing from $0.05r_{500}$ up to $r_{500}$ in steps of $0.05r_{500}$ \citep{2006MNRAS.373..881P,2012MNRAS.421.1583M,2013ApJ...768..177S}. All clusters with a centroid shift larger than 0.01 are considered disturbed. The sample contains a variety of clusters based on these classifications: 17/45 are CC and 16/45 are disturbed, with all but one of the disturbed systems being NCC (see Tab.~\ref{tab:clusterobsinfo}). 

\begin{table*}
	\centering
        \scriptsize
	\caption{Observation Information}
	\label{tab:clusterobsinfo}
	\begin{tabular}{*{11}{c}}
		\toprule
		              &       &                    &  & & &\multicolumn{2}{c}{X-ray} & \multicolumn{2}{c}{SZ} & \\
		\cmidrule(lr){7-8}
		\cmidrule(lr){9-10}
	 	Cluster   & z   &  $r_{500}$  & $M_{500}$                           & NCC/CC & disturbed & ObsID  & Exposure Time &    S/N Peak	    &  Obs Time  & $r_{\rm max}$ \\        
		               &     &  (Mpc) 	    & ($10^{14} M_{\astrosun}$)   &                &                 &             & (ks)                   &                            &   (hrs)       & ($r_{500}$) \\    
		\midrule
Abell 2204        & 0.15 & 	1.46 $\pm$ 0.07  & 10.3  $\pm$ 1.5      & CC     &	&	6104  & \hphantom{1}9.61          & 22.30   & 12.7    & 0.908 \\
Abell 383         & 0.19 & 	1.11  $\pm$ 0.06 & \hphantom{1}4.7 $\pm$ 0.8       & CC     &	&	2321  & 19.51         & \hphantom{1}9.60    & 24.3  & 1.198   \\
Abell 1423        & 0.21 & 	1.35 $\pm$ 0.10  & \hphantom{1}8.7 $\pm$ 2.0         & NCC    &	&	538   & \hphantom{1}9.87          & \hphantom{1}5.80    & 11.5    & 1.182  \\
Abell 209         & 0.21 & 	1.53 $\pm$ 0.08  & 12.6 $\pm$ 1.9       & NCC    &		&3579  & \hphantom{1}9.99          & 13.90   & 17.8    & 1.561 \\
Abell 963         & 0.21 & 	1.35  $\pm$  0.06 & \hphantom{1}6.8 $\pm$ 1.0        & NCC    & 	&	903   & 36.29         & \hphantom{1}8.30    & 11.0  & 1.137   \\
Abell 2261        & 0.22 &  1.59  $\pm$ 0.09 & 14.4 $\pm$ 2.6      & CC     & 	&	5007  & 24.32         & 10.20   & 17.5    & 1.213 \\
Abell 2219        & 0.23 & 	1.74 $\pm$ 0.08 & 18.9 $\pm$ 2.5       & NCC    & 	&	896   & 42.30         & 11.10   & \hphantom{1}6.3     &  0.975 \\
Abell 267         & 0.23 &   1.22 $\pm$ 0.07 & \hphantom{1}6.6 $\pm$ 1.1      & NCC    & \checkmark  &			3580  & 19.88         & \hphantom{1}9.60    & 20.7   & 0.942  \\
RX J2129.6+005    & 0.24 & 1.28 $\pm$ 0.07  & \hphantom{1}7.7 $\pm$ 1.2        & CC     & & 		552   & \hphantom{1}9.96          & \hphantom{1}8.00    & 16.0    & 0.721 \\
Abell 1835        & 0.25 & 1.49 $\pm$ 0.06  & 12.3 $\pm$ 1.4       & CC     & & 			7370    & 39.51           & 15.70   & 14.0    &  1.281\\
Abell 697         & 0.28 & 1.65 $\pm$ 0.09  & 17.1 $\pm$ 2.9        & NCC    & & 			4217  & 19.52         & 22.60   & 14.3    & 1.269 \\
Abell 611         & 0.29 & 1.24 $\pm$ 0.06 & \hphantom{1}7.4 $\pm$ 1.1        & NCC    &  & 			3194  & 36.11         & 10.80   & 18.7  &  1.363  \\
MS 2137       & 0.31 & 1.06 $\pm$ 0.04 & \hphantom{1}4.7  $\pm$ 0.6       & CC     &  & 		4974  & 57.38         & \hphantom{1}6.50    & 12.8    & 1.675 \\
MACSJ1931.8-2634  & 0.35 & 1.34 $\pm$ 0.07  & \hphantom{1}9.9 $\pm$ 1.6        & CC     & & 		9382  & 98.92         & 10.10   & \hphantom{1}7.5   & 0.959   \\
Abell S1063       & 0.35 & 1.76 $\pm$ 0.09 & 22.2 $\pm$ 3.4       & NCC    & & 		4966  & 26.72         & 10.20   & \hphantom{1}5.5    & 1.493  \\
MACS J1115.8+0129  & 0.36 & 1.28 $\pm$ 0.06  & \hphantom{1}8.6 $\pm$ 1,2        & CC     &  & 		9375  & 39.63         & 10.90   & 22.8    &  1.361 \\
MACS J1532.9+3021  & 0.36 & 1.31 $\pm$ 0.08 & \hphantom{1}9.5 $\pm$ 1.7        & CC     & & 		1649  & \hphantom{1}9.36          & \hphantom{1}8.00    & 14.8  &   1.589 \\
Abell 370         & 0.38 & 1.40 $\pm$ 0.08 & 11.7 $\pm$ 2.1       & NCC    & \checkmark  & 			7715  & \hphantom{1}7.09          & 12.80   & 11.8 &   1.117  \\
ZWCL 0024+17    & 0.39 & 1.00 $\pm$ 0.11 & \hphantom{1}4.4  $\pm$ 1.6       & NCC    & \checkmark & 		929   & 39.94         & \hphantom{1}3.30    & \hphantom{1}8.3     & 1.829 \\
MACS J1720.3+3536  & 0.39 & 1.14 $\pm$ 0.07  & \hphantom{1}6.3 $\pm$ 1.1        & CC     &  & 		6107  & \hphantom{1}9.61          & 10.60   & 16.8   & 1.125  \\
MACS J0429.6-0253  & 0.40  & 1.10 $\pm$ 0.05 & \hphantom{1}5.8 $\pm$ 0.8        & CC     &  & 		3271  & 23.17         & \hphantom{1}8.90    & 17.0    & 1.128 \\
MACS J2211.7-0349  & 0.40 & 1.61  $\pm$ 0.07 & 18.1 $\pm$ 2.5       & CC     &   & 		3284  & 17.74         & 14.70   & \hphantom{1}6.5    &  1.281 \\
MACS J0416.1-2403  & 0.42 & 1.27 $\pm$ 0.15  & \hphantom{1}9.1 $\pm$ 2.0        & NCC    & \checkmark & 10446 & 15.83         & \hphantom{1}8.50    & \hphantom{1}7.8   & 0.921   \\
MACS J0451.9+0006  & 0.43 & 1.12  $\pm$ 0.06 & \hphantom{1}6.3 $\pm$ 1.1        & NCC    & \checkmark & 		5815  & 10.21         & \hphantom{1}8.10    & 14.2  &  1.013   \\
MACS J0417.5-1154  & 0.44 & 1.69 $\pm$ 0.07 & 22.1 $\pm$ 2.7       & CC     & \checkmark & 		11759 & 51.36         & 22.70   & \hphantom{1}9.8     & 1.806  \\
MACS J1206.2-0847  & 0.44 & 1.61 $\pm$ 0.08  & 19.2 $\pm$ 3.0       & NCC    & & 		3277  & 23.46         & 21.70   & 24.9  &  1.083  \\
MACS J0329.6-0211  & 0.45 & 1.19 $\pm$ 0.06 & \hphantom{1}7.9 $\pm$ 1.3      & CC     & \checkmark  & 		3582  & 19.85         & 12.10   & 10.3    & 1.103 \\
MACS J1347.5-1144      & 0.45 & 1.67 $\pm$ 0.08  & 21.7  $\pm$ 3.0      & CC     &  & 		3592  & 57.51         & 36.60   & 15.5     & 1.084 \\
MACS J1311.0-0310  & 0.49 & 0.93 $\pm$ 0.04  & \hphantom{1}3.9 $\pm$ 0.5        & CC     &  & 		6110  & 63.21         & \hphantom{1}9.60    & 14.2  &  0.995  \\
MACS J0257.1-2325  & 0.50  & 1.20 $\pm$ 0.06 & \hphantom{1}8.5  $\pm$1.3       & NCC    &  & 		1654  & 19.85         & 10.10   & \hphantom{1}5.0  &   1.293  \\
MACS J0911.2+1746  & 0.50  & 1.22  $\pm$ 0.06 & \hphantom{1}9.0 $\pm$ 1.2        & NCC    &      & 		5012  & 23.79         & \hphantom{1}4.80    & \hphantom{1}6.2  & 1.280   \\
MACS J2214.9-1359  & 0.50  & 1.39 $\pm$ 0.08 & 13.2 $\pm$ 2.3       & NCC    &    & 		3259  & 19.47         & 12.60   & \hphantom{1}7.2   &  1.260  \\
MACS J0018.5+1626      & 0.54 & 1.47 $\pm$ 0.08  & 16.5 $\pm$ 2.5       & NCC    & & 		520   & 67.41         & 15.70   & \hphantom{1}9.8   & 1.294   \\
MACS J1149.5+2223  & 0.54 & 1.53 $\pm$ 0.08  & 18.7 $\pm$ 3.0       & NCC    & \checkmark  & 		3589  & 20.05         & 17.40   & 17.7   & 1.170  \\
MACS J0717.5+3745 & 0.55 & 1.69 $\pm$ 0.06  & 24.9 $\pm$ 2.7       & NCC    & \checkmark & 		4200  & 59.04         & 21.30   & 12.5   & 1.130  \\
MACS J1423.8+2404  & 0.55 & 1.09 $\pm$ 0.05 & \hphantom{1}6.6 $\pm$ 0.9        & CC     & & 		4195  & 115.57        & \hphantom{1}9.40    & 21.7 &    1.796 \\
MACS J0454.1-0300    & 0.55 & 1.31 $\pm$  0.06  & 11.5 $\pm$ 1.5       & NCC    & \checkmark & 		902   & 44.19         & 24.30   & 14.5    & 1.393 \\
MACS J0025.4-1222  & 0.58 & 1.12 $\pm$ 0.04  & \hphantom{1}7.6 $\pm$ 0.9        & NCC    &   & 		10413 & 75.64         & 12.30   & 14.3   & 0.931  \\
MS 2053.7-0449       & 0.58 & 0.82 $\pm$ 0.06  & \hphantom{1}3.0 $\pm$ 0.5        & NCC    & \checkmark & 		1667  & 44.51         & \hphantom{1}5.10    & 14.3 & 0.867    \\
MACS J0647.7+7015  & 0.59 & 1.26 $\pm$ 0.06 & 10.9 $\pm$ 1.6       & NCC    & & 		3584  & 20.00         & 14.40   & 11.7    & 1.377   \\
MACS J2129.4-0741  & 0.59 & 1.25  $\pm$ 0.06 & 10.6 $\pm$ 1.4       & NCC    & \checkmark & 		3595  & 19.87         & 15.20   & 13.2   & 1.283  \\
MACS J0744.8+3927  & 0.69 & 1.26  $\pm$ 0.06 & 12.5 $\pm$ 1.6        & NCC    & \checkmark  & 		6111  & 49.50         & 13.30   & 16.3  &  1.008  \\
MS1054.4-0321       & 0.83 & 1.07 $\pm$ 0.07 & \hphantom{1}9.0  $\pm$ 1.3       & NCC    & \checkmark & 		512   & 89.17         & 17.40   & 18.3     & 0.893 \\
CL J0152.7    & 0.83 & 0.97 $\pm$ 0.26   & \hphantom{1}7.8 $\pm$ 3.0        & NCC    & \checkmark  & 		913   & 36.48         & 10.20   & \hphantom{1}9.3   & 1.098   \\
CL J1226.9+3332    & 0.89 & 1.00 $\pm$ 0.05  & \hphantom{1}7.8 $\pm$ 1.1        & NCC    &    & 		5014  & 32.71         & 13.00   & 11.8 & 0.947   \\
\bottomrule
\end{tabular}
\begin{flushleft}
\normalsize
The BOXSZ cluster sample. The columns give the cluster name, redshift, radius, mass, morphological classification (see Sec.~\ref{sec:ClusterSample}), \chandra\ X-ray and Bolocam SZ observation details, and the maximum radius (see Sec.~\ref{sec:DataIndiviClusterAnalysis}) 
included in this analysis.
\end{flushleft}
\end{table*}

\section{Data Reduction}
\label{sec:DataReduction}

\subsection{X-Ray}
\label{sec:Xray} 

The X-ray data for this analysis were taken from the \chandra\ X-ray Observatory public archive, and all were obtained using the Advanced CCD Imaging Spectrometer (ACIS) focal plane arrays ACIS-I and ACIS-S. The data were reduced according to standard processing using CIAO version 4.7.  Observation information is summarized in Tab.~\ref{tab:clusterobsinfo}.

Raw event files were processed in several steps to make cluster, exposure, and background maps to ensure all sources of non-cluster signal were accounted for.  First, event files were processed to create raw images, which were then filtered to remove cosmic rays and other point sources.  Many of the observations were in VFAINT mode, where a larger pixel kernel was used to identify bad events.  Standard bad pixels and chip boundaries were removed, as well as filtering for good time intervals.  Light curve filtering in the total 0.3-10 keV band of the background was performed to find any observation-specific bad pixels/columns and background flares. We restricted our analysis to only the 0.7-2 keV band in order to minimize the effect of the background on the data and the temperature dependence of the signal.
The background was calculated by renormalizing \chandra\ blank-sky observations of the appropriate epoch to the flux measured in the outskirts of the chips that did not include cluster signal.  These regions were always chosen to be sufficiently far from the cluster center ($R > 1.5r_{500}$).

\begin{figure}
	\includegraphics[width=\columnwidth]{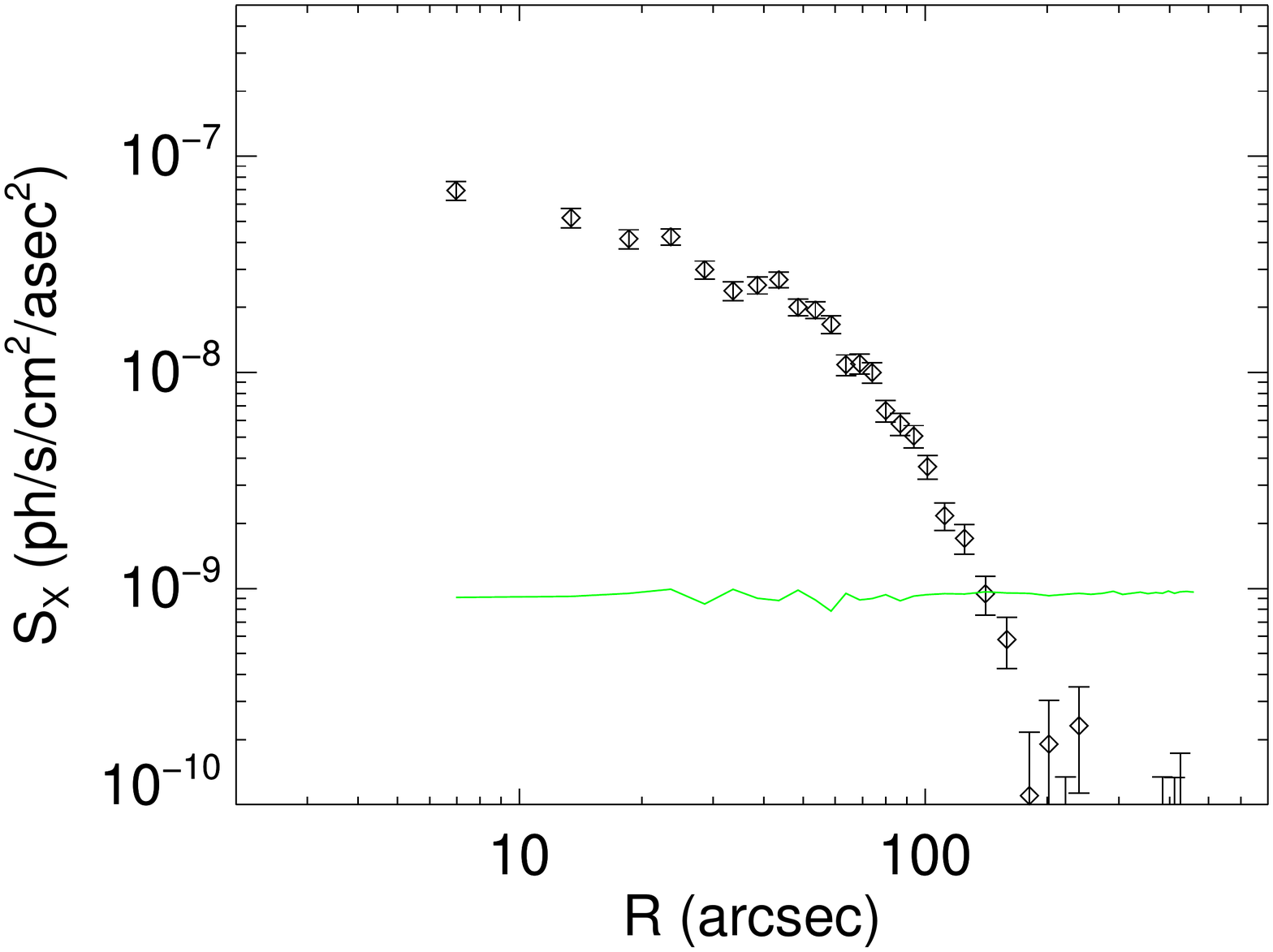}
	\caption{Example X-ray surface brightness profile (MACS J0416.1-2403). The green line represents the background level, and the black points are the the estimated surface brightness after subtracting the background.  
	}
	\label{fig:surf_bright_example}
\end{figure}

The images were then radially binned so that each annulus contained at least 100 total counts and 10 source (i.e. total minus background) counts, and was at least $5''$ in radial width.  Due to the large binning, point-spread-function (PSF) effects could be ignored. Fig.~\ref{fig:surf_bright_example} shows an example surface brightness profile resulting from the data reduction.

To extract the surface brightness we need to compute the cooling function $\Lambda(T)$, which was done using the Mekal plasma modeling code in XSPEC \citep{1996ASPC..101...17A}.  A constant metallicity of $0.3Z_{\astrosun}$ was assumed. The $n_{H}$ column density number was estimated using the program nH in the FTOOLS software from HEASARC \footnote{\href{http://heasarc.gsfc.nasa.gov/ftools/}{http://heasarc.gsfc.nasa.gov/ftools/}} \citep{1995ASPC...77..367B}, and is in general different for each cluster based on its location.

\subsection{SZ}
\label{sec:SZ}

Bolocam was a 144-element bolometric imager stationed on the CSO with an $8'$ circular FOV.  For the BOXSZ observations, Bolocam was configured with an SZ emission-weighted band center of 140 GHz, a PSF with a FWHM of $58''$, and a scan pattern resulting in a final image size of $14' \times 14'$. The observations were conducted between November 2006 and March 2012 (see Tab.~\ref{tab:clusterobsinfo} for an observation summary).  The SZ data reduction technique is described in detail in  \citet{2011ApJ...728...39S}, and the relevant details are listed here. The pointing reconstruction is accurate to $5''$, and the flux calibration is accurate to 5\%. Bright radio sources were removed from the images, including a total of 11 sources in the central cluster region thought to be member galaxies and 6 non-central sources thought to be unrelated to the clusters \citep{2013ApJ...764..152S}. 

An instrument noise model is constructed for each cluster image based on jack-knifed realizations of the data. In addition to instrument noise, the images also contain brightness fluctuations associated with background signals from the primordial CMB anisotropies and dusty star-forming galaxies. These fluctuations are sub-dominant to the instrument noise, and have been included in the noise model based on power spectrum measurements from Planck and the South Pole Telescope \citep{2016A&A...594A..11P,2015ApJ...799..177G}. Brightness fluctuations associated with foreground emission, such as galactic dust and synchrotron, are negligible given the observing frequency, image size, and high galactic latitude of all the clusters, and therefore have not been included in the noise model. In the fits described below, the noise is assumed to be uncorrelated between pixels, which has been demonstrated to be a good approximation \citep{2016arXiv161205377S}, and the validity of this assumption is discussed in more detail in Sec.~\ref{sec:EnsembleResults}.

The data processing, which is primarily to remove atmospheric brightness fluctuations, results in a spatial distortion of the cluster signal. This distortion appears as a high-pass filter in the 2-D images, and has been calculated separately for each cluster to account for the minor differences in filtering based on the observing conditions and the cluster shape. In all of the fits described below, this has been accounted for by applying the cluster-specific high-pass filter to the candidate model prior to comparison with the data.

\section{Method}
\label{sec:Method}

There are two main components of this work: the individual cluster fits and the sample mean fits. For the individual cluster fitting, two types of models are adopted to fit the density and temperature profiles: smooth parametric functions  and concentric shell models.  Two approaches are explored to fit the sample mean profiles, both using
a version of the concentric shell modeling at common radii to constrain average profiles and their intrinsic scatters.

\subsection{Individual Cluster Analysis}
\label{sec:DataIndiviClusterAnalysis}

Both the smooth parametric models and the concentric shell models are separately fit to the data. The former assumes a smooth profile that one can easily compare with other studies and gives a general idea of the cluster as a whole.  The latter assumes constant temperature and density in a set of 5 radial shells. The inner shell spans the radial range 0\arcsec--30\arcsec, which is the approximate angular resolution of Bolocam, and the other 4 shells are logarithmically spaced at increasing radii up to $r_{\rm max}$. While this step-wise modeling is somewhat simplistic, it allows for possible substructures and irregularities in the profiles.  For the individual cluster fitting, the maximum radius, $r_{\rm max}$, is chosen to be the location where the X-ray data reaches a signal-to-noise (i.e. source counts over the square root of total counts, see Sec.~\ref{sec:Xray}) of one. Tab.~\ref{tab:clusterobsinfo} reports $r_{\rm max}$ for all clusters. For most clusters, $r_{\rm max}$ is near $r_{500}$. For both types of fits, we use MCMC to maximize a joint SZ and X--ray likelihood function by varying either the shell deprojection values or the parameters of the smooth parametric profiles.

\subsubsection{Smooth Parametric Profiles}
\label{sec:AnalyticalProfiles}

For the smooth parametric fits, we assume fitting functions similar to those given in \citet{2006ApJ...640..691V}, which are based on X-ray observations of 13 clusters. Specifically, the density is assumed to follow a double-beta profile
\begin{equation}
n_{e}(r)=\left(n_{0,1}^{2}(1+(r/r_{c,1})^{2})^{-3\beta} + n_{0,2}^{2}(1+(r/r_{c,2})^{2})^{-3\beta}\right)^{1/2},
\end{equation}
where $n_{0,\textbf{\textit{i}}}$ and $r_{c,\textbf{\textit{i}}}$ are the central density and the scale radius for the $i$'th component, respectively, and $\beta$ is the slope parameter for the components. The temperature is assumed to follow a profile given by
\begin{equation}
T_{e}(r)=T_{0} \frac{T_{\rm min}/T_{0}+(r/r_{\rm cool})^{1.9}}{1+(r/r_{\rm cool})^{1.9}} \left(1+(r/r_{t})^{2}\right)^{-\alpha},
\end{equation}
where $T_{0}$ is the normalization temperature, $T_{\rm min}$ is the temperature at the center, $r_{\rm cool}$ is the cool-core radius, $r_{t}$ is the outer scale radius, and $\alpha$ is the outer slope.  The inner slope is fixed to $1.9$ \citep{2006ApJ...640..691V}. These profiles are then used to determine a 1-D projected X-ray surface brightness profile and a 2-D SZ image that can be compared to the observational data. Specifically,
\begin{equation}
S_{X-ray}(R) = \frac{1}{4\pi (1+z)^4} \int n^2_e(\ell) \Lambda(T_e(\ell)) d\ell
\end{equation}
gives the X-ray surface brightness at the projected radius $R$, where the integral is over the full line of sight at $R$.   For these spherical models we make the substitutions $\ell = \sqrt{r^2-R^2}$ and $d\ell = rdr / \sqrt{r^2-R^2}$.  $\Lambda(T_e(\ell))$ is the cooling function (which is computed numerically as outlined in Sec.~\ref{sec:Xray}, and scales approximately as $T^{1/2}$).

The change in CMB brightness due to the SZ effect is given by
\begin{equation}
S_{SZ}(R) = g(x) S_{CMB} \frac{k_B \sigma_T}{m_e c^2} \int n_e(\ell) T_e(\ell) d\ell,
\end{equation}
where $S_{CMB}$ is the average CMB brightness, $k_B$ is the Boltzmann constant, $\sigma_T$ is the Thomson cross-section, $m_e$ is the electron mass, $c$ is the speed of light, and the integral is again over the full line of sight at $R$. The prefactor $g(x)$ contains the frequency dependence of the SZ effect (e.g., see \citet{2002ARA&A..40..643C}). $S_{SZ}(R)$ is used to obtain a 2-D image of the SZ signal, which is directly compared to the observational data after accounting for the effects of the image filtering and PSF described in Sec.~\ref{sec:SZ}.

The fit for each cluster proceeds as follows. First, a basic single-beta isothermal model is assumed, with the parameters $n_{0,1}$, $r_{c,1}$, $\beta$, and $T_{0}$ allowed to vary. Next, a second fit is performed with additional free parameters that describe a second density component, generally due to the steep inner profile often found in cool-core clusters ($n_{0,2}$, $r_{c,2}$). The $\chi^2$ value from each of these fits is used to compute a probability to exceed (PTE), and the fit with the higher PTE is then selected as the better description of the data for that cluster. The PTE values are again used to compare fits with additional parameters to allow for a temperature change at large radii ($r_{t}$, $\alpha$) and with additional parameters associated with a temperature drop towards to core ($T_{\rm min}$, $r_{\rm cool}$). Out of all the possible permutations given above, the fit with the highest PTE is selected for each cluster.

\subsubsection{Concentric Shell Deprojections}
\label{sec:OnionskinDeprojections}

\begin{figure}
	\centering
	\includegraphics[width=\columnwidth]{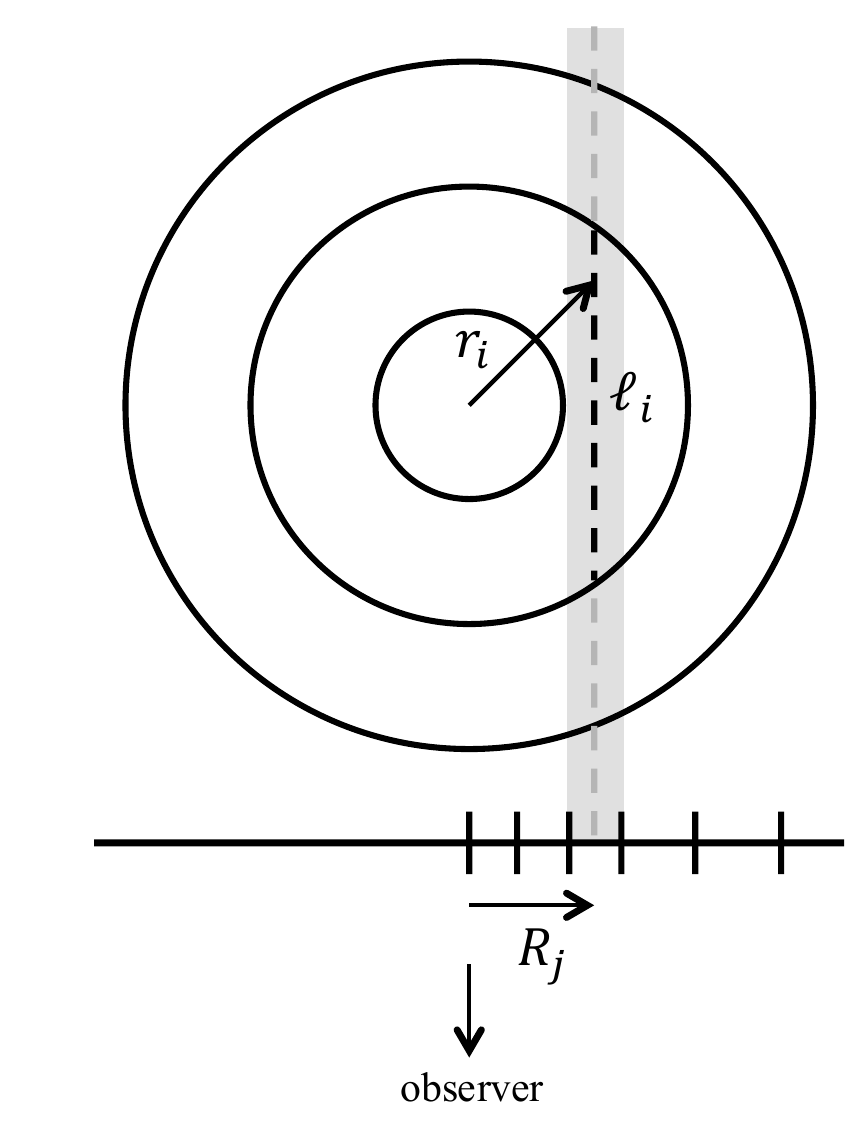}
	\caption{The concentric shell model, adapted from \citet{1999AJ....117.2398M}.  The observer sees a 2-D map with radial bins denoted by $R_{\textbf{\textit{j}}}$, while the 3-D shells have radii $r_{\textbf{\textit{i}}}$.  The radial coordinates for the 2-D bins do not necessarily match those for the 3-D shells.  $\ell_{\textbf{\textit{i}}}$ is the distance along the line of sight that radial shell $\textbf{\textit{i}}$ contributes to the 1D bin at radius $R_{\textbf{\textit{j}}}$.}
	\label{fig:sphere}
\end{figure}

For the deprojections, a concentric shell model is assumed, which assigns constant densities and temperatures within concentric 3-D shells \citep{1999AJ....117.2398M} - see Fig.~\ref{fig:sphere} for the geometry. For the individual cluster analysis, we  use 5 shells: the first shell corresponds to the approximate resolution of Bolocam, and spans the radial range 0\arcsec--30\arcsec.  At increasing radii, the shells are logarithmically spaced up to the cut-off radius, $r_{\rm max}$. This logarithmic spacing results in more uniform signal-to-noise within each shell compared to a linear spacing. Note that the azimuthally-averaged 1-D projected X-ray surface brightness bins do not need to have the same radii as the 3-D model shells. In fact, the spacing between the 1-D projected data bins is much smaller than the spacing between the 3-D shells, especially in the center of the cluster where  there are many counts.  

The X-ray surface brightness in the projected bin \textbf{\textit{j}} centered at radius $R_{\textbf{\textit{j}}}$ is modeled as:
\begin{equation}
S_{X-ray}(R_{\textbf{\textit{j}}})= \frac{1}{4\pi (1+z)^4} \sum_{\textbf{\textit{i}}=1} n_{e,\textbf{\textit{i}}}^2\Lambda(T_{e,\textbf{\textit{i}}}) \cdot l_{\textbf{\textit{i}}}(R_{\textbf{\textit{j}}})
\label{eq:XraySB}
\end{equation}
where \textbf{\textit{i}} is the index of the 3-D shell, $n_{e,\textbf{\textit{i}}}$ is the electron density within the shell, $n_{H,\textbf{\textit{i}}}$ is the hydrogen density within the shell, $T_{e,\textbf{\textit{i}}}$ is the electron temperature within the shell, and $l_{\textbf{\textit{i}}}(R_{\textbf{\textit{j}}})$ is the line of sight length of the shell. The cooling function $\Lambda(T_{e,\textbf{\textit{i}}})$ is the same as in the smooth parametric fits.
The SZ signal in the projected bin centered at radius $R_{\textbf{\textit{j}}}$ is modeled as:
\begin{equation}
S_{SZ}(R_{\textbf{\textit{j}}})= g(x) S_{CMB} \frac{k_B\sigma}{m_{e}c^{2}} \sum_{\textbf{\textit{i}}=1} n_{e,\textbf{\textit{i}}}T_{e,\textbf{\textit{i}}} \cdot l_{\textbf{\textit{i}}}(R_{\textbf{\textit{j}}}).
\label{eq:SZSB}
\end{equation}

Without modification, the concentric shell model only accounts for signal from within $r_{\rm max}$. Particularly for the SZ signal, this can lead to a bias in the results because a significant amount of the projected signal originates from line-of-sight distances beyond $r_{\rm max}$.\footnote{For example, near the cluster center approximately 3\% of the SZ signal originates beyond $r_{\rm max}$, while at a projected radius of $r = 0.5 \times r_{500}$ approximately 25\% of the SZ signal originates beyond $r_{\rm max}$. In practice, not accounting for the projected signal beyond $r_{\rm max}$ can bias the recovered temperature in the final deprojection shell by up to 25\%.} We have therefore adopted the following   iterative procedure to estimate this emission. First, we perform deprojections assuming there is no emission beyond the last shell.  We then use the resulting shell values in temperature and density to compute the corresponding squared density and pressure shell values. These are then fit by the smooth parametric curves given in \citet{2011A&A...534A.109P} (for density) and and \citet[from now on referenced as S13]{2013ApJ...768..177S} (for pressure); where the overall amplitude is allowed to vary. Once the normalization is constrained, these profiles are then integrated between $r_{\rm max}$ and $10 \times r_{500}$ to estimate the signal contributed by the line of sight beyond the last concentric shell, and new deprojections are computed after accounting for this additional signal. This process is repeated until the deprojections converge, typically after 3-4 iterations. The results are not sensitive to the exact model used to describe the smooth density and pressure profiles, and, for example, assuming either the S13, \citet{2010A&A...517A..92A}, or \citet{2013A&A...550A.131P} parameterization for the pressure results in negligible changes to the deprojected values.

Most results in this work are obtained  with the general approach presented in  \citet{2007MNRAS.382..397A}. However, this work  contains several modifications to that approach: no temperature regularization is applied, the assumption of the emission beyond the last radial shell is modified (see above), and the radial bins used to azimuthally average the X-ray surface brightness images no longer need to be at the same radii as the 3-D concentric shells.

\subsubsection{Individual Cluster Likelihood}
\label{sec:IndividualClusterLikelihood}

Both the smooth parametric profiles and the concentric shell deprojections are fit to the data using a MCMC to maximize a single joint likelihood:
\begin{equation}
\mathcal{L}=\mathcal{L}_{X-ray} \cdot \mathcal{L}_{SZ}.
\end{equation}
For the SZ data, the log-likelihood is computed according to
\begin{equation}
ln(\mathcal{L}_{SZ})=\sum_{\textbf{\textit{j}}} - \frac{1}{2} \frac{(O_{\textbf{\textit{j}}}-M_{\textbf{\textit{j}}})^{2}}{\sigma_{\textbf{\textit{j}}}^{2}}
\end{equation}
where $O_{\textbf{\textit{j}}}$ is the observed projected SZ brightness in 2-D map pixel \textbf{\textit{j}}, $M_{\textbf{\textit{j}}}$ projected brightness based on either the smooth parametric profile or the concentric shell model (after accounting for the effects of the PSF and the image filtering), and $\sigma_{\textbf{\textit{j}}}$ is the rms noise in each pixel \textbf{\textit{j}}. The noise in the SZ images is assumed to be uncorrelated between individual map pixels and to follow a Gaussian distribution, which was shown to be a good approximation in \citet{2016arXiv161205377S}.
For the X-ray data, the log-likelihood is computed according to
 \begin{equation}
ln(\mathcal{L}_{X-ray})=\sum_{\textbf{\textit{j}}} - \frac{1}{2} \frac{(O_{\textbf{\textit{j}}}-M_{\textbf{\textit{j}}})^{2}}{M_{\textbf{\textit{j}}}} - \frac{1}{2}ln(2\pi M_{\textbf{\textit{j}}}).
\end{equation}
The naming conventions are identical to the SZ log likelihood, except that the index \textbf{\textit{j}} corresponds to a projected 1-D radial bin rather than a 2-D map pixel. The extra term is needed since the variance (the model) is fit for as well, and so the normalization cannot be ignored. Although the X-ray noise follows a Poisson distribution, we approximate the likelihood with a Gaussian distribution with variance equal to the mean value of the model. By requiring each radial bin to have a sufficiently large number of counts, at least 100, this Gaussian approximation provides a good description of the noise statistics.

\subsection{Ensemble Cluster Analysis}
\label{sec:EnsembleClusterAnalysis}

\subsubsection{Joint Cluster Likelihood}
\label{JointClusterLikelihood}

Under the assumption of purely gravitational collapse, clusters' physical properties obey simple scaling relations. However, the other physical effects that impact cluster formation can produce some bias and scatter relative to these relations. So, considering the entire population of clusters in the Universe, we expect that any given physical property can be described by an average value along with a scatter relative to this average value. Any representative sample of observed clusters can then be used to constrain such universal mean values and their intrinsic scatters. Therefore, in addition to characterizing the individual profiles of the observed clusters, we utilized our data to constrain the universal mean temperature and density profiles and their scatters. In order to perform these ensemble analyses, we first rescaled the data according to the self-similar relations from \citet{2007ApJ...668....1N}, which account for the gravity-only differences in the clusters' physical properties given their masses and redshifts. The radial coordinate was scaled with respect to the value of $r_{500}$. The density was scaled according to
\begin{equation}
n_{e,500}=500 \frac{\Omega_{b}}{\Omega_{m}} \frac{\rho_{crit}(z)}{\mu_{e} m_{p}},
\label{eq:nn}
\end{equation}

where $\Omega_{m}$ is the total matter density, $\Omega_{b}$ is the baryonic density, $\rho_{crit}$ is the critical density of the universe, $\mu_{e}$ is the mean molecular weight for electrons, and $m_{p}$ is the mass of the proton. The temperature was scaled according to
\begin{equation}
T_{500}=8.71 \: keV \: \left(\frac{M_{500}}{10^{15} \, M_{\astrosun}}\right)^{2/3} \: E(z)^{2/3},
\label{eq:TT}
\end{equation}
where $E(z)=\sqrt{\Omega_{m}(1+z)^{3}+\Omega_{\Lambda}}$ represents the evolution of the Hubble parameter. The values of $r_{500}$, $M_{500}$, and $z$ are listed in Tab.~\ref{tab:clusterobsinfo}, and were taken from \citet{2013ApJ...768..177S} based on the analysis methods of \citet{2010MNRAS.406.1773M}.

Given this common rescaling, we modeled the density and temperature as constant values within each radial shell. Specifically, we assumed that within each radial shell of each cluster, the actual density and temperature values are a random realization based on a single universal mean value and its intrinsic scatter for that shell. The clusters were modeled assuming five radial shells,  so 20 parameters were included in the fit: 5 universal mean densities, 5 universal mean temperatures, 5 density intrinsic scatters, and 5 temperature intrinsic scatters. The global likelihood aims to determine  these 20 parameters  by simultaneously fitting to the SZ and X-ray data from all of the clusters within a given sample. In the  likelihood, the stochastic nature  of the cluster modeling  is implemented within the MCMC approach for the likelihood determination. Specifically, whenever the MCMC code determines the likelihood for a  given set of these 20 parameters, the model-predicted value within each radial shell for each individual cluster is randomly drawn based on a Gaussian distribution centered on the universal mean value with a standard deviation equal to the intrinsic scatter value. More explicitly, the modeling of the physical properties for cluster $\textbf{\textit{k}}$ at each iteration of the MCMC code practically  reads:
 
\begin{equation}
\mathbb{T}_{\textbf{\textit{i}},\textbf{\textit{k}}} =  \bar{\mathbb{T}_\textbf{\textit{i}}} + \textrm{ran}_\textbf{\textit{k}}(\sigma_{\mathbb{T}\textbf{\textit{i}}})
\label{eq:temp}
\end{equation}
and
\begin{equation}
\mathbb{N}_{\textbf{\textit{i}},\textbf{\textit{k}}}=  \bar{\mathbb{N}_\textbf{\textit{i}}} + \textrm{ran}_\textbf{\textit{k}}(\sigma_{\mathbb{N}\textbf{\textit{i}}})
\label{eq:dens}
\end{equation}
where $\textbf{\textit{i}}$ is the radial shell, $\textbf{\textit{k}}$ represents the cluster, $\bar{\mathbb{T}_\textbf{\textit{i}}}$ and $\bar{\mathbb{N}_\textbf{\textit{i}}}$ are the universal  mean values of the temperature and density in radial shell $\textbf{\textit{i}}$ (in units of $T_{500}$ and $n_{e,500}$), and $\textrm{ran}_\textbf{\textit{k}}(\sigma_{\mathbb{T}\textbf{\textit{i}}})$ and $\textrm{ran}_\textbf{\textit{k}}(\sigma_{\mathbb{N}\textbf{\textit{i}}})$ indicate random draws for each cluster $k$ from Gaussian distributions with the given standard deviation for that MCMC draw (e.g., $\sigma_{\mathbb{T}\textbf{\textit{i}}}$). 

The joint likelihood of all clusters using both the SZ and X-ray observations reads:
\begin{equation}
\begin{split}
ln(\mathcal{L}) &= \sum_{\textbf{\textit{k}}} \left( \sum_{\textbf{\textit{j}}} -\frac{1}{2} \frac{(O_{SZ,\textbf{\textit{j,k}}}-M_{SZ,\textbf{\textit{j,k}}})^{2}}{\sigma_{SZ,\textbf{\textit{j,k}}}^2} \right. \\
  & - \left. \sum_{\textbf{\textit{i}}} \frac{1}{2} \frac{(O_{Xray,\textbf{\textit{i,k}}}-M_{Xray,\textbf{\textit{i,k}}})^{2}}{M_{Xray,\textbf{\textit{i,k}}}} -\frac{1}{2}ln(2\pi M_{Xray,\textbf{\textit{i,k}}}) \right)
\end{split}
\end{equation}
where $\textbf{\textit{k}}$ again indicates a given cluster, $\textbf{\textit{j}}$ indicates a single SZ map pixel, and $\textbf{\textit{i}}$ indicates a single projected radial shell for the X--ray data. $O$ denotes the observed SZ or X--ray brightness, and $M$ gives the model-predicted SZ or X-ray brightness based on the temperature and density from Eqns.~\ref{eq:temp} and \ref{eq:dens} and the relations provided in Eqns.~\ref{eq:XraySB}, \ref{eq:SZSB}, \ref{eq:nn}, and \ref{eq:TT}. Given Eqns.~\ref{eq:temp} and \ref{eq:dens}, $M$ naturally contain information about the universal intrinsic scatters in temperature and density. Note  that, although the likelihood is assumed to be Gaussian in the observables, it is not Gaussian in the parameters (the X--ray emission  has a quadratic dependence on density and a  non--linear dependence on temperature while the SZ brightness is proportional to the product of density and temperature). The weight of each individual cluster in the  global likelihood reflects the quality of the observational data, and clusters with higher SZ noise $\sigma_{SZ}$ and/or fewer X-ray source photons may have very little influence on the overall constraints. Indeed, for our sample, the total weighting factors vary from cluster to cluster by approximately two orders of magnitude. Furthermore, in determining the uncertainties on the intrinsic scatter values, this expression correctly accounts for the finite cluster sample size. Finally, uncertainties in the X-ray measured values of $M_{500}$ and $r_{500}$ could easily be accounted for in the likelihood expression, although they were not included in the present analysis.

We implemented this global likelihood in an MCMC,  and analyzed CC clusters separately from NCC ones due to the expected difference in behavior of the density and temperature profiles in the clusters' centers. This split also reduced the number of clusters in a given MCMC from 45 to $\sim 20$, and allowed the fits to converge in a reasonable amount of time, several weeks, with our available computing resources. For these fits, the only quantities that are meaningfully constrained are the ensemble mean density profiles and their intrinsic scatter, which are reported in Sec.~\ref{sec:EnsembleResults}. The ensemble mean temperature profiles and their intrinsic scatter have large uncertainties and are therefore poorly constrained. This cause of these poor constraints is not fully understood, but we suspect it may be related to the relatively small samples sizes used in our MCMC fits combined with the large differences in cluster weighting factors, which effectively reduce the sample sizes even further.

This approach, while it has a clean interpretation and full retention of information in the data, did not produce satisfactory results for the temperature analysis. Furthermore, it was not computationally tractable to apply this approach to the full cluster sample. Therefore, an alternative method was also adopted, based on a meta--analysis of the individual  clusters' deprojections, and this approach is described below and referred to as the ``meta-analysis'' method.

\subsubsection{Meta-Analysis Based on an Ensemble Fit to the Individual Cluster Deprojections}
\label{sec:Ensemblefitondeprojections}

For the meta-analysis method to obtain ensemble-average constraints on the density and temperature profiles, we first compute concentric shell deprojections for all of the clusters in a set of five logarithmically spaced radial shells, scaled according to $r_{500}$ and extending to $r_{\rm max} = 1.25 \times r_{500}$. The innermost shell extends to $0.15 \times r_{500}$, which is larger than the Bolocam PSF for 34/45 of the BOXSZ clusters. 

Due to the differing signal-to-noise of the X-ray maps, not all of the clusters were deprojected in all five shells. Out of the BOXSZ sample of 45 clusters, 17 were deprojected in all 5 shells, 27 were deprojected in the four innermost shells, and 1 cluster (RX J2129.6+005) was deprojected in the three innermost shells.  

Once the scaled individual deprojections for the clusters were obtained at the set of common scaled radii, they were combined to obtain ensemble mean profiles, along with the intrinsic scatters about these mean profiles, using the Gaussian process formalism described in \citet{2013ApJ...768..177S}. In this approach, the binned density and temperature profiles of the  clusters  were assumed to be Gaussian distributed around a   mean value, with  a width that includes  the individual clusters' covariances (as computed from the  deprojections) and an  intrinsic scatter for the ensemble.  The log-likelihood used to constrain the model is
\begin{equation}
ln\mathcal{L}=\sum_{\textbf{\textit{k}}} -\frac{1}{2}\left[\tilde{x}_{\textbf{\textit{k}}}^{T}(S+U_{\textbf{\textit{k}}})^{-1}\tilde{x}_{\textbf{\textit{k}}} + ln | S + U_{\textbf{\textit{k}}} | \right]
\end{equation}
where $\tilde{x}_{\textbf{\textit{k}}}$ is a vector containing ten elements (i.e., one element for each of the five radial shells for both the density and the temperature), with $\tilde{x}_{\textbf{\textit{k}}} = \overline{x}-x_{\textbf{\textit{k}}}$ for the ensemble mean profile $\overline{x}$ and individual cluster profile $x_{\textbf{\textit{k}}}$, \textbf{\textit{k}} is the cluster index (e.g., ranging from 1 to 45 for the full BOXSZ sample), $U_{\textbf{\textit{k}}}$ is the covariance matrix for the individual cluster deprojection (which can be treated as a 'measurement' covariance matrix), and $S$ is the intrinsic scatter matrix. The parameters returned from the fit are the mean profile $\overline{x}$ and the intrinsic scatter about this profile $S$. Note that, as with $\tilde{x}_{\textbf{\textit{k}}}$, both $U_{\textbf{\textit{k}}}$ and $S$ include terms associated with both the density and the temperature. Note also that the intrinsic scatter matrix $S$ introduced here does not, in general, have a direct mapping into the Gaussian errors introduced in  eq.~\ref{eq:dens}. The two procedures are different, and the scatters are not necessarily the same. Finally, we have assumed the intrinsic scatters to be Gaussian, rather than log-Gaussian. This choice was motivated by the shape of the measured distributions for $U_{\textbf{\textit{k}}}$ along with the fact that noise fluctuations in the SZ data can sometimes produce negative values of $\tilde{x}_{\textbf{\textit{k}}}$ which are incompatible with the assumption of a log-Gaussian distribution.

\begin{figure}
	\centering
	\includegraphics[width=\columnwidth]{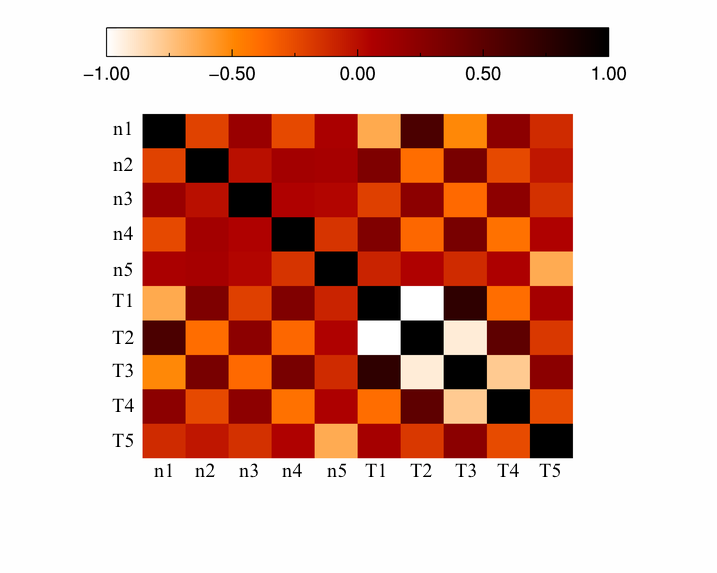}
	\caption{A typical correlation matrix $U_{\textbf{\textit{k}}}$ of a concentric shell deprojection (the cluster is MACS J1423.8+2404).  $n_{i}$ denotes the $i$'th density shell while $T_{i}$ denotes the $i$'th temperature shell.}
	\label{fig:covar_a697}
\end{figure}

Fig.~\ref{fig:covar_a697} shows a typical example of correlations among the elements in the covariance matrix $U_{\textbf{\textit{k}}}$.   There are significant anticorrelations between adjacent temperatures.  In particular, the two inner shells have a correlation close to $-1$.  As expected there are also anticorrelations between densities and temperatures of the same shell. Furthermore, there are positive correlations between temperature values separated by two shells.

Given the quality/quantity of the observational data, it was not possible to constrain all values of the $S$ matrix.
To minimize the number of free parameters, we assume a diagonal-only $S$ matrix, so there are no covariances between the intrinsic scatter values. This results in fitting for 5 mean densities, 5 mean temperatures, 5 density intrinsic scatter elements, and 5 temperature intrinsic scatter elements, for a total of 20 parameters. 

The procedure was repeated to fit for the ensemble-average pressure profiles.  First, the MCMC chains containing the scaled densities and temperatures were converted to scaled pressures according to $P(P_{500}) \propto n(n_{e,500}) T(T_{500})$.  Explicitly, the pressure scaling is:
\begin{equation}
P_{500}=3.68 \times 10^{-3} \: \frac{keV}{cm^{-3}} \: \left(\frac{M_{500}}{10^{15} \, M_{\astrosun}}\right)^{2/3} \: E(z)^{8/3},
\end{equation}
Individual cluster pressure deprojections and their covariance matrices were then calculated from these chains. The meta-analysis procedure described above was then used to determine the ensemble-average pressure profile and its intrinsic scatter based on these individual deprojections.

\section{Analysis of Mock Cluster Observations}
\label{sec:AnalysisofMockClusterObservations}

In order to search for potential biases in our fitting procedures, we performed the complete analysis on a set of mock observations of smooth cluster models generated from a set of known parameters. Although the smooth models used to generate the clusters in the mock observations do not capture the full complexity of real clusters, they were chosen because they allow for simpler comparisons between the input cluster parameters and those generated by our fitting procedures.

\subsection{Individual Cluster Analysis}
\label{sec:IndividualClusterAnalysis}

A set of four clusters was constructed to test our individual cluster fitting routines. These clusters were generated using either a single- or double-beta density profile and a constant temperature profile. The parameters used to produce the test clusters were chosen based on the characteristics of a representative set of real clusters from the BOXSZ sample: Abell 1835, MACS J0417.5-1154, MS 2053.7-0449, and MACS J0744.8+3927, denoted as clusters A, B, C, and D, respectively, and are given in Tab.~\ref{tab:mock_a_cluster_parameters}. Based on the fits to the real clusters, clusters A, B, and D were generated using double-beta density profiles while cluster C was generated using a single-beta density profile.

Given the input density and temperature profiles, along with the exposure times ($\sim 50$ ksec) and effective areas of the real images, mock X-ray observations were produced to match the \chandra\ ACIS-I FOV and angular resolution \citep{2004MNRAS.351..505G,2008ApJ...674..728R}.  The mock maps did not include any background signal.  An analogous procedure was used to generate the mock SZ observations, such that they match the noise properties, image size, and angular resolution of the real Bolocam observations. In addition, the image filtering of the Bolocam data processing was also applied to the cluster in the mock observation. For each of the four test clusters, 100 mock X-ray and SZ observations were produced, each with a different random realization of the noise.

\begin{table*}
\centering
\caption{Cluster Models used for the Mock Observations}\label{tab:mock_a_cluster_parameters}
\begin{tabular}{*{10}{c}} 
\hline
	Cluster & z & Real Counterpart & $M_{2500}$  & T  & $n_{e0,1}$  & $r_{c,1}$  & $n_{e0,2}$ & $r_{c,2}$ & $\beta$ \  \\ 
	 & & & ($10^{14} M_{\astrosun}$) & (keV) & ($10^{-2} cm^{-3}$) & (kpc) & ($10^{-2} cm^{-3}$) &  (kpc) & \\ \hline
	A & 0.25 & Abell 1835 & 5.11 & 7.0 & 23.48 & \hphantom{1}43.0 & 2.13 & 190.0 & 0.70 \\ 
	B & 0.44 & J0417.5-1154 & 9.50 & 9.5 & \hphantom{1}6.80 & \hphantom{1}55.7 & 0.52 & 454.8 & 0.86 \\ 
	C & 0.58 & MS 2053.7-0449 & 0.59 & 5.0 & \hphantom{1}0.90 & 105.3 & - & - & 0.61 \\ 
	D & 0.69 & MACS J0744.8+3927 & 3.50 & 8.0 & \hphantom{1}6.20 & \hphantom{1}42.6 & 1.05 & 184.7 & 0.68 \\ 
\bottomrule
\end{tabular}
\begin{flushleft}
The input parameters used to generate the smooth cluster models for the mock observations.  Each of these four test clusters was based on the fit to a real cluster in the BOXSZ sample. In all cases, the cluster was assumed to be isothermal. The density profiles for clusters A, B, and D were based on the double-beta model, while the density profile for cluster C was based on the single-beta model.
\end{flushleft}
\end{table*}

We then applied our MCMC fitting code to the X-ray/SZ pair of mock observations for each given cluster type in order to obtain single-cluster parametric fits. The smooth parametric profiles assumed for the fits matched those used to generate the clusters (i.e., isothermal with a single or double-beta profile for the density). Due to degeneracies between the fitted parameters, the noise fluctuations in the different mock observations often produce results with different fit parameters yet similar profile shapes. Therefore, rather than comparing the fitted parameter values with the inputs used to generate the test clusters, we instead compared the actual shapes of the fitted profiles. Using the output of the MCMC, we thus plotted the best-fit density for each of the 100 mock observation pairs at a set of closely spaced radii spanning the approximate range constrained by the data. Then, at each radius the overall median fitted density values, along with the inner 68\% of the fitted density values, were computed. This output was then  compared with the input profile used to generate the cluster. 

\begin{figure*}
\centering
\subfloat[]{\includegraphics[width=3.5in]{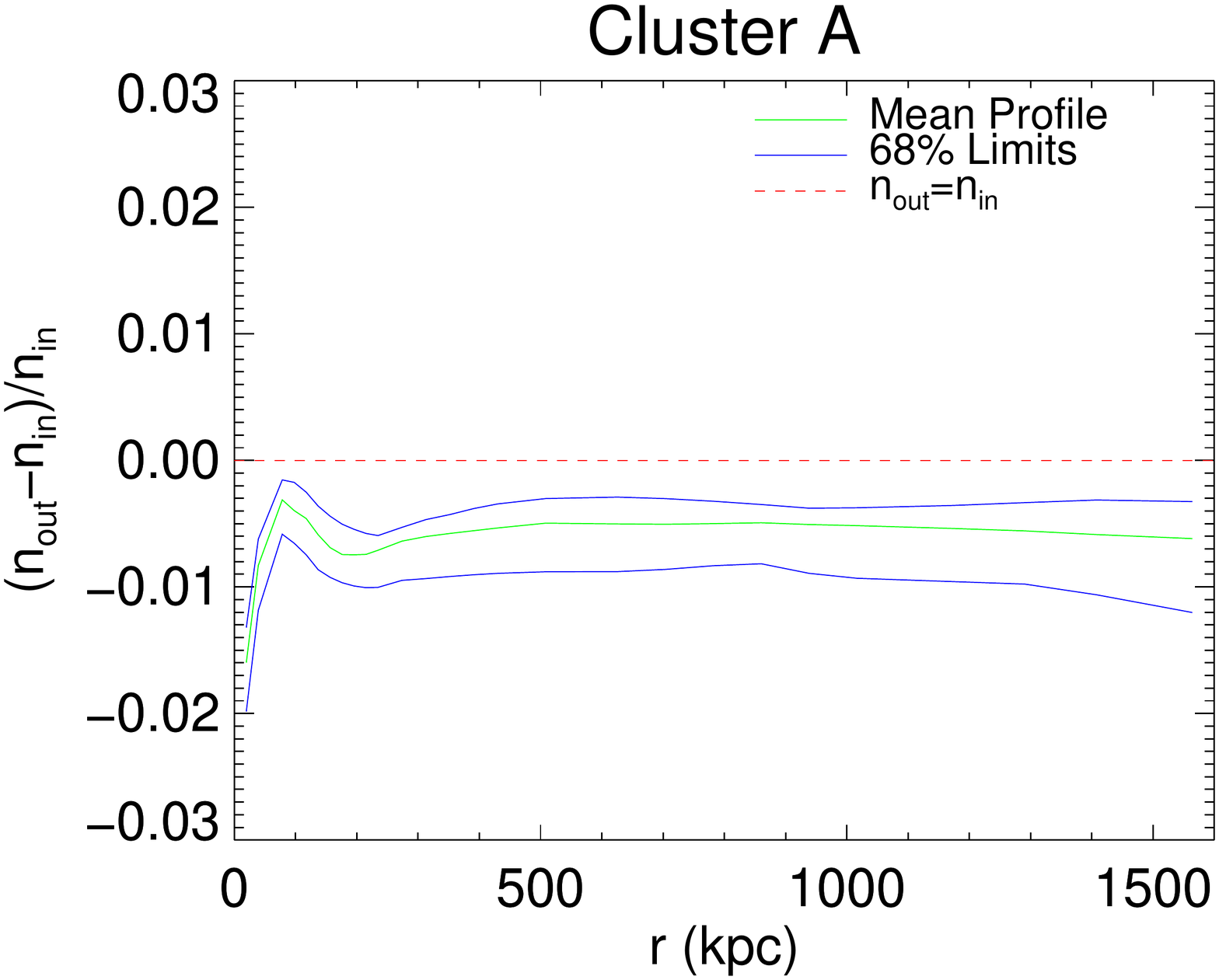}}
\subfloat[]{\includegraphics[width = 3.5in]{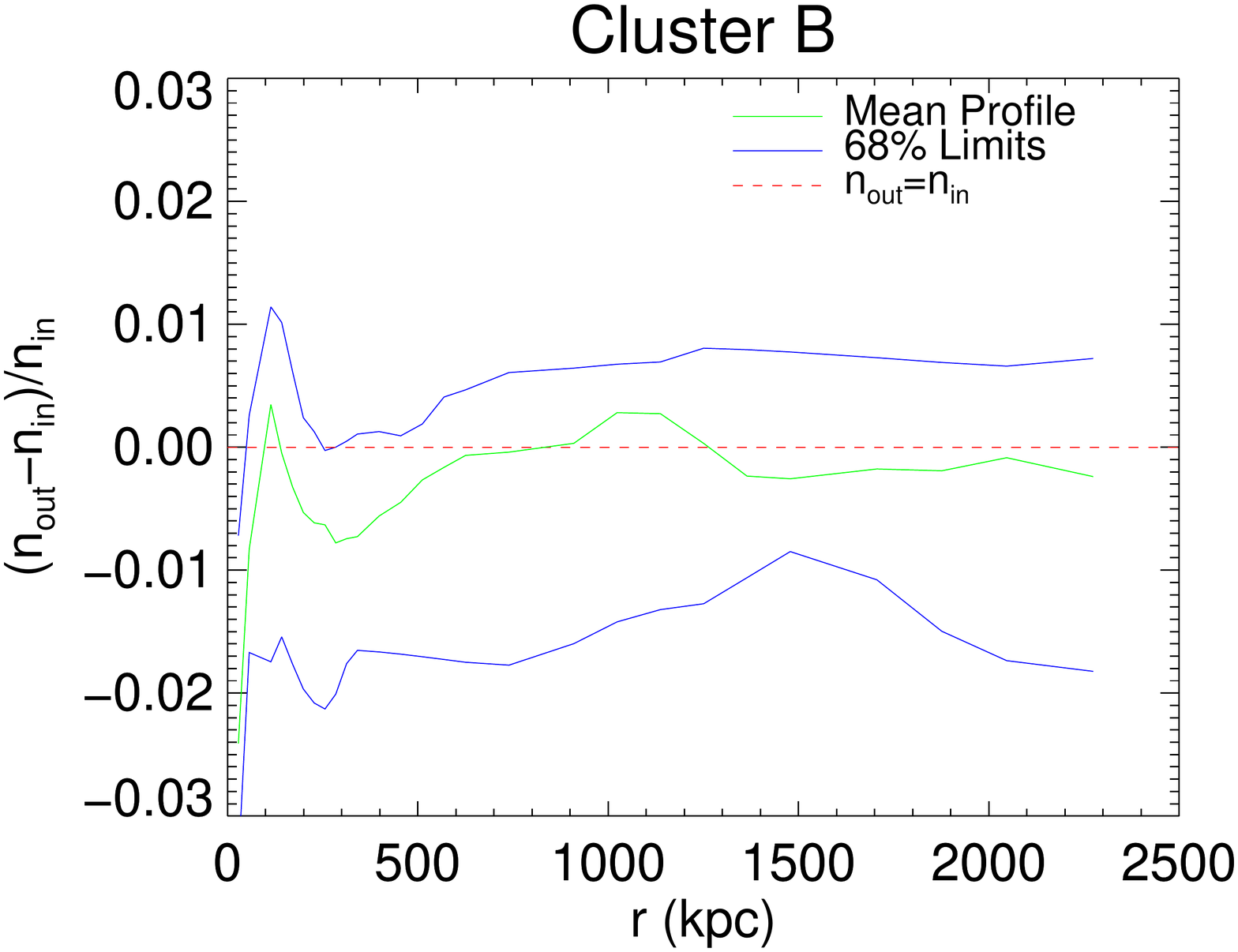}}\    \subfloat[]{\includegraphics[width = 3.5in]{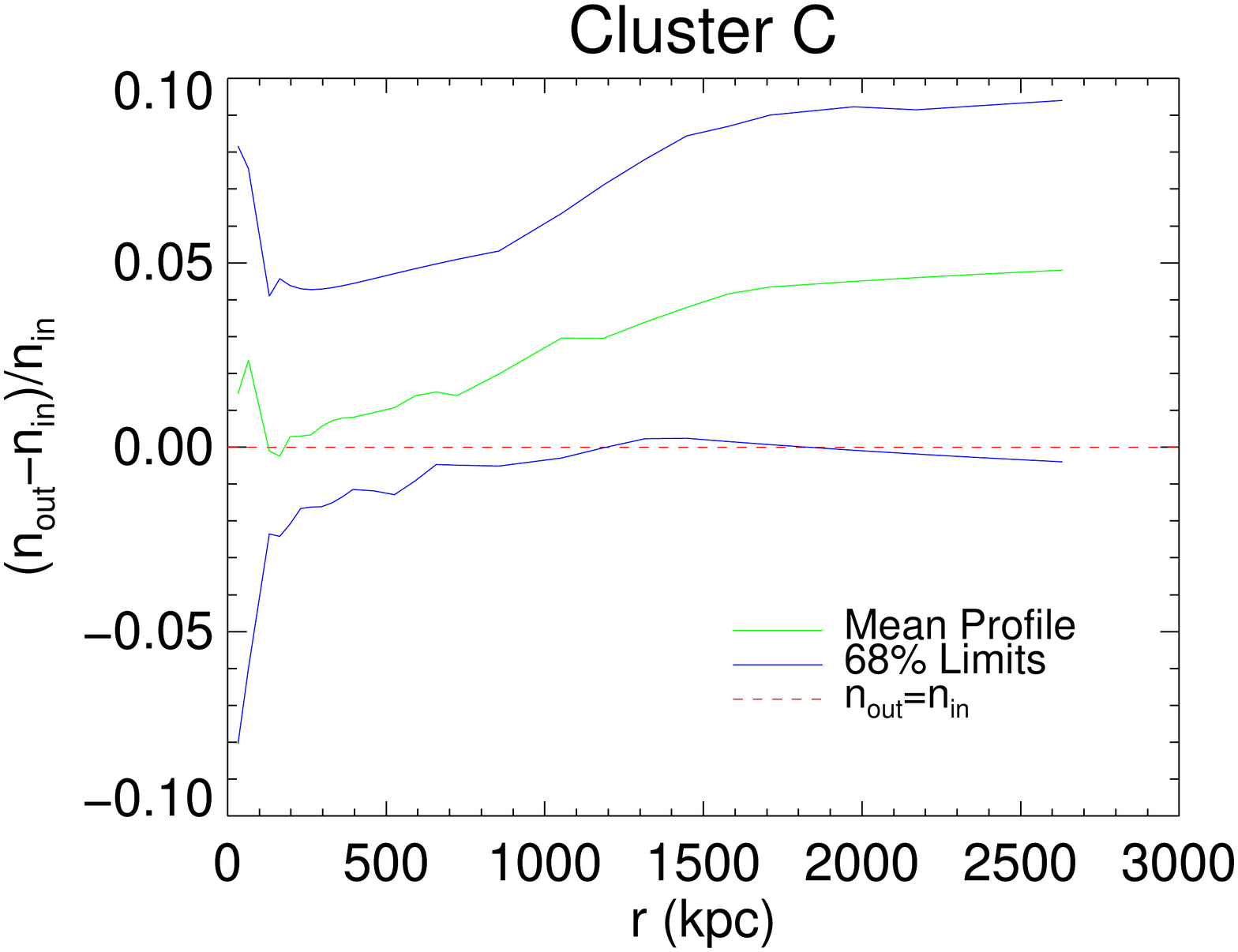}}
\subfloat[]{\includegraphics[width = 3.5in]{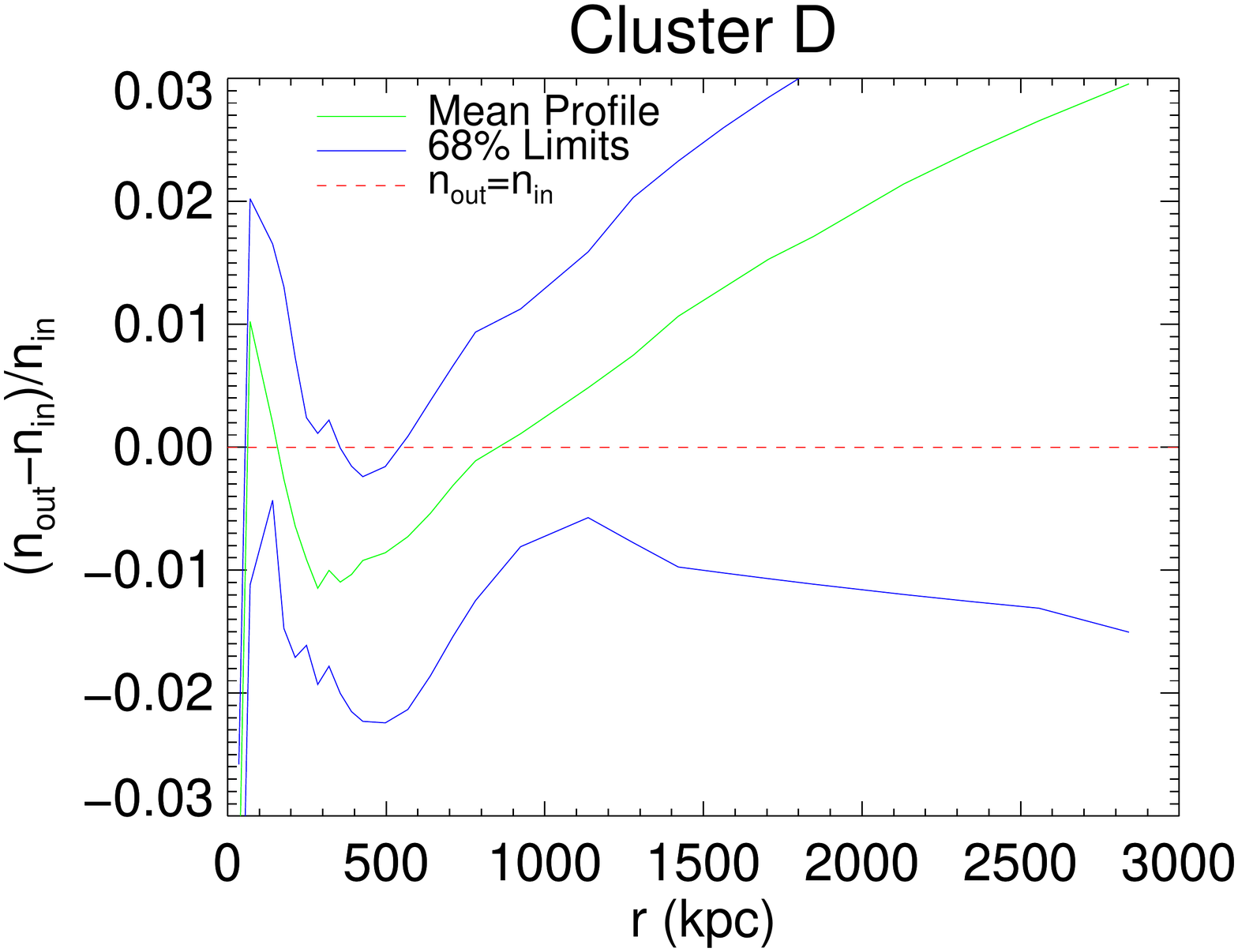}} 
\caption{Normalized difference between the input deprojected density profile used to generate the test clusters and the fitted deprojected density profile recovered from the mock observations. Note the y-axis scale is larger for Cluster C than for the others.  The green line represents the mean fitted profile, while the blue limits represent the upper and lower 68\% limits for the fitted density at that particular radius.}
\label{fig:analytical_cluster_sims}
\end{figure*}

Fig.~\ref{fig:analytical_cluster_sims}  shows the fractional difference between the fitted and input density profiles computed using the above procedure. The span encompassing 68\% of the recovered density values is different for each of the four test clusters, due to the clusters' varying distances, masses,  noise properties, and the number of parameters that were fit for. The only statistically significant bias appears in cluster A, where the profile is on average 2$\sigma$ lower than the input profile. However, the absolute magnitude of this bias is quite small ($<1$\%), and it appears to be random between the four test clusters (i.e., the direction of the bias is not constant). In addition, other sources of error, such as calibration and measurement noise, are $\simeq5$--10\%, significantly larger than this potential bias in our reconstructions. There are further indications of a possible bias near the extreme centers of clusters A, B, and D, although this is likely an artifact of the minimum radial width allowed when creating the binned projected profiles ($5''$). In other words, this bias is outside of the region nominally constrained by our fitting procedure. This test therefore demonstrates that our fitting method recovers individual cluster density profiles with a bias that is negligible compared to our measurement noise and calibration uncertainties.

For each X-ray/SZ pair of mock observations, a single isothermal value for the temperature was also fitted for. For each test cluster, the 100 resulting temperature values were used to determine the mean and standard deviation of the fitted values, and the results are presented in Tab.~\ref{tab:temp_analytical_sim_results}. Overall, the fitted temperatures agree reasonably well with the input values, and the only statistically significant difference is found in cluster A at $\simeq 3\sigma$. However, as with the slight bias in the recovered density profile for this cluster, the absolute magnitude of the bias is small compared to other sources of error in our analysis.

Based on the above results, we also infer that the profiles obtained from the concentric shell deprojections do not contain a significant bias. The statistical uncertainties on the deprojections are larger than those for the smooth parametric fits, and so the bias would need to be significantly larger to noticeably impact those fits. Further supporting this conclusion, as described in the following section, the ensemble deprojection analysis of the mock cluster observations results in profile shapes that are unbiased with respect to the input values.

\begin{table}
\centering
\caption{Cluster Temperatures for the Mock Observations}\label{tab:temp_analytical_sim_results}
\begin{tabular}{*{4}{c}} 
\hline
	Cluster & Input T (keV) & Output T (keV) & Input Scatter\  \\ \hline
	A & 7.0 & 6.43 $\pm$ 0.16 & 0.25 \\ 
	B & 9.5 & 9.35 $\pm$ 0.13 & 0.25 \\ 
	C & 5.0 & 4.69 $\pm$ 0.21 & 0.30 \\ 
	D & 8.0 & 7.93 $\pm$ 0.12 & 0.25 \\ 
\bottomrule
\end{tabular}
\begin{flushleft}
The input temperatures used to generate the test clusters, along with the fitted temperatures recovered from the mock observations of those clusters (individual cluster fitting, see Sec.~\ref{sec:IndividualClusterAnalysis}). The error bars denote the uncertainty on the mean recovered temperature from 100 mock observations. The last column reports the fractional input scatter for the ensemble cluster temperature profiles in Sec.~\ref{sec:EnsembleClusterAnalysissim}. 
\end{flushleft}
\end{table}

 \subsection{Ensemble Cluster Analysis}
  \label{sec:EnsembleClusterAnalysissim}
 
In addition to analyzing mock observations of individual clusters, we also fit for mean profiles, and the intrinsic scatters about these mean profiles, using mock observations of larger cluster samples. This procedure allowed us to search for potential biases in the ensemble fits presented in Sec.~\ref{sec:Ensemblefitondeprojections}. For this test, we produced 4 sets of 40 mock cluster observations, with each set thus containing a similar number of clusters to the full BOXSZ sample. Within each of these 4 sets, all 40 of the test clusters were generated using the parameters for a single cluster from Tab.~\ref{tab:mock_a_cluster_parameters} (e.g., set one was based on cluster A, set two on cluster B, etc.) Within a given set, all 40 of the test clusters were generated based on the same density profile, but the isothermal temperature of each cluster was scattered about its nominal value with an rms equal to the value given in Tab.~\ref{tab:temp_analytical_sim_results}. 

\begin{figure}
	\centering
	\includegraphics[width=\columnwidth]{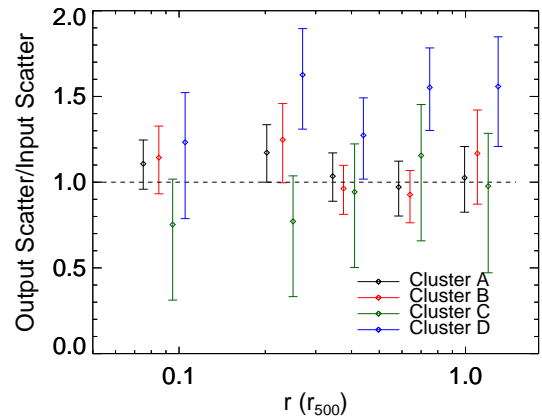}
	\caption{ Output intrinsic temperature  scatters with errors for the four mock cluster samples. Since the input scatter used for Cluster C was different from the other three test clusters, the recovered scatters are plotted relative to the input values.}
	\label{fig:intrins_mock}
\end{figure}

Individual deprojections were then computed from each mock observation in each sample. Next, the joint fitting code 
presented in sec~\ref{sec:Ensemblefitondeprojections}
 was applied to those 40 deprojections to obtain the fitted values of the mean profiles and the intrinsic scatters about those profiles, with the results shown in Fig.~\ref{fig:intrins_mock}. Although the test clusters are isothermal, and therefore the scatter is completely correlated between all of the deprojection shells, we performed the fits using our baseline procedure which sets all of the off-diagonal scatter values to 0. Even with this simplifying assumption, the fitted intrinsic scatter values, along with the mean profile shapes, are consistent with the input values used to generate the test clusters, implying there is no significant bias in our procedure. There are hints of a bias in the sample constructed based on cluster D, where 3/5 deprojection shells have fitted scatters $\simeq 2\sigma$ higher than the input values. However, given that this is the only set to show any such hints, we suspect the possible bias is more likely due to random noise fluctuations and/or something specific to the properties of cluster D. We therefore conclude that our ensemble fits do not contain any significant bias.

\section{Results} 
\label{sec:Results}
\subsection{Individual Cluster Results}
\label{sec:IndividualClusterResults}

The reader should refer to the Appendix for the parameters obtained from the smooth parametric fits (Tab.~\ref{analytical_fit_parameters}) and the concentric shell models  (Tab.~\ref{dpj_fit_params}).  The individual cluster profiles are plotted in the Appendix along with descriptions and previously published results that exists for each cluster. The individual cluster results may be summarized as follows: 
 
\begin{itemize}
\item 
We are able to fit for density and temperature profiles to or beyond $r_{500}$ in 34/45 clusters. For the clusters in common with the ACCEPT spectroscopic X-ray analysis, our results extend to larger radii in all cases, typically by a factor of $\simeq 2$ (see the plots in the Appendix).  
\item 
For the clusters that have an overlap with the ACCEPT X-ray sample, 30/36 show good agreement in the profiles obtained by the two analyses. Of the 6 that show some inconsistencies, one has a known error in the ACCEPT sample (Abell 611, M. Donahue, priv. comm.). The other five clusters (MACS J1115.8+0129, MACS J2211.7-0349, MACSJ2214.9-1359, MACS J0257.1-2325, MACS J0454.1-0300) have density and temperature profiles that are systematically higher in our analysis than the ACCEPT analysis. However, our results for these clusters are consistent with those obtained in several other works \citep{2006ApJ...647...25B,2016A&A...590A.126A,2016MNRAS.456.4020M} that are based on the same X-ray data used by ACCEPT (although some contain more recent additional observations). Of these clusters, MACSJ2211.7-0349 is the only cluster where there is no independent analysis to compare to.
\item  
The PTEs of the concentric shell deprojections are generally very low, $<0.001$, indicating a poor fit quality.  The assumption of flat densities within the radial shell bounds (which are at least $30''$ wide in radius) is not a good representation of the projected X-ray data which has been binned to a much finer resolution ($\sim 5''$ in width). The PTEs of the smooth parametric fits are also very low on average. This is again driven mainly by the X-ray data, which are of sufficient quality to detect sub-structures and other features that are not well modeled by a smooth profile.
\item 
Based on the parametric fits, the data show $>3\sigma$ evidence of a cool core in five clusters (29\% of the cool-core subset of 17 clusters defined based on the X-ray luminosity ratio described in Sec.~\ref{sec:ClusterSample}): Abell 2204, MS 2137, MACS J0417.5-1154, MACS J1347.5-1144, and MACS J1423.8+2404. The significance of the cool core was determined based on the difference of $T_{0}$ and $T_{\rm min}$, with the uncertainty on this difference determined using the full covariance matrix between the two parameters. The relatively low detection rate was driven by a combination of measurement noise and coarse angular resolution in the SZ data, the latter of which is insufficient to resolve a potential cool core for many of the BOXSZ clusters.
\item  
Several X-ray spectroscopic analyses indicate that cluster temperature profiles generally peak near $0.3 \times r_{500}$ and then decrease at larger radii \citep[e.g.,][]{2006ApJ...640..691V,2008A&A...486..359L,2010A&A...517A..92A}. The combined X-ray/SZ analysis of \citet{2013A&A...551A..22E} shows a similar trend. Based on our smooth parametric fits, ten clusters show $>3\sigma$ evidence for a temperature decrease at large radii: Abell 2204, Abell 209, Abell 1423, Abell 963, Abell 2261, Abell 2219, Abell 697, MS 2137, MACS J1720.3+3536, and MACS J0717.5+3745. One cluster, Abell 1835 shows $>3\sigma$ evidence for an increase in the temperature profile at large radii. This implies that most of the clusters are consistent with an isothermal profile, which is expected based on the signal-to-noise ratio obtained in our smooth parametric fits at large radii. Specifically, previous results indicate a typical temperature drop of $\simeq 3$--5~keV between the peak and $r_{500}$. The noise uncertainties on our smooth parametric profile fits over that radial range are $\simeq 1$--2~keV, thus implying that only a small fraction of our sample is likely to show a temperature drop with a statistical significance of $>3\sigma$.
\end{itemize}

\subsection{Ensemble Results}
\label{sec:EnsembleResults}
\begin{figure*}
\centering
\subfloat[]{\includegraphics[width=2.5in]{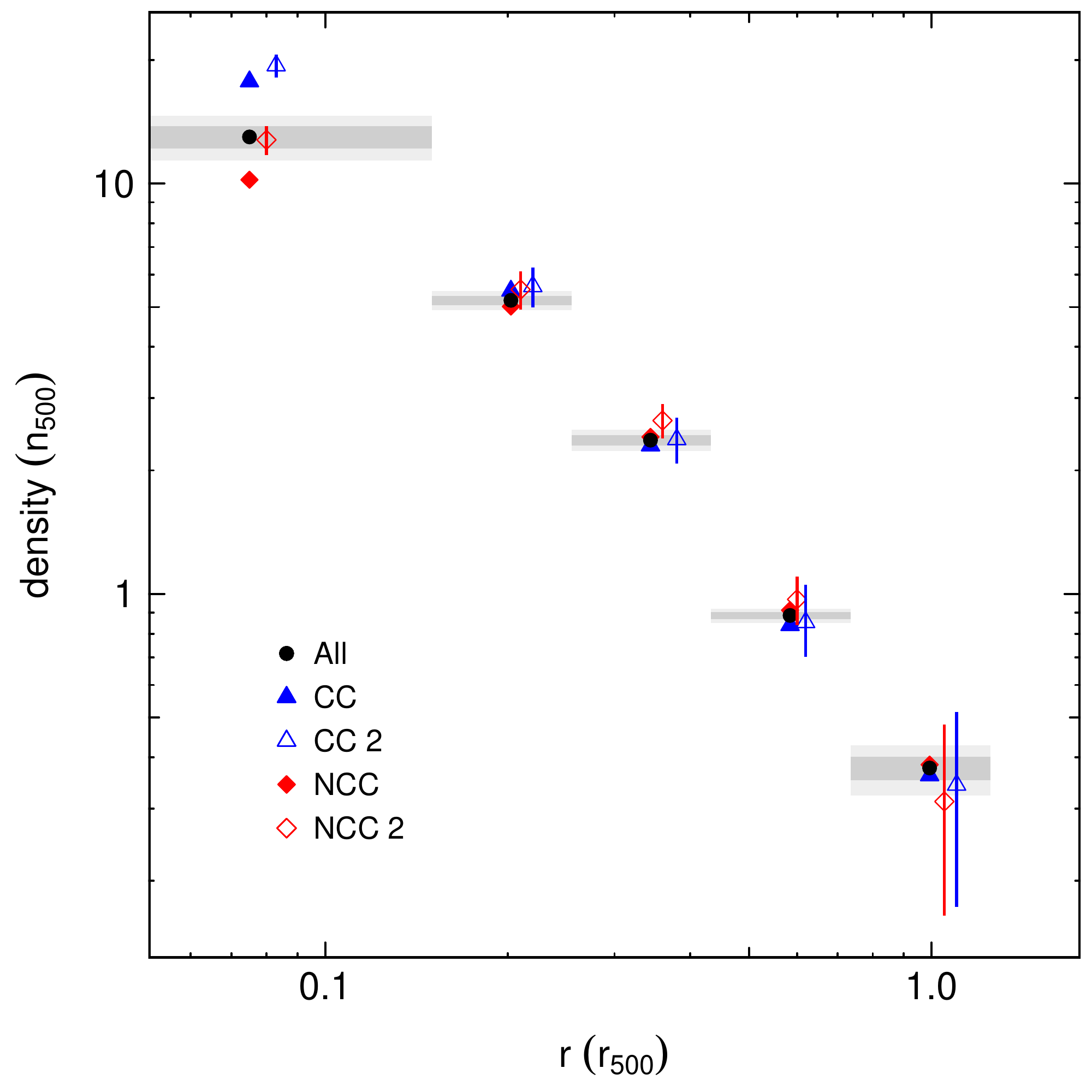}}
\subfloat[]{\includegraphics[width = 2.5in]{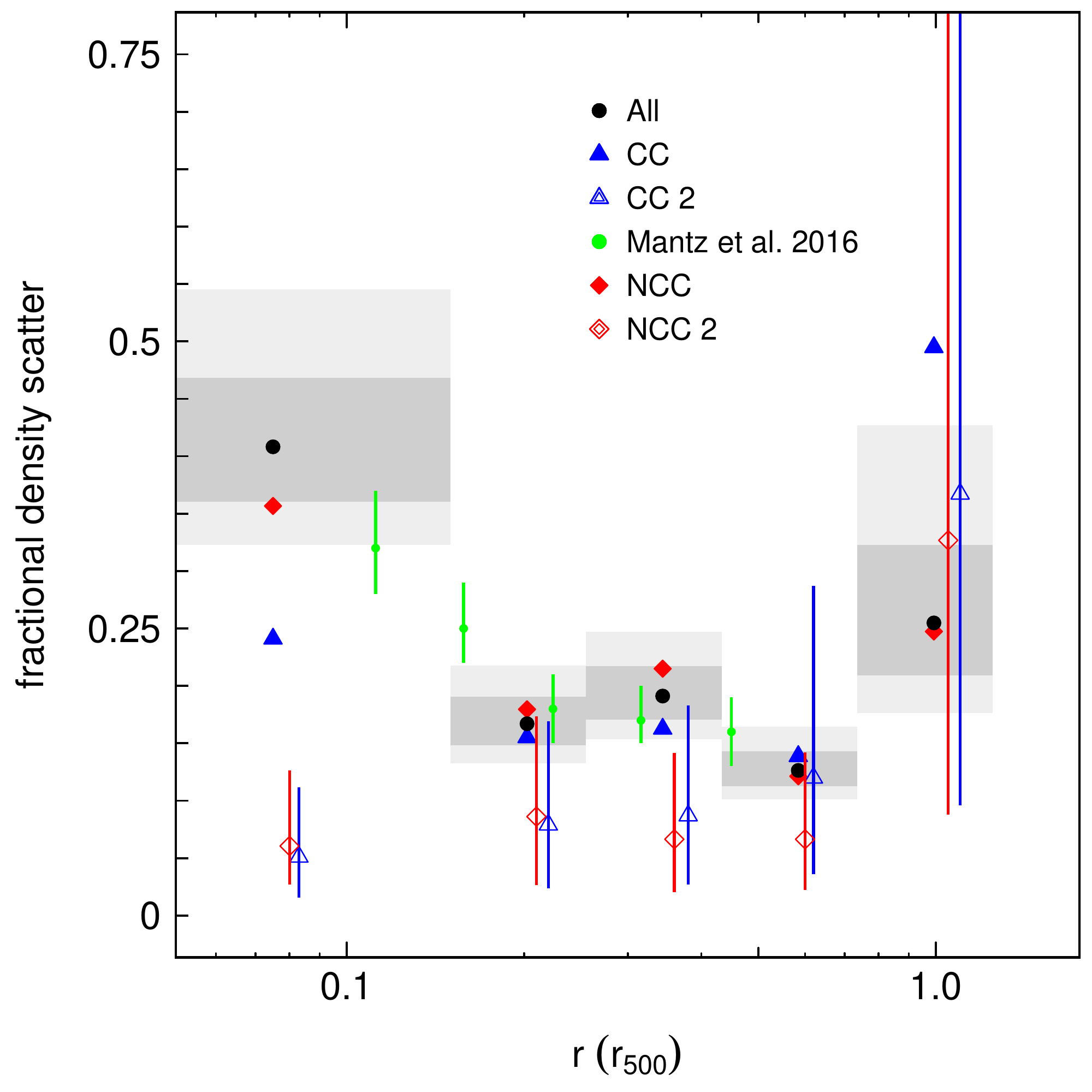}}\    \subfloat[]{\includegraphics[width = 2.5in]{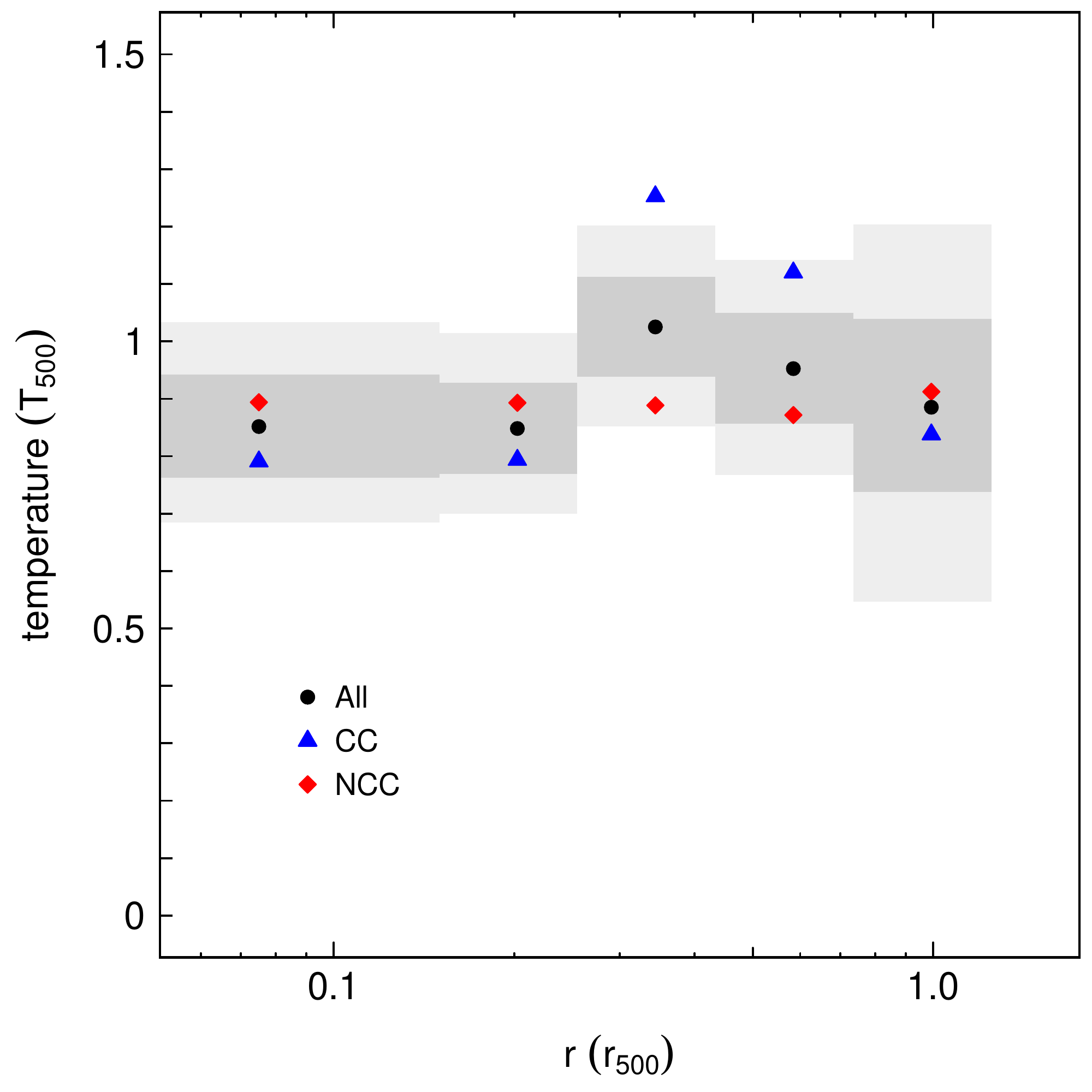}}
\subfloat[]{\includegraphics[width = 2.5in]{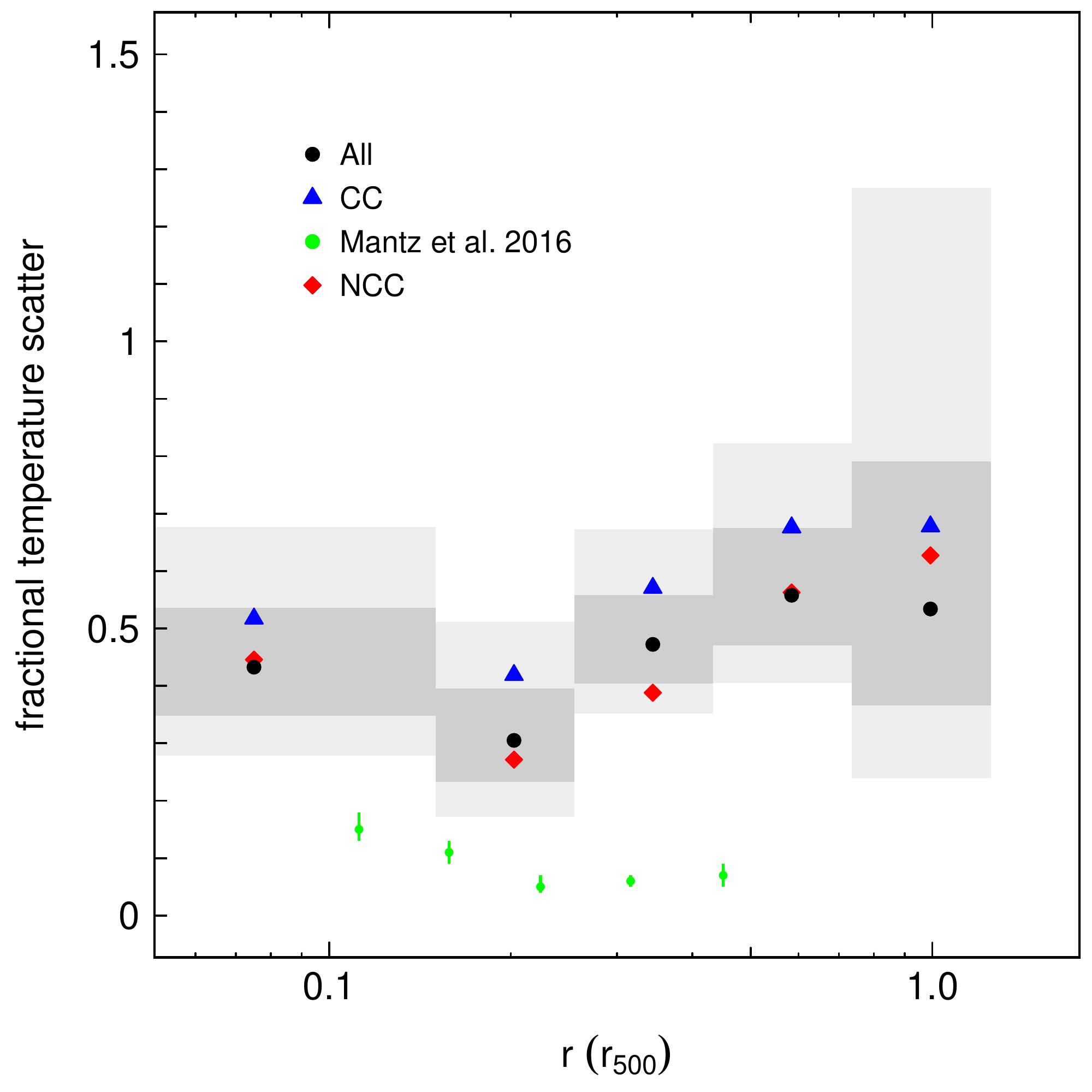}}\    \subfloat[]{\includegraphics[width = 2.5in]{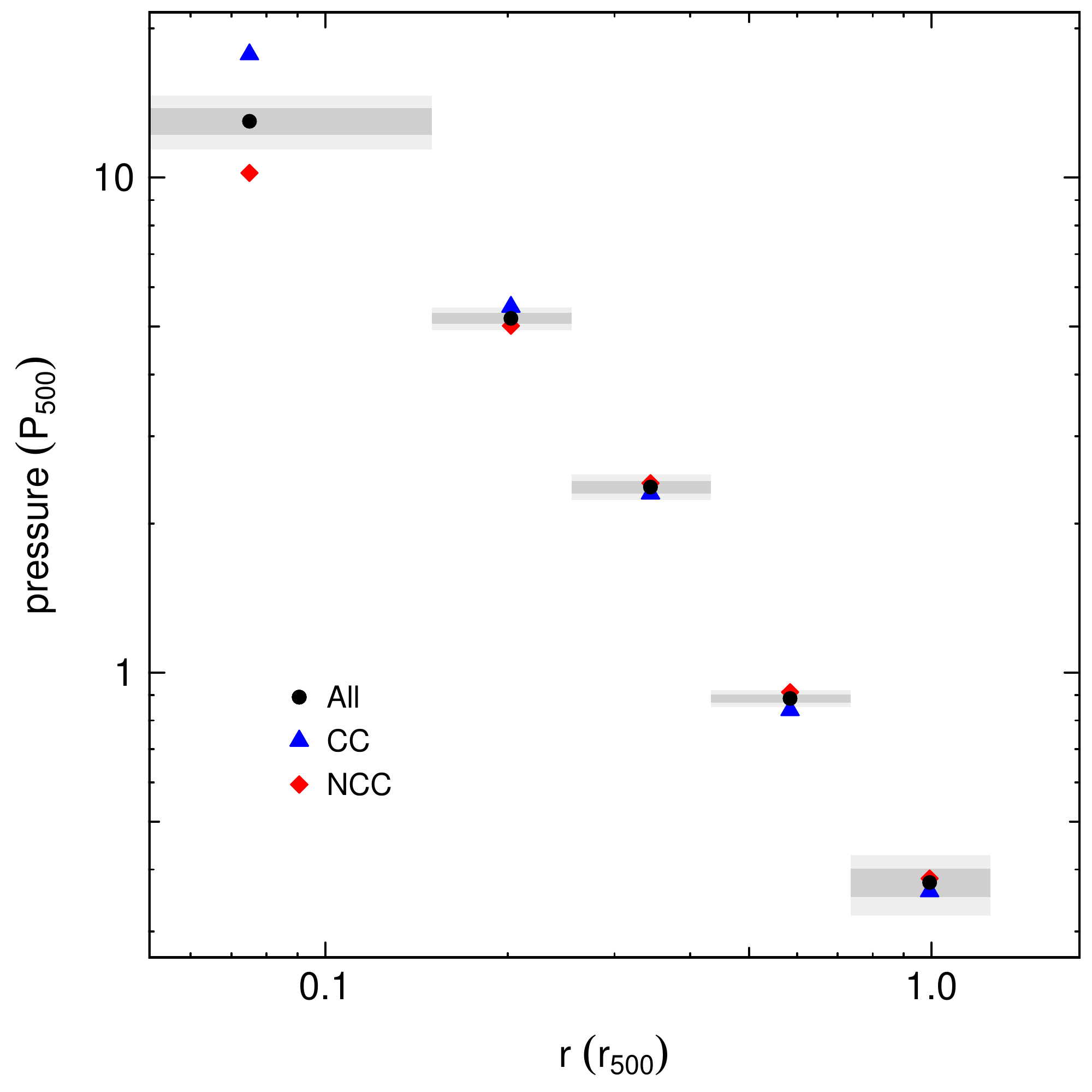}}
 \subfloat[]{\includegraphics[width = 2.5in]{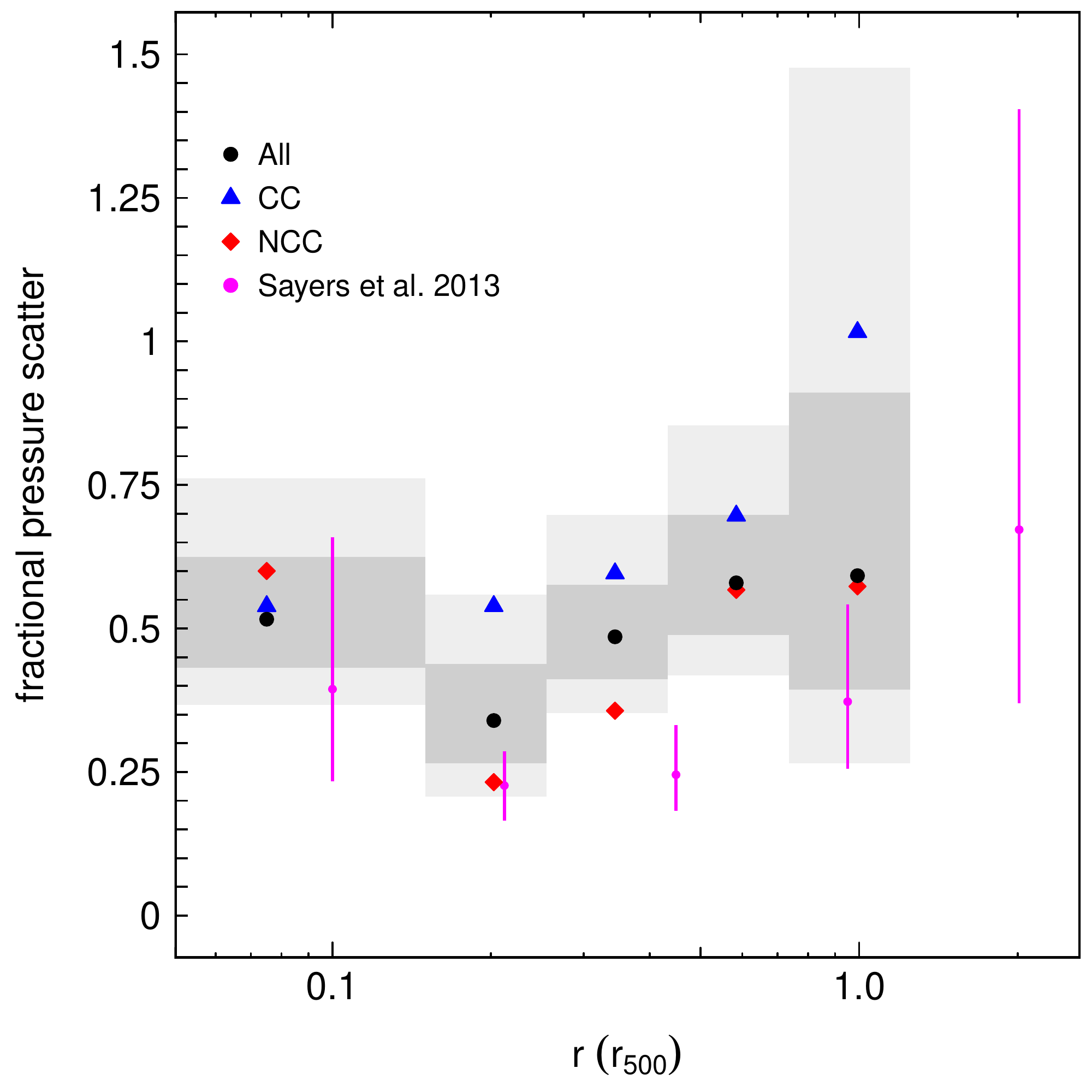}}
\caption{The left column shows the mean density (a), temperature (c), and pressure (e) profiles for the full sample, CC, and NCC subsets (scaled according to the self-similar values given in Sec.~\ref{JointClusterLikelihood}). The solid points correspond to the results obtained from the meta--analysis of Sec.~\ref{sec:Ensemblefitondeprojections}, and the hollow points correspond to the results obtained from full likelihood method of Sec.~\ref{JointClusterLikelihood}.  The gray shaded regions are the 68\% and 95\% confidence regions for the meta--analysis of the full sample. The confidence regions for the CC and NCC subsets are slightly larger in size, and have been omitted for clarity. The right column shows the fractional intrinsic scatter about the mean profiles. The green points are from \citet{2016MNRAS.456.4020M}, while the magenta points are from \citet{2013ApJ...768..177S}.}
\label{mean_scatter_profiles}
\end{figure*}

The density profiles for CC and NCC clusters obtained with the joint likelihood of Sec.~\ref{JointClusterLikelihood} are reported in Fig.~\ref{mean_scatter_profiles} (top panels), together with their uncertainties. The mean density profiles show good agreement with the results obtained with the meta--analysis of Sec.~\ref{sec:Ensemblefitondeprojections}, which are also plotted in the same panels. Furthermore, both methods find a  higher density in the innermost shell for the CC clusters compared to the NCC clusters. However, the intrinsic scatters about the mean density obtained from the joint likelihood are generally smaller than those obtained from the meta--analysis, although the difference is only significant in the innermost shell. Furthermore, the density scatter values are consistent for both the CC and NCC samples (and for the full sample in the meta--analysis). For the meta--analysis, the fractional intrinsic scatter values recovered for the density profiles fall from approximately 0.4 in the innermost shell to 0.1 at intermediate radius, before showing an increase near $r_{500}$, which is largely consistent with previous works using a similar ensemble fitting approach  \citep[e.g.,][]{2016MNRAS.456.4020M}. 

The mean temperature and pressure profiles, along with their associated intrinsic scatters, obtained from the meta--analysis of Sec.~\ref{sec:Ensemblefitondeprojections} are also shown in Fig.~\ref{mean_scatter_profiles}. Overall, at larger radii, the mean temperatures are relatively flat, and we do not detect the clear decrease found in other studies \citep[e.g.,][]{2006ApJ...640..691V,2008A&A...486..359L,2010A&A...517A..92A,2013A&A...551A..23E}. Specifically, those studies imply that the temperature should drop to a value of $0.5$--$0.7 \times T_{500}$ at a radius of $\simeq r_{500}$, or approximately 1--$2\sigma$ lower than our results based on our uncertainties. Given the modest statistical significance of this difference, it may simply be due to noise fluctuations. However, inaccurate subtraction of the X-ray background may also play a role. For example, over-subtraction of the hard background would tend to bias spectroscopic X-ray temperatures low while biasing our derived temperatures high (via the reduction in X-ray-derived density). The difference could also be related to cluster physics. For instance, clumping within the ICM, which is expected to increase with radius, will tend to bias spectroscopically derived temperatures low compared to the X-ray/SZ values derived in our analysis \citep[e.g.,][]{2004MNRAS.354...10M,2005ApJ...618L...1R,2013MNRAS.429..799V}. Elongation of the cluster along the line of sight, which will increase the SZ brightness compared to the X-ray brightness, could also artificially boost the temperatures recovered in our analysis \citep[e.g.,][]{2000MNRAS.313..783C}. Another possibility for the slightly higher than expected temperatures at large radii may be biases in the SZ data. Indirect evidence for such an effect can be found by comparing the pressure profiles obtained by \citet{2013ApJ...768..177S}, based solely on Bolocam data, to those obtained by \citet{2016ApJ...832...26S} based on a joint analysis of Bolocam and Planck for a nearly identical set of clusters. Beyond $\simeq 0.5 \times r_{500}$, the latter work found a lower value for the pressure profile, with the difference increasing to a factor of $\simeq 1.5$ at $r_{500}$. This implies that the Bolocam data alone may be overestimating the pressure at large radii, which would bias our temperature profiles high in that region. In addition, the mean temperature profiles recovered for the CC and NCC subsets in the innermost shell are not different at a statistically significant level (i.e., the CC subset does not show a significant drop towards the cluster center). This is likely a result of the coarse binning required by the SZ data. Specifically, the innermost shell extends to $0.15r_{500}$, which is generally not small enough to resolve the cool core. 

The fractional temperature intrinsic scatters are approximately constant with radius, with a value of 0.4--0.5. Spectroscopic X-ray studies, such as \citet{2006ApJ...640..691V} and \citet{2016MNRAS.456.4020M} have found similarly constant profiles, although they have found significantly lower fractional scatters of approximately 0.1. The reason for this discrepancy is not fully understood, although it appears to be largely due to the combination of: line of sight differences in the SZ/X-ray signals, cluster selection, and non-idealities in the SZ measurement noise. First, \citet{2007MNRAS.382..397A}, using an SZ/X-ray deprojection method nearly identical to ours, found that, for a given cluster, the recovered temperature profiles differ with a fractional rms of $\simeq 0.15$ when different observational lines of sight are considered. This variation is due to deviations from spherical symmetry in the ICM combined with the differing density/temperature dependance of the X-ray and SZ signals, and would therefore not appear in a single-probe X-ray analysis but would appear in our combined SZ/X-ray analysis. Second, the simulations of \citet{2017MNRAS.465..213B} indicate that the temperature profiles of massive clusters of diverse morphologies, such as our sample, show strong redshift evolution ($\sim 20$\% between $z=0$ and $z=1$) which would appear as a fractional intrinsic scatter with an rms of $\simeq 0.1$ in our analysis.\footnote{Temperature profiles of high mass relaxed clusters, such as those in the \citet{2016MNRAS.456.4020M} sample, are not expected to noticeably evolve with redshift, and therefore would not include this additional scatter. Furthermore, the simulations of \citet{2017MNRAS.465..213B} indicate little evolution in the density profiles of massive clusters at $r \gtrsim 0.2 \times r_{500}$, and we therefore do not expect a similar increase in the intrinsic scatter of the density profiles in our analysis.} Finally, our assumption that the noise in the SZ map pixels is uncorrelated is not strictly correct. This results in a slight underestimate of the SZ measurement noise in the derived temperature profiles, which directly translates to an overestimate of the intrinsic scatter. By comparing to the results of \citet[][see Fig.~\ref{mean_scatter_profiles}]{2013ApJ...768..177S}, which are based on identical Bolocam SZ data and do account for the noise non-idealities, we estimate that the unaccounted for noise is equivalent to a fractional intrinsic scatter of 0.3 in our derived temperature profiles.\footnote{Fully accounting for the SZ noise correlations requires a bootstrap-like sampling of noise realizations, which is not computationally tractable for our MCMC fits. Therefore, to obtain a rough estimate of the impact of the unaccounted for noise on our scatter results, we artificially increased the per-cluster SZ noise until the pressure intrinsic scatter values were approximately equal to those obtained by \citet{2013ApJ...768..177S}. We then recomputed the temperature intrinsic scatters using the artificially increased SZ noise estimates, and the quadrature difference compared to the intrinsic scatter values obtained from our nominal analysis corresponds to a fractional scatter of 0.3.} In sum, given that \citet{2016MNRAS.456.4020M} measure a fractional intrinsic scatter of 0.1 in their X-ray analysis, we expect to measure a value of $\sqrt{0.1^2 + 0.15^2 + 0.1^2 + 0.3^2} \simeq 0.35$ after accounting for the three effects described above, in reasonable agreement with our actual measurements.

\section{Conclusions}
\label{sec:Conclusions}

We have obtained individual cluster density and temperature profiles extending to $\simeq r_{500}$ for the BOXSZ sample of 45 clusters using a combination of X-ray surface brightness and SZ measurements.  The profiles extend to larger radii than can typically be probed using X-ray spectroscopy (e.g., a factor of $\simeq 2$ compared to the ACCEPT analysis), although with modest signal-to-noise limited mainly by the SZ data. In general, the recovered profiles agree well with those obtained in previous X-ray spectroscopic studies. However, there are some differences that appear with modest statistical significance, most notably the general lack of a temperature drop at large radii. While this may be due to noise fluctuations, it could also be related to systematic errors in X-ray background subtraction, clumping in the ICM, cluster elongation, and/or a slight bias in the Bolocam SZ data at large radius. 

We applied two different methods to obtain ensemble-average scaled deprojections for density and temperature and their intrinsic scatters: a global--likelihood analysis and a meta--analysis based on individual cluster deprojections.  The former is a direct fit to the projected X-ray profiles and the SZ maps, while the latter method involves fitting the individual cluster deprojections.  In both cases, these were computed at a set of common radii extending to $1.25 r_{500}$. In general, our results for the ensemble-average density profiles agree with previous works. For example, we find that CC clusters, as defined by the X-ray luminosity ratio described in Sec.~\ref{sec:ClusterSample}, have higher densities in the inner regions than NCC clusters. In addition, we find (in the meta--analysis) that the fractional intrinsic scatter in the density profiles is highest near the center of the cluster, where a range of non-gravitational processes occur, lowest in the intermediate regions where the cluster is more regular, and then increasing at larger radii ($r_{500}$) where active accretion of new material onto the cluster is occurring. 

The ensemble-average temperature profile we recover is also in good agreement with previous results. We did not detect a significant drop in temperature near the cluster center, even for the CC subset, due mainly to the coarse angular resolution of the SZ data which limit the innermost shell to a radius of $0.15 \times r_{500}$. We also did not detect a temperature drop at large radii, where our recovered profile is 1--$2\sigma$ higher compared to previous results like \citet{2006ApJ...640..691V} and \citet{2013A&A...551A..22E}. While this may be a noise fluctuation given its modest statistical significance, the slightly higher temperature we obtained may also be due to the range of effects detailed above such as X-ray background subtraction, ICM clumping, cluster elongation, and/or a slight bias in the SZ data at large radii.

The 0.4--0.5 fractional intrinsic scatter we obtained for the temperature profile is significantly larger than previous results (e.g., $\lesssim 0.1$ by \citet{2016MNRAS.456.4020M}. As detailed in Sec.~\ref{sec:Results}, this difference is not fully understood but is thought to be caused by a combination of several effects. In particular, line-of-sight differences in the SZ and X-ray signals, combined with departures from spherical symmetry, is estimated to increase the fractional intrinsic scatter by $\simeq 0.15$ in our analysis compared to X-ray spectroscopic analyses. Furthermore, redshift evolution in the average temperature profile of diverse cluster populations, such as the BOXSZ sample, is estimated to contribute an additional fractional intrinsic scatter of $\simeq 0.1$ in our analysis. Finally, our assumption that the noise in the SZ maps is uncorrelated between pixels is not strictly valid, and results in a slight underestimate of the SZ measurement noise, which in turn produces an overestimate of the fractional intrinsic scatter. In sum, these effects largely explain the difference between our temperature intrinsic scatter measurements and those based on X-ray spectroscopy.

\section*{Acknowledgements}

Computation for the work described in this paper was supported by the University of Southern California's Center for High-Performance Computing (\href{http://hpcc.usc.edu}{http://hpcc.usc.edu}).  JAS would like to thank the Women in Science and Engineering (WiSE) Program at USC for financial support. Silvia Ameglio was supported by the USC WiSE postdoctoral fellowship and the NSF ADVANCE AST-0649899.
EP wishes to thank the Aspen center for Physics for hospitality during part of the preparation of this work. JS was partially supported by NSF/AST-1617022.




\bibliography{references} 
\bibliographystyle{mnras}



\appendix

\section{Individual Cluster Results}
 
\begin{table*}
\centering
\setlength\tabcolsep{3pt}
\scriptsize
\caption{Goodness of Fit Details}
\label{pvalues}
\begin{tabular}{ccc|cccccc|cccccc}            &                &             & \multicolumn{6}{c|}{Concentric Shell Deprojections}                & \multicolumn{6}{c}{Smooth Parametric Profiles}                       \\ \cline{2-15} 
            & X-ray proj. bins & SZ map pixels & $\chi^{2}_{XR}$ & $\chi^{2}_{SZ}$ & $\chi^{2}_{tot}$ & dof & $\chi^{2}_{red}$ & PTE   & $\chi^{2}_{XR}$ & $\chi^{2}_{SZ}$ & $\chi^{2}_{tot}$ & dof & $\chi^{2}_{red}$ & PTE   \\ \hline
Abell 2204      & 77             & 1677        & 1239.55   & 1864.14 & 3103.69  & 1744      & 1.782    & 0.000 & \hphantom{1}82.23     & 2116.02 & 2198.25  & 1745      & 1.260    & 0.000 \\
Abell 383        & 58             & 1388        & \hphantom{1}742.58    & 1453.50 & 2196.08  & 1437      & 1.531    & 0.000 & \hphantom{1}76.86     & 1481.79 & 1558.65  & 1440      & 1.082    & 0.015 \\
Abell 1423       & 44             & 1577        & \hphantom{1}170.90    & 1650.79 & 1821.69  & 1612      & 1.132    & 0.000 & \hphantom{1}84.87     & 1666.52 & 1751.39  & 1616      & 1.084    & 0.010 \\
Abell 209        & 61             & \hphantom{1}905         & \hphantom{1}903.42    & 1046.86 & 1950.28  & \hphantom{1}957       & 2.044    & 0.000 & \hphantom{1}94.05     & 1044.52 & 1138.57  & \hphantom{1}961       & 1.185    & 0.000 \\
Abell 963         & 71             & 1518        & 1318.75   & 1574.04 & 2892.79  & 1580      & 1.834    & 0.000 & 145.30    & 1580.28 & 1725.58  & 1582      & 1.091    & 0.006 \\
Abell 2261       & 76             & 1739        & \hphantom{1}821.29    & 1809.37 & 2630.66  & 1806      & 1.459    & 0.000 & 119.87    & 1809.71 & 1929.58  & 1808      & 1.067    & 0.023 \\
\end{tabular}
\end{table*}

\begin{landscape}
\begin{table}
\setlength\tabcolsep{3pt}
\centering
\caption{Best Fit Parameters for the Smooth Parametric Profiles}
\label{analytical_fit_parameters}
\begin{tabular}{cccrrccrrccc}
Cluster     & n Profile & T Profile    & \multicolumn{1}{c}{$n_{e0,1}$ ($10^{-2} cm^{-3}$)} & \multicolumn{1}{c}{$r_{c,1}$ ('')} & $n_{e0,2}$ ($10^{-3} cm^{-3}$)               & $r_{c,2}$ ('')                               & \multicolumn{1}{c}{$\beta$} & \multicolumn{1}{c}{$T_{0}$}  & $T_{\rm min}$                                   & $r_{\rm cool}$                                     & $r_{t}$                                          \\ \hline
Abell 2204        & double-B  & CC-full      & 22.88$_{-0.67}^{+0.67}$                            & 7.27$_{- 0.23}^{+ 0.23}$           & \multicolumn{1}{r}{15.40$_{-0.74}^{+0.74}$}  & \multicolumn{1}{r}{50.06$_{-2.07}^{+2.01}$}  & 0.641$_{-0.008}^{+0.008}$   & 16.82$_{-  3.40}^{+  3.93}$  & \multicolumn{1}{r}{3.51$_{-2.01}^{+1.89}$}  & \multicolumn{1}{r}{62.85$_{-24.47}^{+25.92}$}  & \multicolumn{1}{r}{248.42$_{-114.84}^{+127.01}$} \\
Abell 383        & double-B  & iso          & 13.50$_{-0.62}^{+0.61}$                            & 5.58$_{- 0.30}^{+ 0.31}$           & \multicolumn{1}{r}{19.73$_{-17.98}^{+1.73}$} & \multicolumn{1}{r}{24.77$_{-23.17}^{+1.58}$} & 0.601$_{-0.006}^{+0.006}$   & 7.21$_{-  0.48}^{+  0.48}$   & -                                           & -                                              & -                                                \\
Abell 1423       & single-B  & NCC          & 2.61$_{-0.20}^{+0.20}$                             & 14.42$_{- 0.99}^{+ 1.00}$          & -                                            & -                                            & 0.497$_{-0.005}^{+0.005}$   & 16.52$_{-  5.35}^{+  6.43}$  & -                                           & -                                              & \multicolumn{1}{r}{54.31$_{-42.176}^{+30.85}$}                                                \\
\end{tabular}
\end{table}
\end{landscape}

\begin{landscape}
\begin{table}
\setlength\tabcolsep{3pt}
\centering
\caption{Best Fit Parameters for the Concentric Shell Deprojections}
\scriptsize
\label{dpj_fit_params}
\begin{tabular}{c|ccccc|crrrrr|rrrrr}
Cluster     & $r_{1}$ & $r_{2}$ & $r_{3}$ & $r_{4}$ & $r_{5}$ &  & $n_{e,1}$               & \multicolumn{1}{c}{$n_{e,2}$} & \multicolumn{1}{c}{$n_{e,3}$} & \multicolumn{1}{c}{$n_{e,4}$} & \multicolumn{1}{c|}{$n_{e,5}$} & \multicolumn{1}{c}{$T_{e,1}$} & \multicolumn{1}{c}{$T_{e,2}$} & \multicolumn{1}{c}{$T_{e,3}$} & \multicolumn{1}{c}{$T_{e,4}$} & \multicolumn{1}{c}{$T_{e,5}$} \\ \hline
Abell 2204       & 0.0537  & 0.109   & 0.221   & 0.448   & 0.908   &  & 65.07$_{-1.66}^{+1.71}$ & 22.52$_{-0.45}^{+0.45}$       & 8.36$_{-0.09}^{+0.09}$        & 2.40$_{-0.03}^{+0.03}$        & 0.64$_{-0.01}^{+0.01}$         & 0.75$_{-0.75}^{+0.18}$        & 1.16$_{-0.3}^{+0.3}$          & 1.12$_{-0.09}^{+0.09}$        & 0.72$_{-0.13}^{+0.13}$        & 0.23$_{-0.23}^{+0.05}$        \\
Abell 383        & 0.0856  & 0.166   & 0.320   & 0.619   & 1.200   &  & 33.40$_{-0.87}^{+0.85}$ & 12.12$_{-0.2}^{+0.19}$        & 4.21$_{-0.06}^{+0.06}$        & 1.41$_{-0.02}^{+0.02}$        & 0.44$_{-0.02}^{+0.02}$         & 1.53$_{-1.53}^{+0.36}$        & 0.60$_{-0.6}^{+0.16}$         & 1.63$_{-0.25}^{+0.25}$        & 1.93$_{-0.3}^{+0.3}$          & 0.52$_{-0.52}^{+0.11}$        \\
Abell 1423       & 0.0762  & 0.151   & 0.300   & 0.595   & 1.180   &  & 20.43$_{-0.81}^{+0.86}$ & 7.79$_{-0.31}^{+0.31}$        & 3.28$_{-0.09}^{+0.09}$        & 1.31$_{-0.03}^{+0.03}$        & 0.42$_{-0.01}^{+0.01}$         & 1.26$_{-1.26}^{+0.3}$         & 1.39$_{-1.39}^{+0.35}$        & 1.04$_{-0.32}^{+0.32}$        & 0.81$_{-0.3}^{+0.3}$          & 0.34$_{-0.34}^{+0.07}$        \\
Abell 209         & 0.0672  & 0.148   & 0.324   & 0.711   & 1.560   &  & 14.98$_{-0.87}^{+0.87}$ & 8.95$_{-0.23}^{+0.23}$        & 4.20$_{-0.07}^{+0.07}$        & 1.37$_{-0.03}^{+0.03}$        & 0.41$_{-0.01}^{+0.01}$         & 1.14$_{-1.14}^{+0.29}$        & 0.68$_{-0.28}^{+0.27}$        & 1.15$_{-0.13}^{+0.13}$        & 0.96$_{-0.24}^{+0.24}$        & 0.48$_{-0.48}^{+0.1}$         \\
\hline
\end{tabular}

All of the parameter values in this table have been scaled relative to the self-similar values $r_{500}$, $n_{e,500}$, and $T_{500}$ (see Equations~\ref{eq:nn} and \ref{eq:TT}). The value of $r_i$ corresponds to the center of the shell (see Figure~\ref{fig:sphere}).

\end{table}
\end{landscape}

Individual cluster profiles, shown in order of increasing redshift. From top to bottom, the three plots for each cluster show the density, temperature, and pressure profiles. The results from this work are shown as black crosshairs for the concentric shell deprojections and as blue lines are the smooth parametric profiles +/- 1$\sigma$. Also included as green points are the ACCEPT X-ray-only study from \citep{2009ApJS..182...12C}. In each plot a vertical line denotes $r_{500}$. Where available, the joint X-ray/SZ fits from \citet{2006ApJ...652..917L} are plotted in magenta. 
\begin{figure}
	\includegraphics[width=\linewidth]{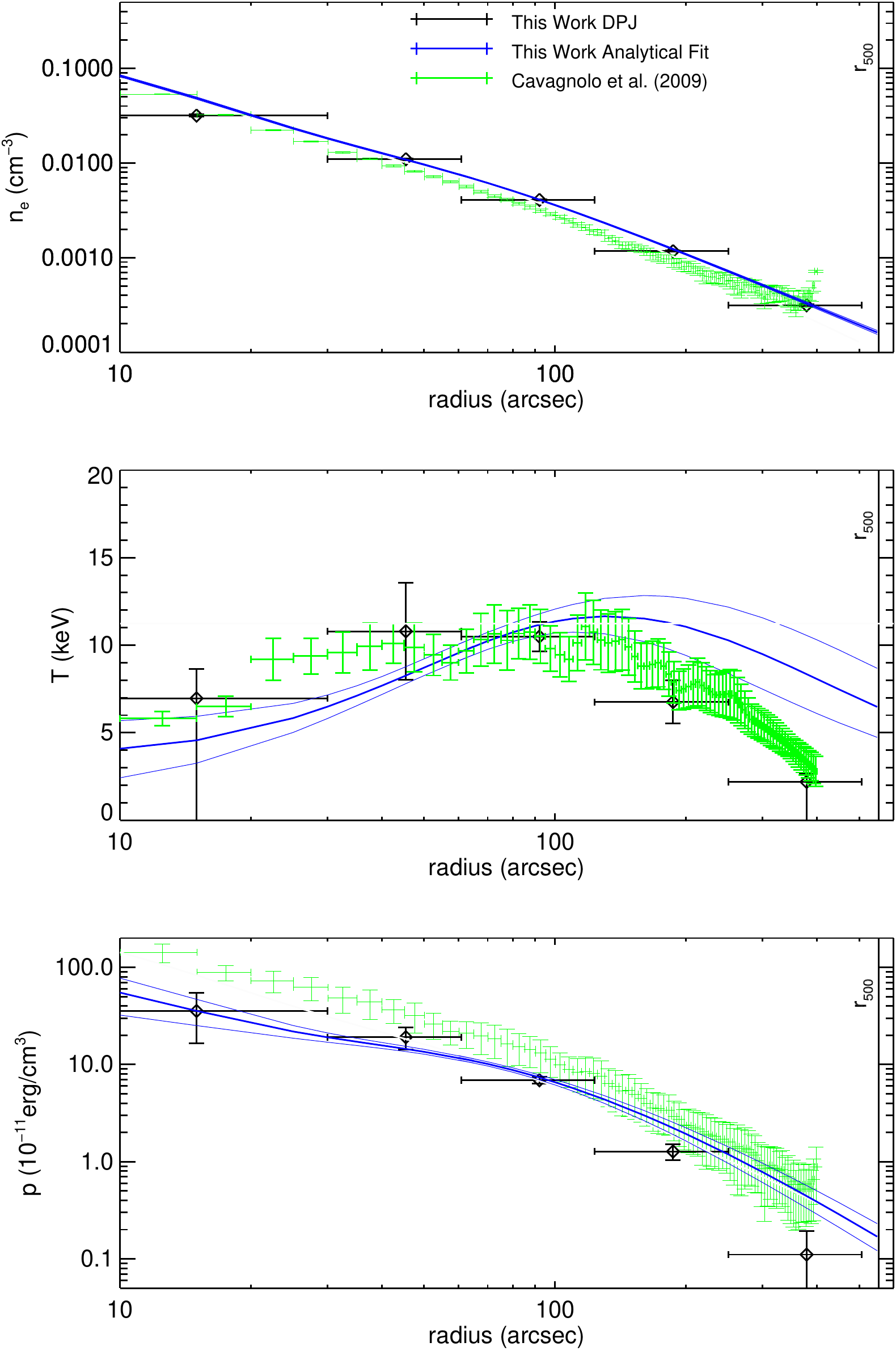}
	\caption{{\bf Abell 2204.} Our smooth parametric profile fit detects a cool core at a significance of $3.8\sigma$, and the temperature drop in the outskirts is detected at a significance $5\sigma$. In general, both our smooth parametric fits and deprojections of both the density and temperature are in good agreement with ACCEPT and other previous X-ray results (e.g., \citealt{2005MNRAS.356.1022S} and \citealt{2009A&A...501..899R}). Although we classify this cluster as relaxed, \citet{2005MNRAS.356.1022S}, using X-ray data, found a complicated core consisting of several cold fronts that may be attributed to the cluster recovering from a merger. \citet{2009ApJ...700.1161G} also found such temperature substructures, and speculate that they may be caused by AGN.  \citet{2009MNRAS.393...71S}, again using X-ray data, found dips in the surface brightness that may be caused by radio bubble formation in the core of the cluster.  \citet{2009A&A...501..899R} used \suzaku\ X-ray measurements to probe the cluster out to large radii, approximately 12\arcmin, where the temperature was found to be 4 keV. This is in good agreement with both the extrapolation of our smooth parametric profile and with the predictions from the hydrodynamical simulations of \citet{2006MNRAS.373.1339R}.} 
	\label{fig:a2204}
\end{figure}

\begin{figure}
	\includegraphics[width=\linewidth]{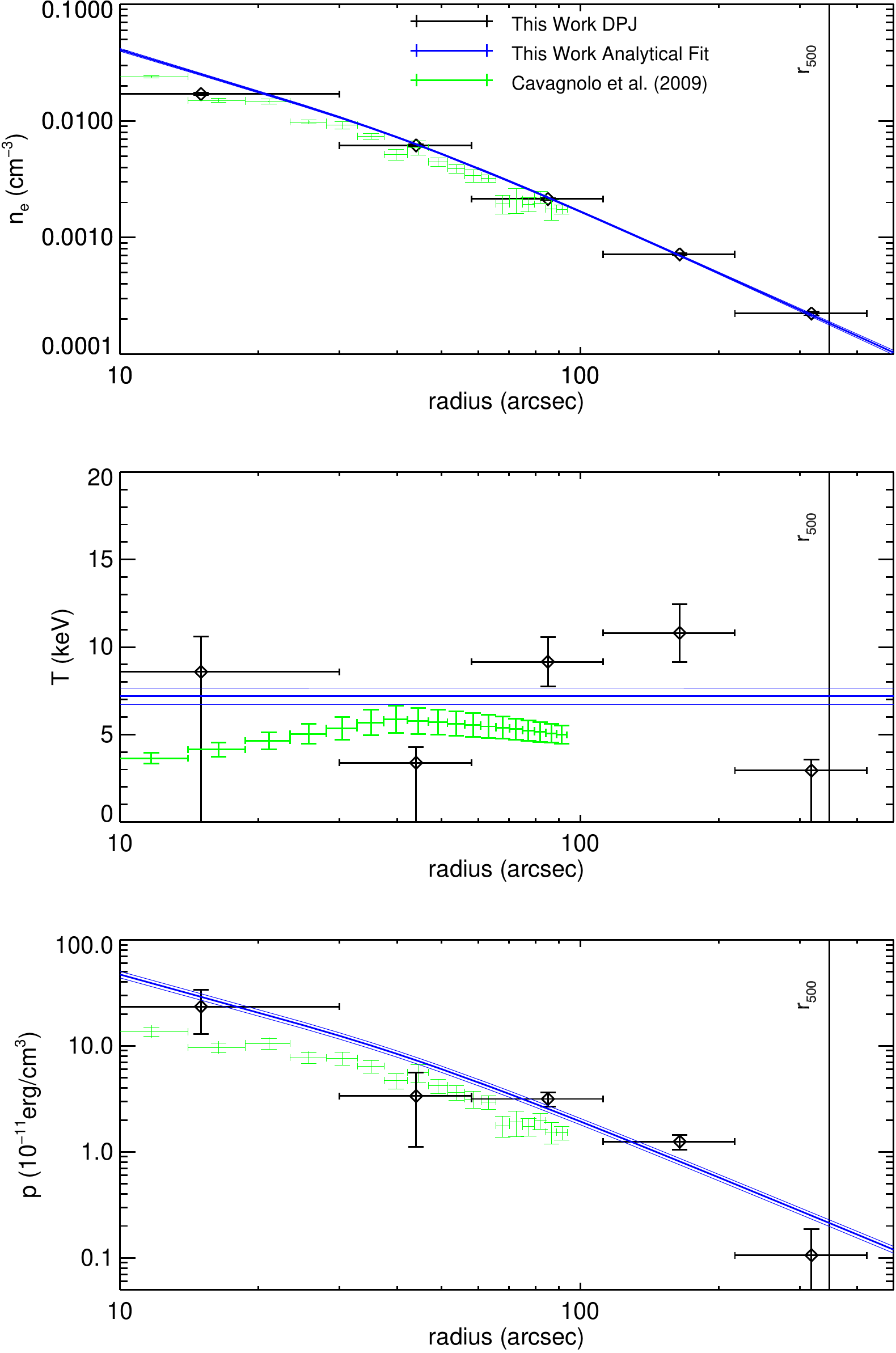}
	\caption{{\bf Abell 383.} Our density and temperature profiles are reasonably consistent with the ACCEPT results at overlapping radii, although our isothermal smooth parametric fit to the temperature is somewhat higher. This cluster is thought to be prolate and elongated along the line of sight \citep{2013ApJ...765...24N}, and a triaxial geometry is required to reconcile masses determined from X-rays and lensing \citep{2012MNRAS.421.3147M}.}
	\label{fig:a383}
\end{figure}

\begin{figure}
	\includegraphics[width=\linewidth]{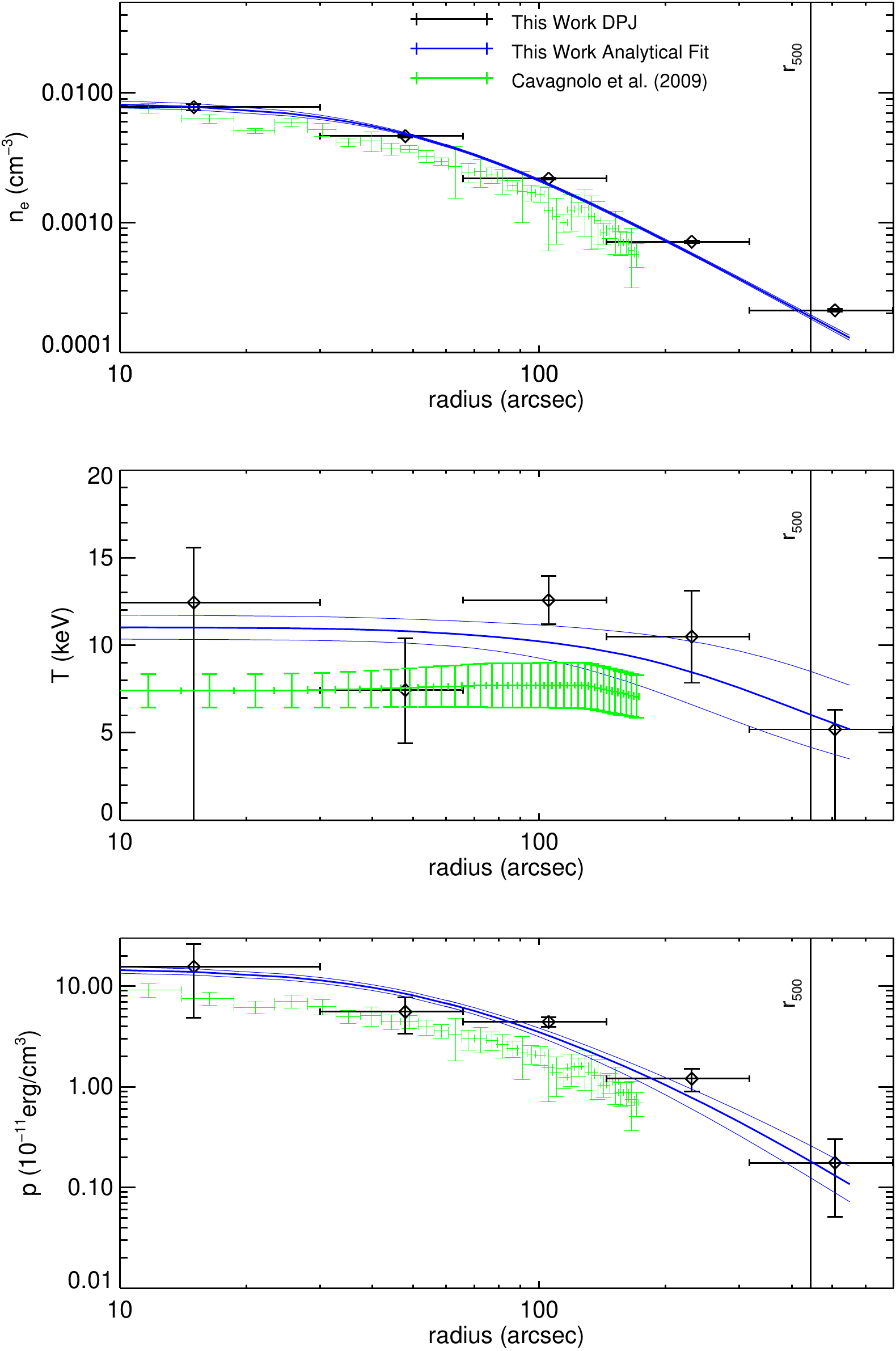}
	\caption{{\bf Abell 209.} In our smooth parametric fit, we detect a temperature drop in the outskirts at a significance of $3.5\sigma$. Although our temperature is somewhat higher the the ACCEPT results, we note that \citet{2003A&A...397..431M}  found a temperature of $10.2_{-1.2}^{+1.4}$ keV using X-ray spectroscopy within 180\arcsec, in good agreement with our results. We classify this cluster as non-disturbed, but \citet{2003A&A...397..431M} found a slightly elongated structure in the X-ray map arising from two X-ray peaks. Adding to the evidence that this cluster is at least somewhat disturbed, \citet{2007A&A...467..427P} also found elongation in the mass distribution through weak lensing.}
	\label{fig:a209}
\end{figure}

\begin{figure}
	\includegraphics[width=\linewidth]{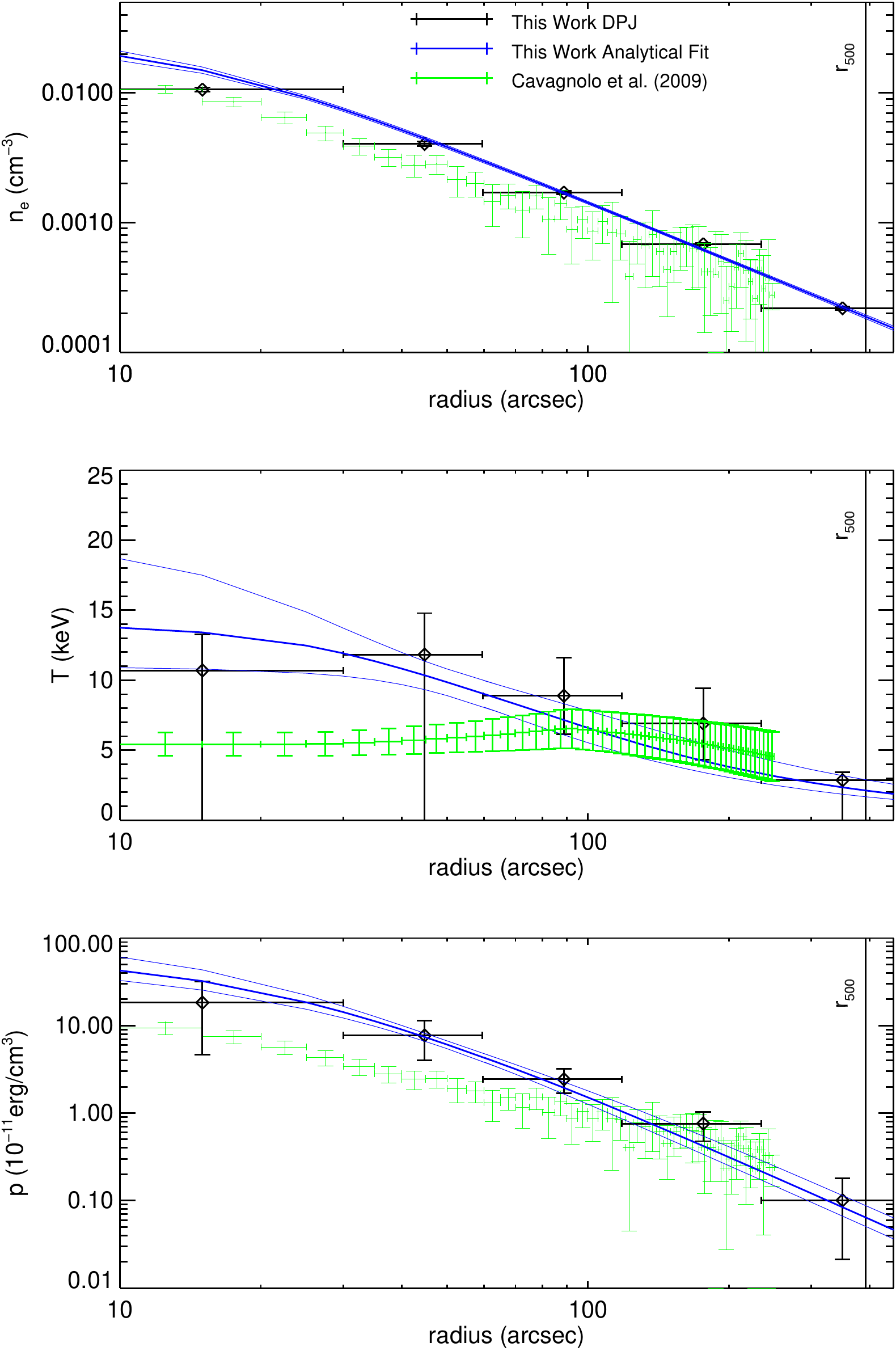}
	\caption{{\bf Abell 1423.} We detect a decrease in the temperature profile at large radius at a significance of $6.9\sigma$. \citet{2012MNRAS.425..162A}, using SZ data rather than X-ray spectroscopic measurements, fit an isothermal model with a temperature of $3.0 \pm 0.8$~keV, somewhat low compared to our measurements. However, the X-ray spectroscopic measurements from ACCEPT indicate a higher temperature (5.2~keV), which is in better agreement with our results, at least outside of the core region. }
	\label{fig:a1423}
\end{figure}

\begin{figure}
	\includegraphics[width=\linewidth]{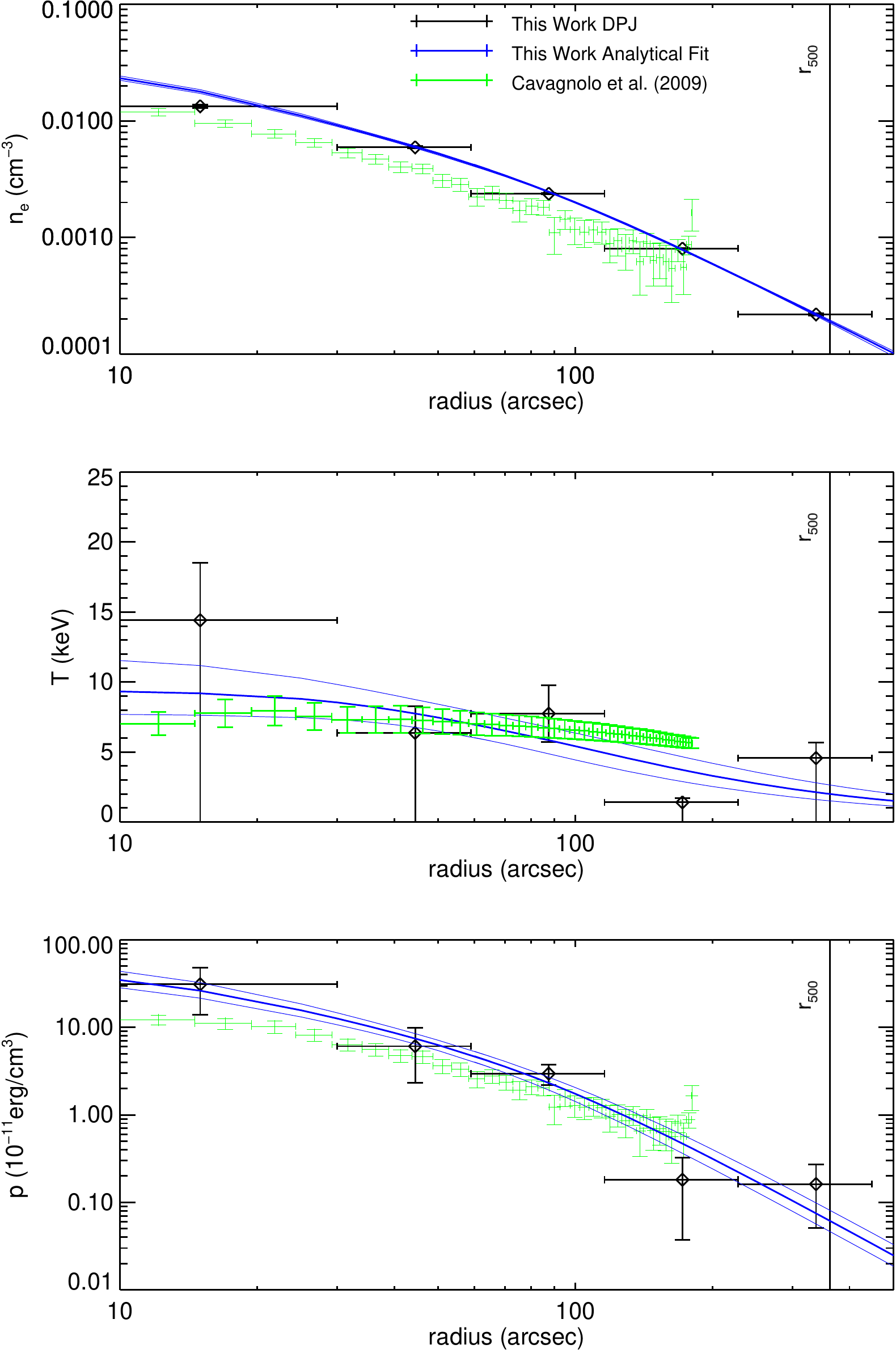}
	\caption{{\bf Abell 963.} We detect a decrease in the temperature profile at large radius at a significance of $6\sigma$. Our temperature profile is in good agreement with the ACCEPT results, although our density profile is somewhat higher. We do not detect the presence of a cool-core, but others have found a slight temperature decrease in the center \citep{2007ApJ...666..835B,2009ApJS..182...12C}.  \citet{2008A&A...486..359L}, using \xmm, specifically made note of this and identified it as an 'intermediate' cluster since the temperature decrease in the center is modest.}
	\label{fig:a963}
\end{figure}

\begin{figure}
	\includegraphics[width=\linewidth]{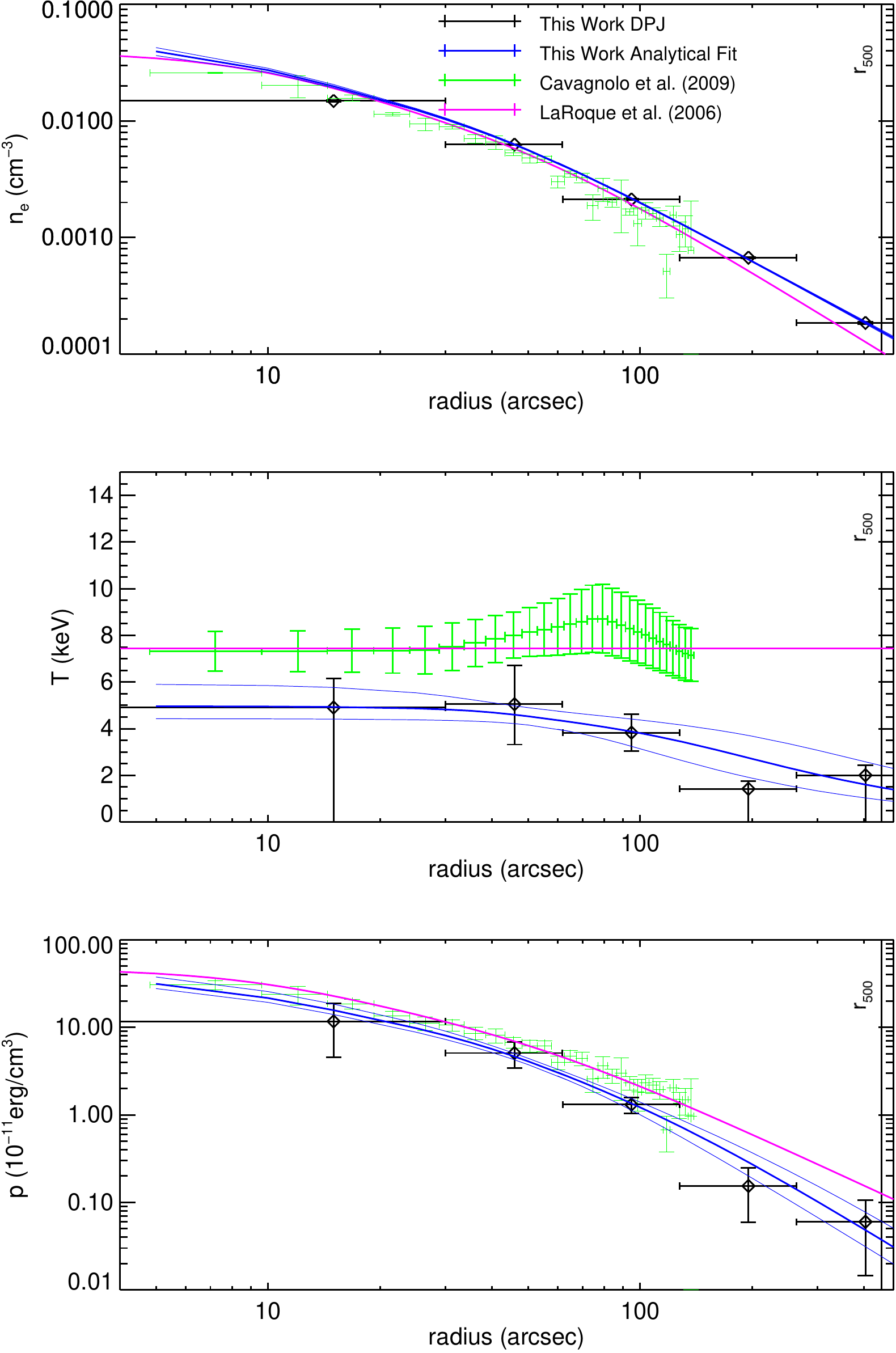}
	\caption{{\bf Abell 2261.} Based on our smooth parametric fit, we detect a temperature drop in the outskirts at a significance of $5.5\sigma$. Comparing to X-ray spectroscopy, ACCEPT \citep{2009ApJS..182...12C} shows a slight temperature drop in the center which is not detected in our analysis, along with an overall temperature somewhat higher than our results. In good agreement with the ACCEPT results, \citet{2007ApJ...666..835B} found a slight cool-core, with a central temperature of $7.7 \pm 0.4$~keV and peak of $9.0 \pm 0.4$~keV. \citet{2006ApJ...647...25B}, also using X-ray spectroscopy, similarly found a temperature profile in good agreement with the ACCEPT results. \citet{2012ApJ...757...22C} found a density profile slope that is shallower than the usual slope at small radii for cool-core clusters, thus they defined it as a "borderline" relaxed and cool core cluster (our analysis classifies it as a CC non-disturbed cluster). Finally, Abell 2261 is a CLASH cluster \citep{2012ApJS..199...25P}, and \citet{2012ApJ...757...22C} used the CLASH data to obtain a detailed mass measurement, deducing that the DM halo is elongated along the line of sight. }
	\label{fig:a2261}
\end{figure}

\begin{figure}
	\includegraphics[width=\linewidth]{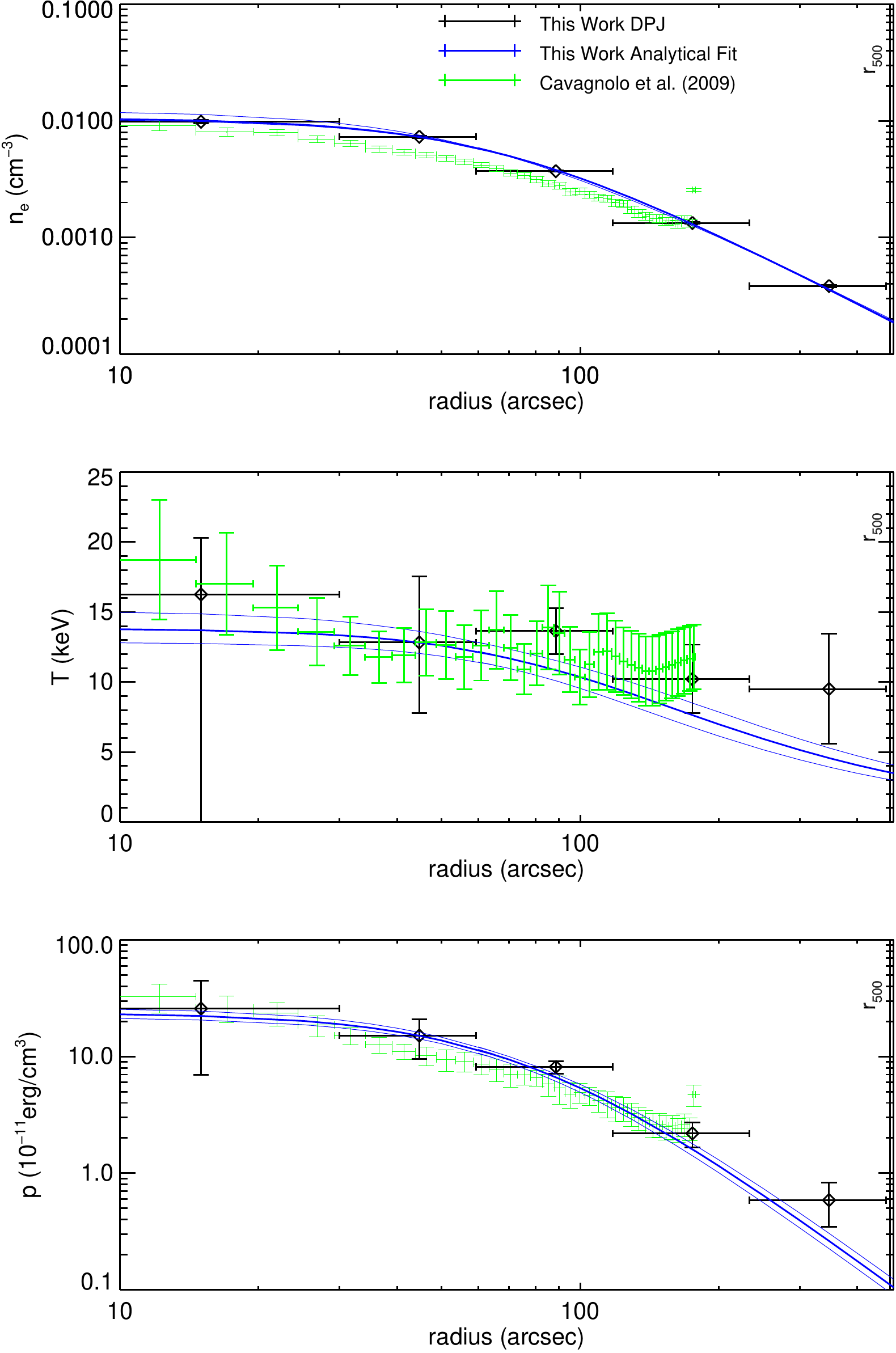}
	\caption{{\bf Abell 2219.} Our smooth parametric fit shows a decrease in temperature in the outskirts at a significance of $11.8\sigma$, and both our density and temperature profiles are in excellent agreement with the ACCEPT values. \citet{2013AN....334..377J} found a shock front with $>$16keV gas $2\arcmin$ NW of cluster, and this elongated cluster is possibly undergoing a large merger event \citep{2004A&A...416..839B}. Further evidence for a major merger was detailed in \citet{2015arXiv150505790C}, who used \chandra\ data to identify numerous cold and shock fronts, along with a large temperature spike $\sim25\arcsec$ away from the core of the cluster.}
	\label{fig:a2219}
\end{figure}

\begin{figure}
	\includegraphics[width=\linewidth]{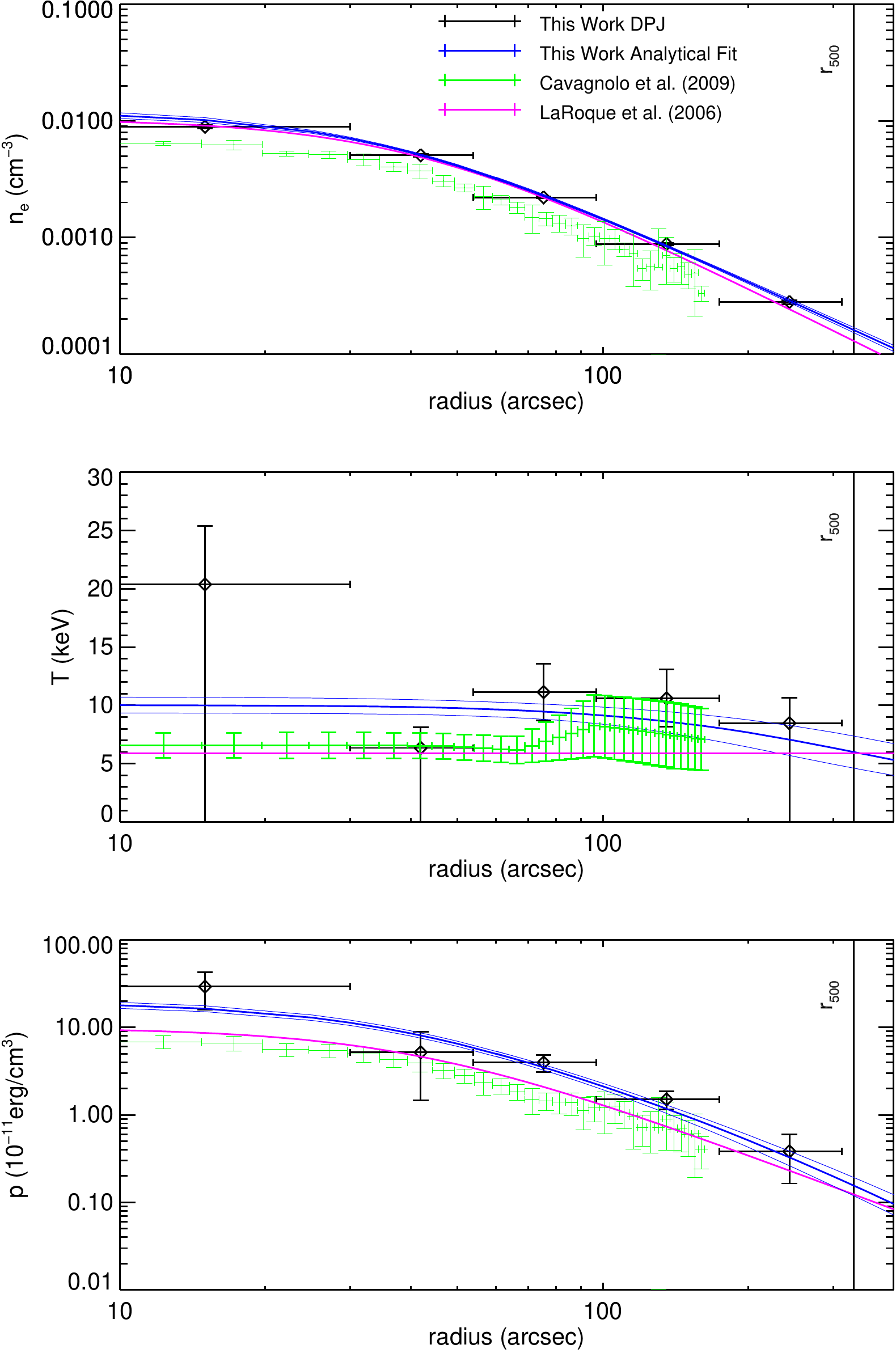}
	\caption{{\bf Abell 267.} Our deprojections and smooth parametric fits show a mostly flat temperature profile. Overall, the temperature is consistent with ACCEPT's relatively flat temperature profile, although our data indicate a somewhat higher temperature in the core region. Our smooth parametric fit shows some evidence for a drop in temperature at large radius ($2.8\sigma$). \citet{2013AN....334..377J} used \xmm\ spectroscopy to deproject a temperature profile, and they also found an approximately flat profile with a temperature of $\sim$6 keV.  They consider A267 a 'fossil' system, which is dominated by a massive elliptical galaxy.  Fossil systems in general have not merged recently, and usually host a cool-core. The lack of a cool core in this cluster is perhaps related to heating from an AGN.}
	\label{fig:a267}
\end{figure}

\begin{figure}
	\includegraphics[width=\linewidth]{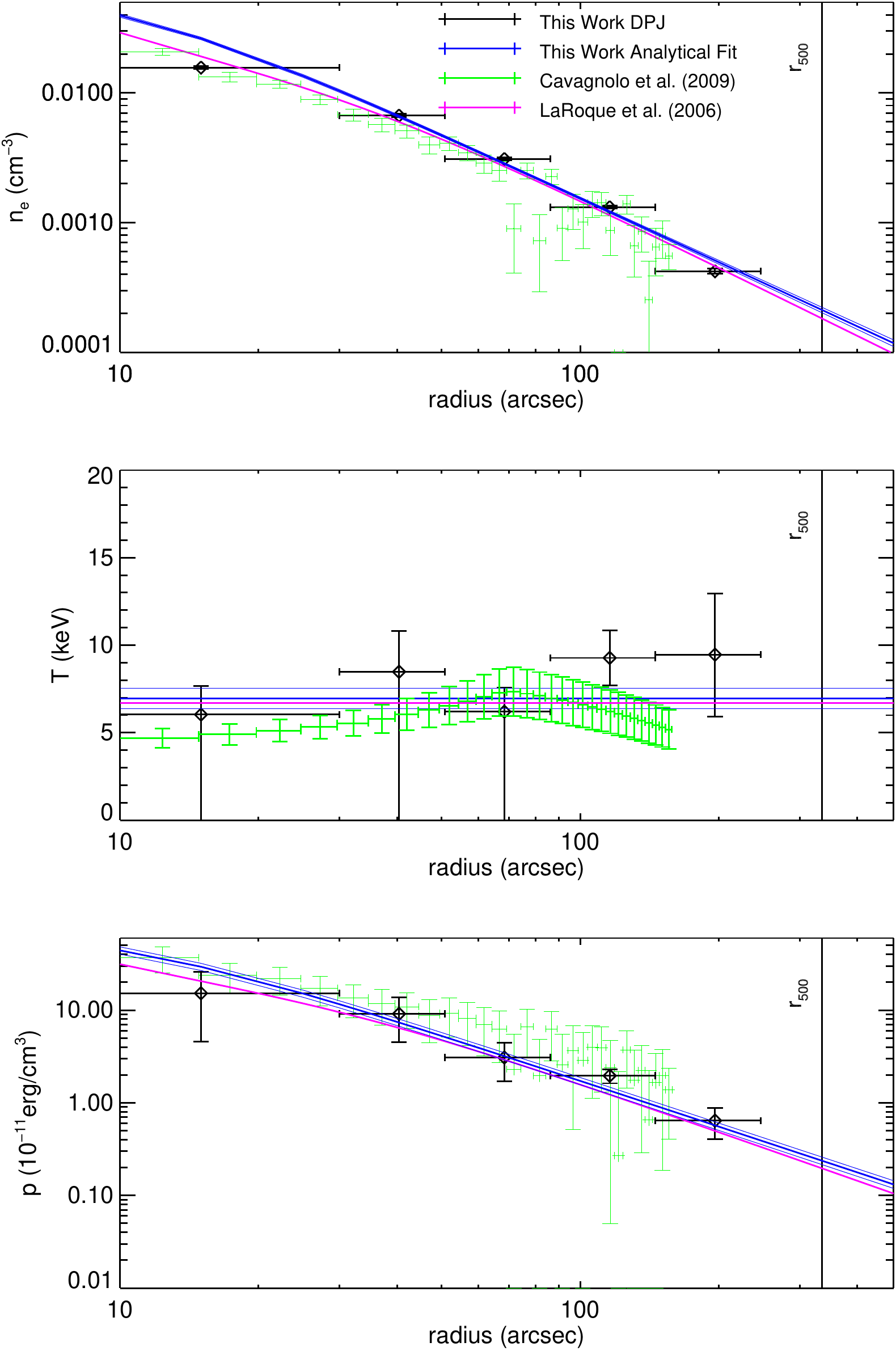}
	\caption{{\bf RX J2129.6+005.} Both our density and temperature profiles are in fairy good agreement with the ACCEPT results. There is some deviation between our density deprojection and the smooth parametric fit near the center, which may suggest that the single-beta profile is not fully sufficient in describing the data. This is one of the few clusters in the sample where the fits do not reach $r_{500}$, due to the poor quality of the X-ray data. \citet{2013MNRAS.433.2790L} used \chandra\ observations to identify this as a relaxed cluster based on the centroid shift. Furthermore, unlike the relatively isothermal temperature profile found in both our analysis and the ACCEPT analysis, they found a central temperature of 4 keV, a peak temperature of 8 keV near 90\arcsec and a decrease in temperature back to 4 keV in the outskirts (near 240\arcsec).}
	\label{fig:rxj21296}
\end{figure}

\begin{figure}
	\includegraphics[width=\linewidth]{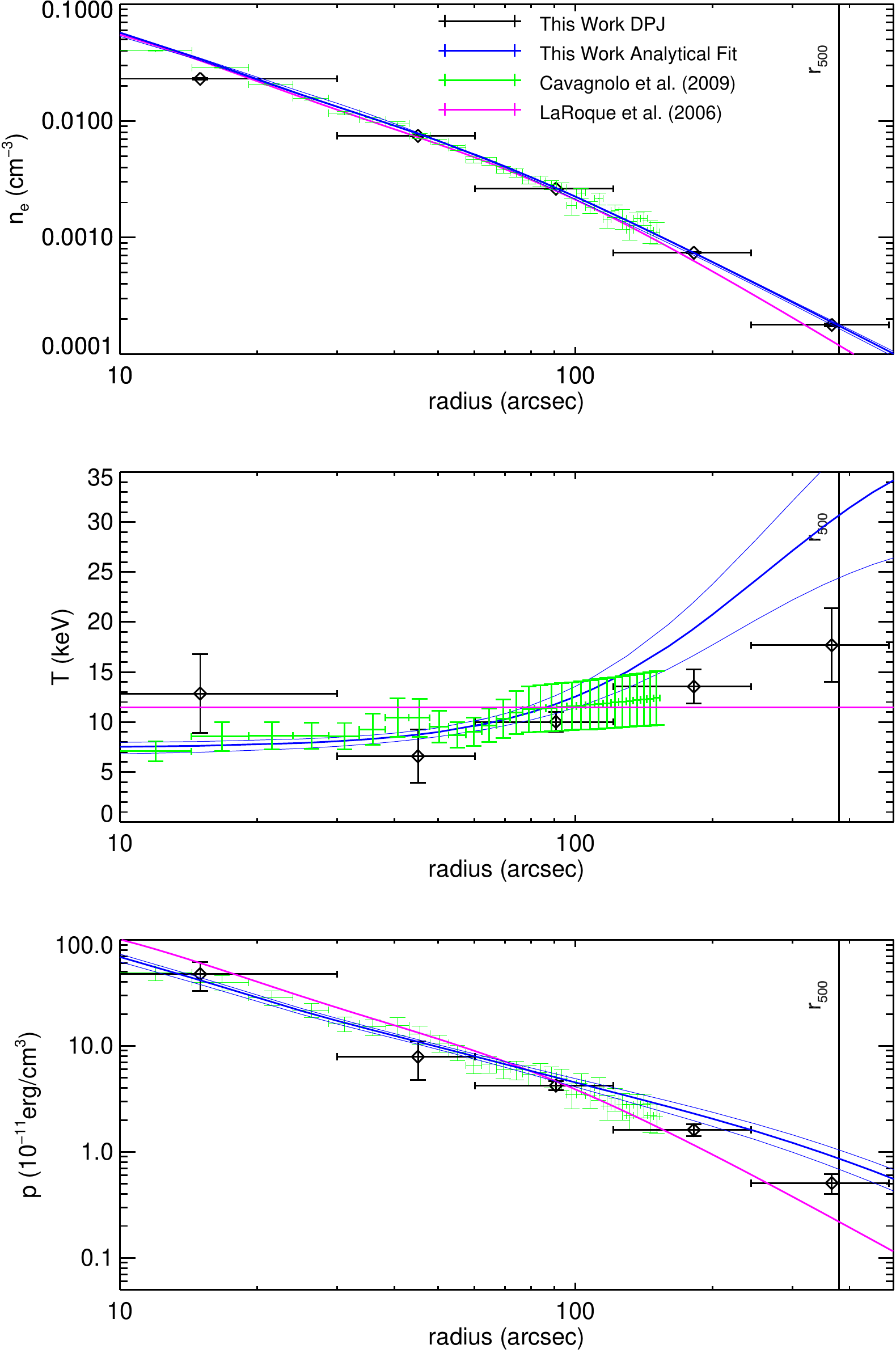}
	\caption{{\bf Abell 1835.}
In the regions of overlap, both our density and temperature profile fits are in excellent agreement with the ACCEPT results, and our smooth parametric fits show modest evidence for a temperature drop in the core ($2.8\sigma$). In the outer regions of the cluster, our temperature and pressure profiles are somewhat higher compared to the joint X-ray/SZ study presented in \citet{2006ApJ...652..917L}, although this may be related to the isothermal assumption used in that analysis. In contrast to the rising temperature at large radius seen in our fits, \citet{2013ApJ...766...90I} found that the temperature slowly decreases from 8 keV in the inner region to around 2 keV near the virial radius using \suzaku\ X-ray spectroscopy. Based on \chandra\ observations, \citet{2013MNRAS.428.2812B} also found a similarly large drop in temperature towards the cluster outskirts. We note that many studies have found this cluster to be spherically symmetric without any significant substructure \citep{2011ApJ...734...10K,2013ApJ...766...90I}, although \citet{2012MNRAS.425.2069M} find that the cluster is elongated in the line of sight, which may be the cause of the apparent temperature increase at large radius in our fits.  In addition, \citet{2007ApJ...666..835B} found that the temperature drops by a factor of 2 in core region relative to the peak, in contrast to both our results and the ACCEPT results. }
	\label{fig:a1835}
\end{figure}

\begin{figure}
	\includegraphics[width=\linewidth]{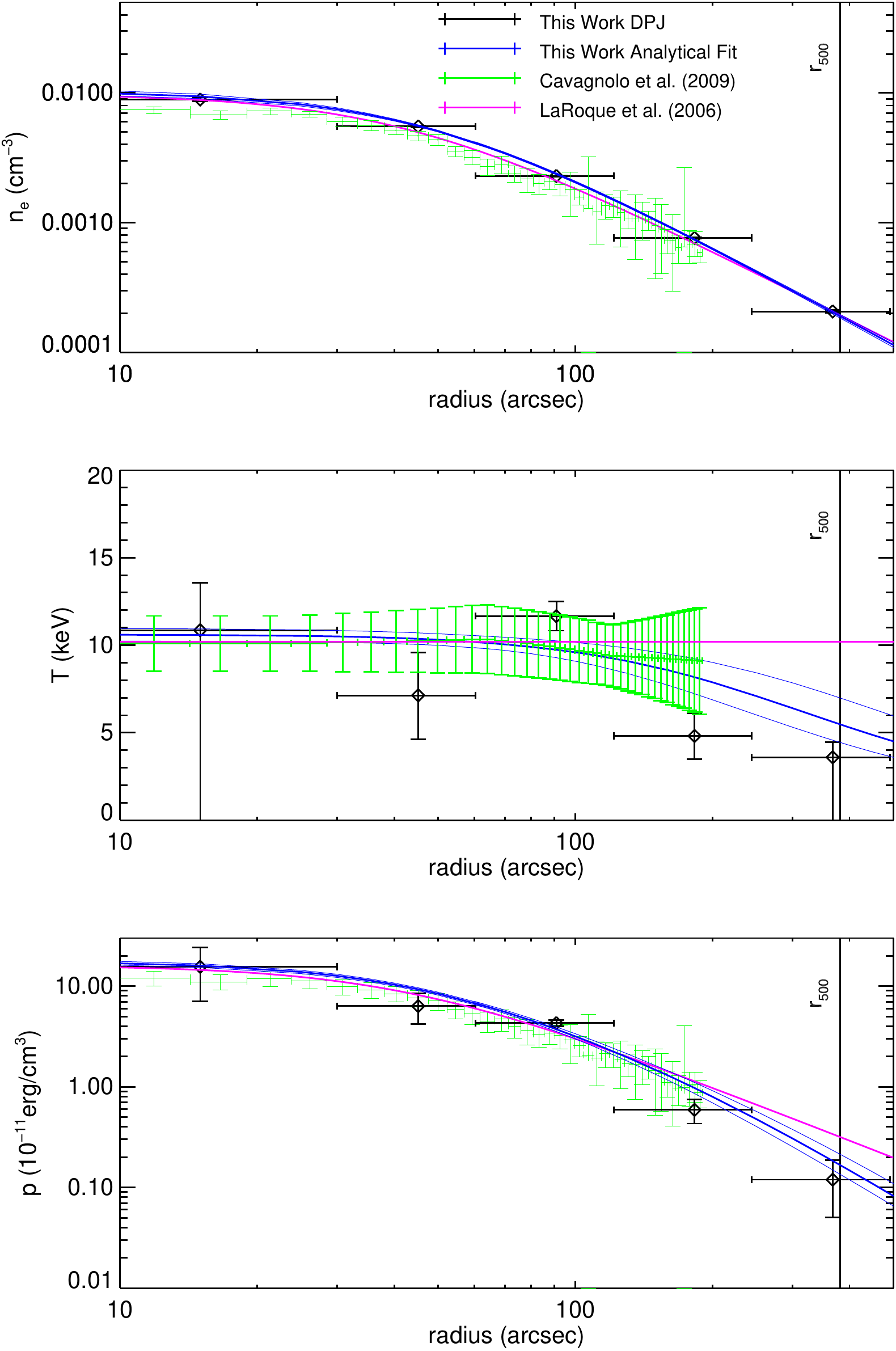}
	\caption{{\bf Abell 697.} We find a relatively flat inner temperature profile, and we detect a drop in temperature near $r_{500}$ at a significance of $4.5\sigma$.  Our Pressure profiles are consistent with the joint X-ray/SZ study of \citet{2006ApJ...652..917L}, and there is good agreement between our density, temperature, and pressure profiles and the corresponding ACCEPT profiles.  Although we characterized the cluster as non-disturbed, several other studies have shown that the cluster has undergone recent merging. \citet{2006A&A...455...45G} conducted a multi wavelength study using optical and X-ray data, where line of sight galaxy velocity dispersions and spectroscopic temperatures were calculated, respectively, and found that A697 is not relaxed.  With elongated X-ray emission and substructures near the center, Abell 697 likely went through many mergers.  Through lensing, \citet{2000AJ....120.2879M} also concluded that the cluster has undergone a recent merger.  }
	\label{fig:a697}
\end{figure}

\begin{figure}
	\includegraphics[width=\linewidth]{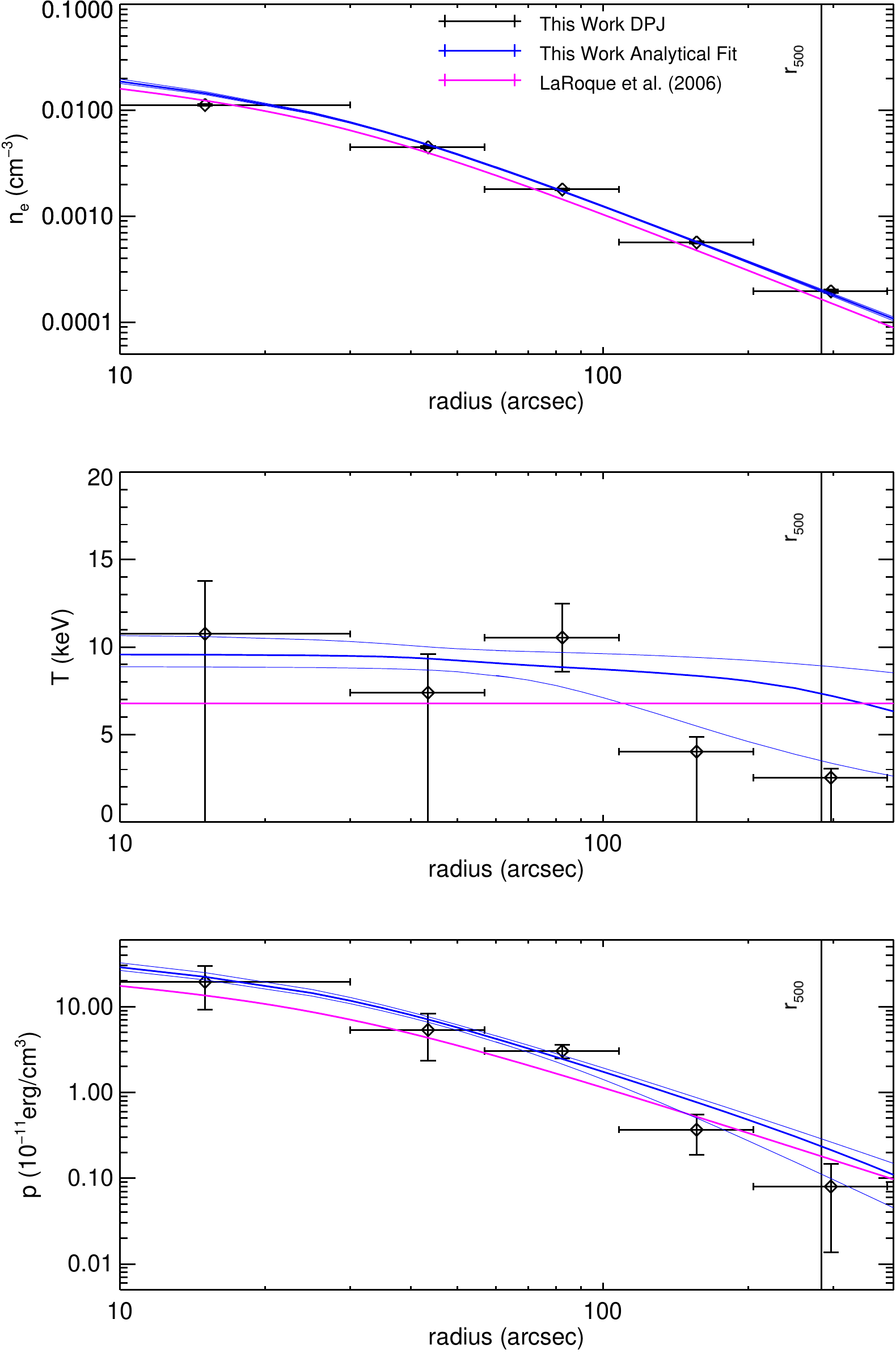}
	\caption{{\bf Abell 611.} We find a relatively flat temperature profile, in good agreement with the spectroscopic X-ray measurements from \citet{2006ApJ...647...25B}. In addition, \citet{2011A&A...528A..73D} performed an analysis of this cluster using X-ray data from \chandra\ and lensing results from \hubble.  Unlike our analysis, they identify it as a cool-core cluster. However, their overall temperature profile is in good agreement with our results. They found a central temperature of 6 keV, a peak of 8 keV at 100 kpc (20\arcsec), and a slow drop in temperature to 5keV at 600 kpc (150\arcsec). There is a known error in the ACCEPT results for this cluster, and so they are not included in the plot.  }
	\label{fig:a611}
\end{figure}

\begin{figure}
	\includegraphics[width=\linewidth]{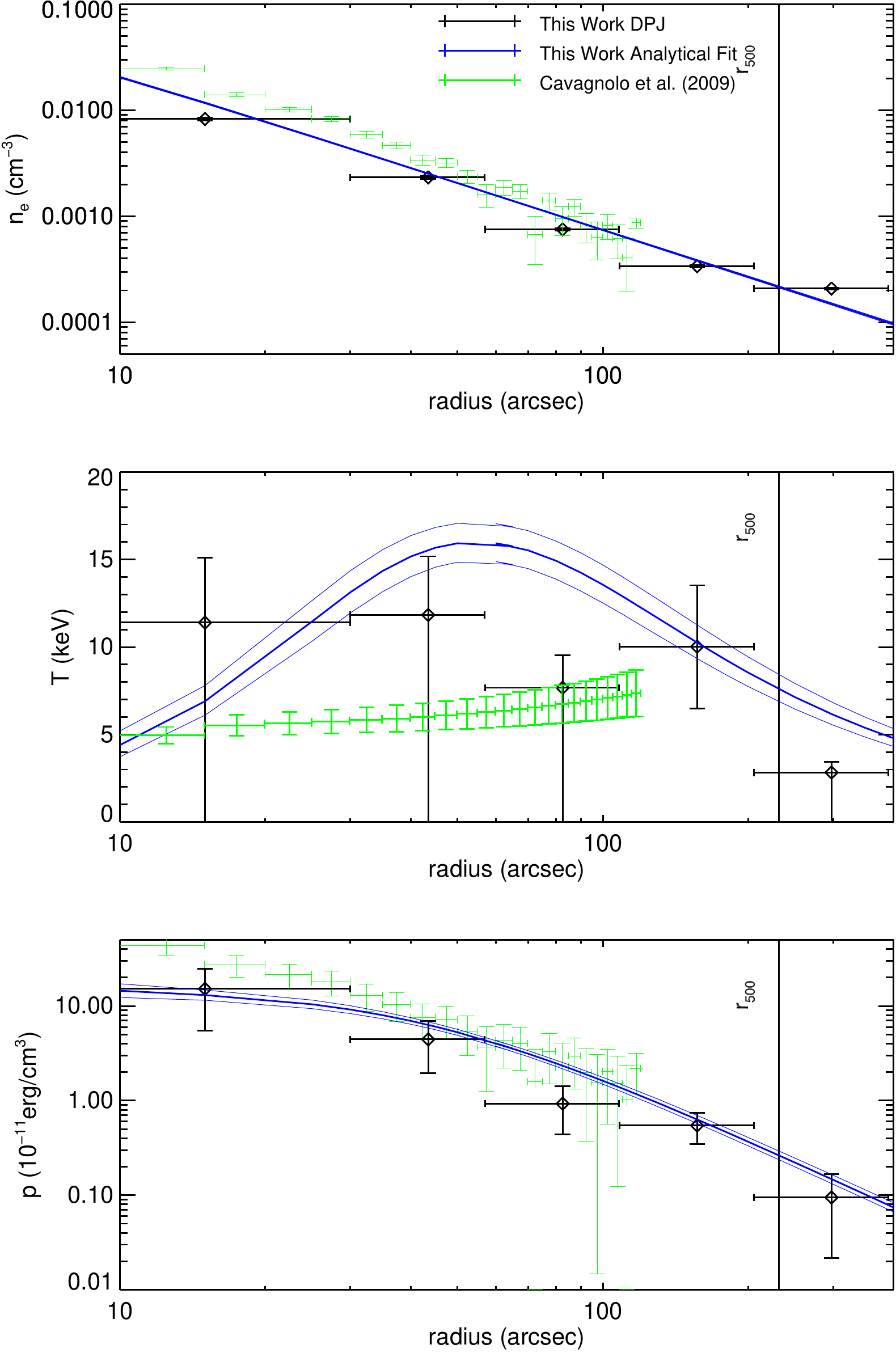}
	\caption{{\bf MS 2137.} In our smooth parametric fits, we detect a temperature drop in the center at a significance of $5.3\sigma$ and a temperature drop in the outskirts at a significance of $11.3\sigma$. Our fits are mostly consistent with the ACCEPT results, although our smooth parametric fit indicates a significantly higher temperature at intermediate radii. In reasonable agreement with the ACCEPT results, \citet{2009MNRAS.398..438D} obtained a spectroscopic temperature profile using \chandra\ with a 4 keV central temperature, a peak of 5.5 keV at 100 kpc (around 20\arcsec), and decreasing to 3.5 keV in the outskirts at 500 kpc (around 110\arcsec).}
	\label{fig:ms2137}
\end{figure}

\begin{figure}
	\includegraphics[width=\linewidth]{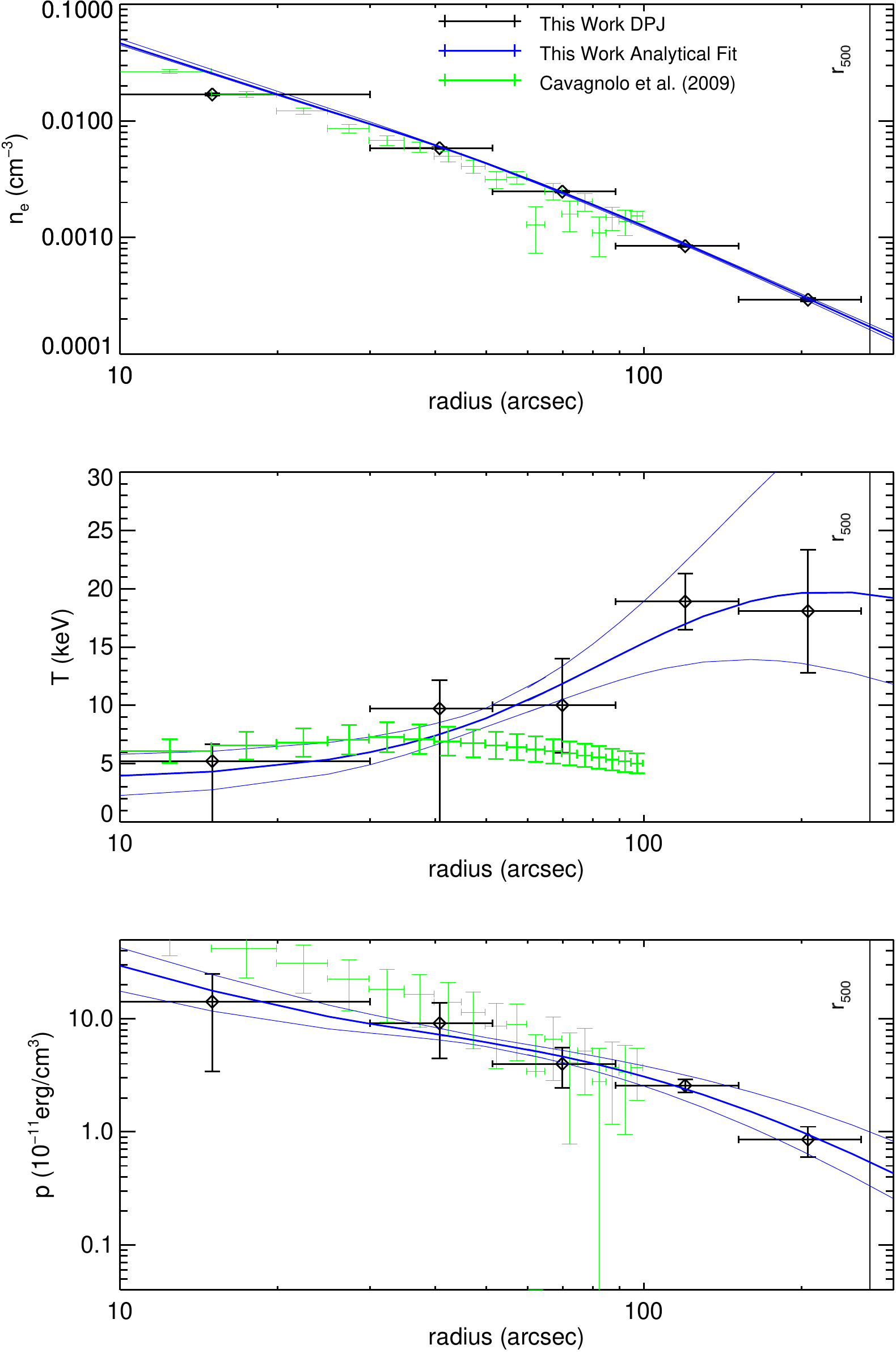}
	\caption{{\bf MACSJ1931.8-2634.} At overlapping radii, our results are fairly consistent with the ACCEPT results, except for a slight increase in our derived temperatures towards large radius. Based on our smooth parametric fits, there is some evidence for a drop in temperature at the cluster center ($\sim 2\sigma$). A detailed multi-wavelength study of this cluster was performed by \citet{2011MNRAS.411.1641E} using X-ray (\chandra), optical (Subaru), and radio (Very Large Array) data.  They found a cool core with AGN feedback (seen in the X-ray and in the radio), along with evidence of merging, suggesting a cool core that is currently being destroyed. They created a temperature map of the cluster from X-ray spectroscopy out to radii of 500 kpc (100\arcsec), and found an asymmetric temperature distribution.  Their azimuthally averaged temperature profile has a central temperature of 5 keV and peak of 10 keV, roughly consistent with our results.}
	\label{fig:macsj19318}
\end{figure}

\begin{figure}
	\includegraphics[width=\linewidth]{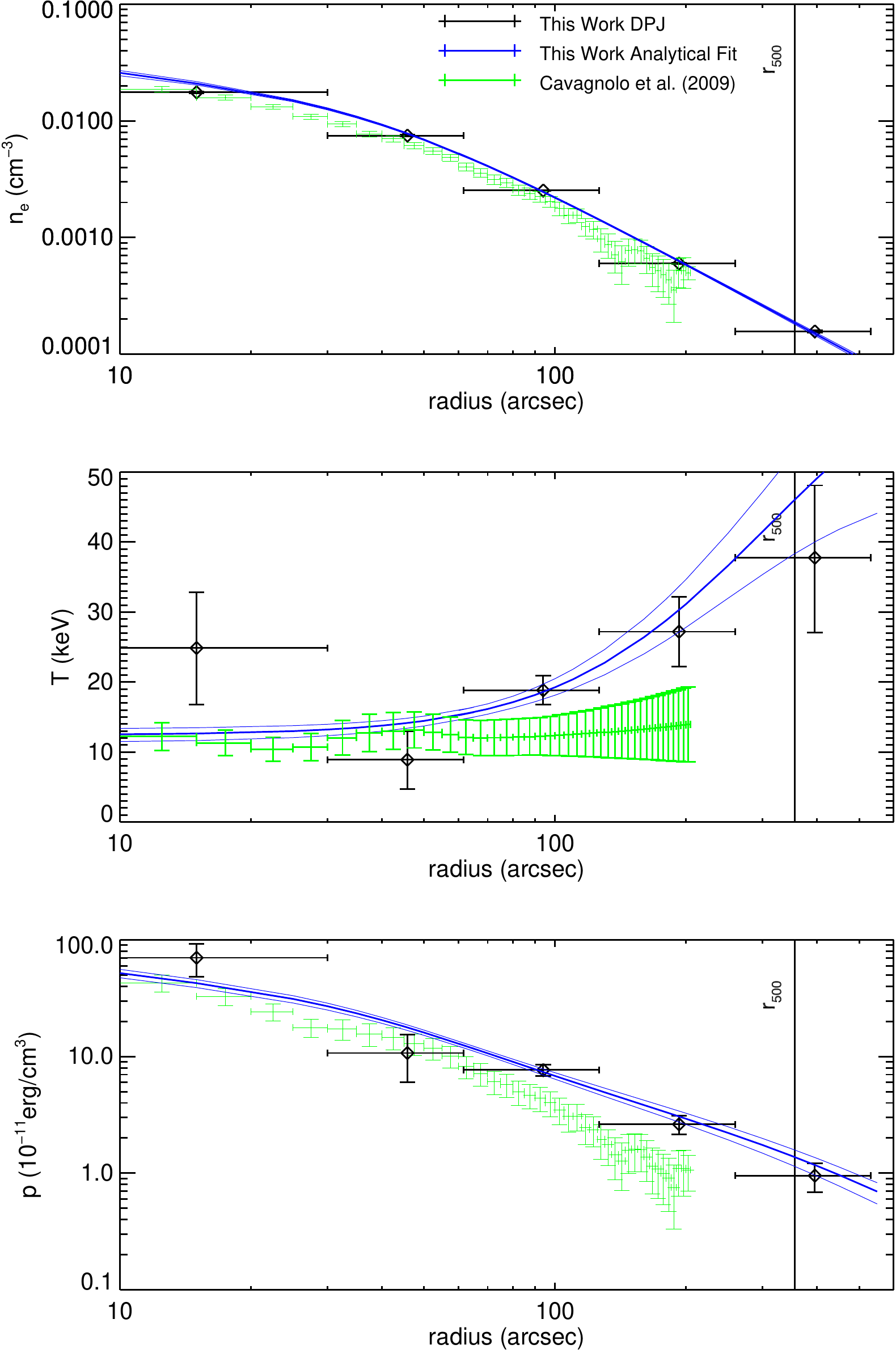}
	\caption{{\bf Abell S1063.} As with MACSJ1931.8-2634, our results are fairly consistent with the ACCEPT results other than a slight increase in temperature at large radius.  \citet{2012AJ....144...79G} found evidence of merging through both X-ray and optical observations, with an elongated X-ray emission feature in the same direction as two regions of high galaxy density.  They found high cluster temperatures of 12-17 keV from the center to $\sim$800 kpc (160\arcsec), in reasonable agreement with our results over the same range.}
	\label{fig:as1063}
\end{figure}

\begin{figure}
	\includegraphics[width=\linewidth]{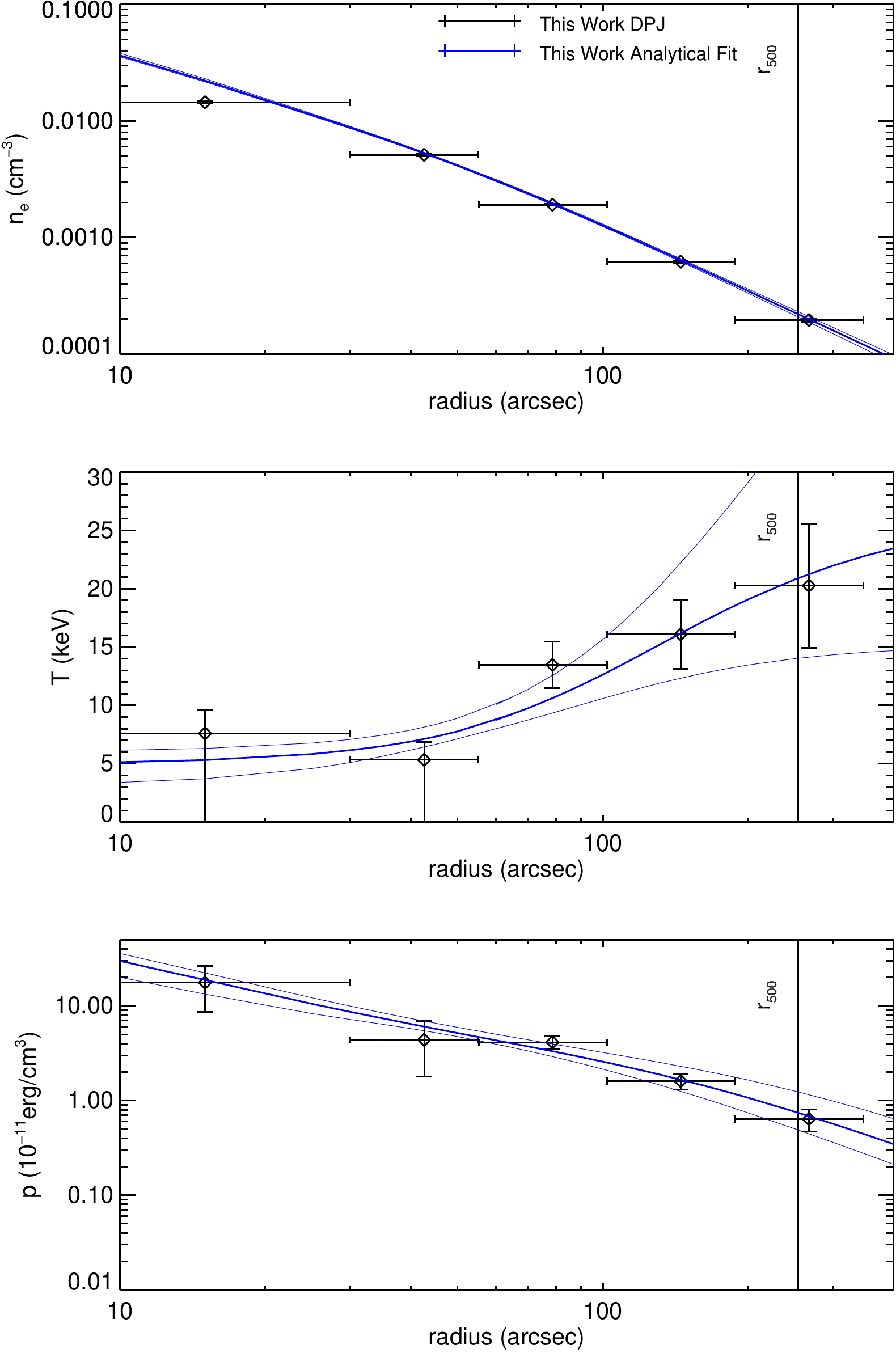}
	\caption{{\bf MACS J1115.8+0129.} The ACCEPT analysis finds densities systematically lower than ours, but our results are consistent with other studies (e.g., \citealt{2016MNRAS.456.4020M}), and so the ACCEPT profiles are not plotted. \citet{2014ApJ...794..136D} found temperature profiles from X-ray spectroscopy (\xmm\ and \chandra), showing a central temperature of around 3 keV rising to 8-9 keV, roughly matching our results over the same region. However, the \xmm\ data then show a decrease in the temperature profile near 800 kpc (160\arcsec), which is not seen in our results. }
	\label{fig:macsj11158}
\end{figure}

\begin{figure}
	\includegraphics[width=\linewidth]{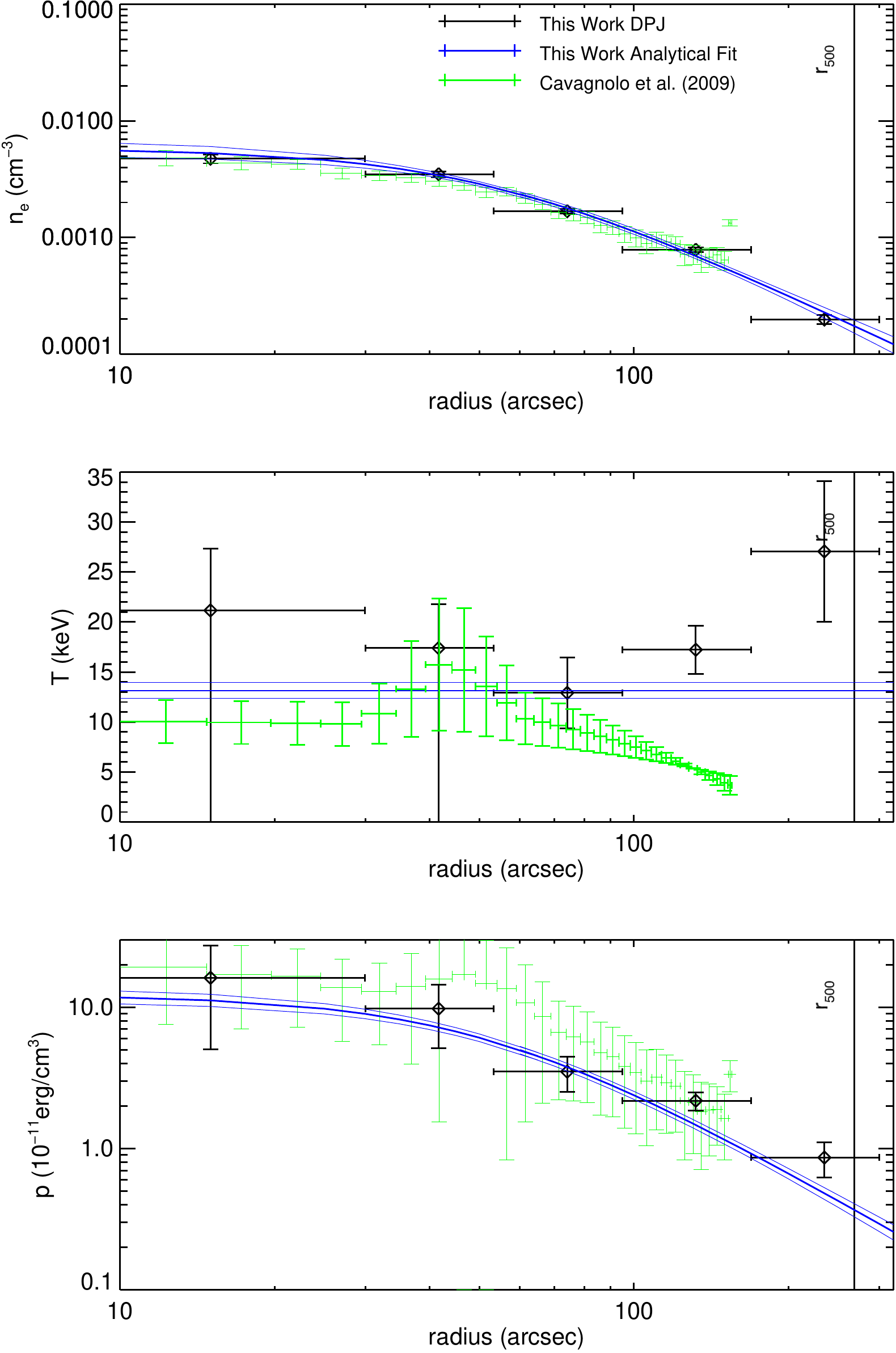}
	\caption{{\bf Abell 370.} Our results for this cluster are fairly consistent with the ACCEPT results, although they do not show a temperature drop at large radius. Several studies at different wavelengths have found that this cluster is not spherical.  This is a very popular lensing cluster, and many such studies have been performed \citep{2010MNRAS.405..257M}.  \citet{2005ApJ...625..108D} combined X-ray and SZ data to find the 3d shapes of galaxy clusters, including Abell 370, and found a triaxial morphology, elongated along the line of sight.   \citet{2010MNRAS.402L..44R} conducted a strong lensing analysis based on \hubble/ACS observations and reconstructed the mass distribution, which indicates an elongated, bimodal, mass distribution aligned with the \chandra\ X-ray luminosity maps.  The galaxy distribution also shows bimodality, suggesting a merging cluster.  \citet{2000ApJ...539...39G}, using OVRO SZ observations of this cluster, also found a smooth but non-spherical distribution.  }
	\label{fig:a370}
\end{figure}
\clearpage

\begin{figure}
	\includegraphics[width=\linewidth]{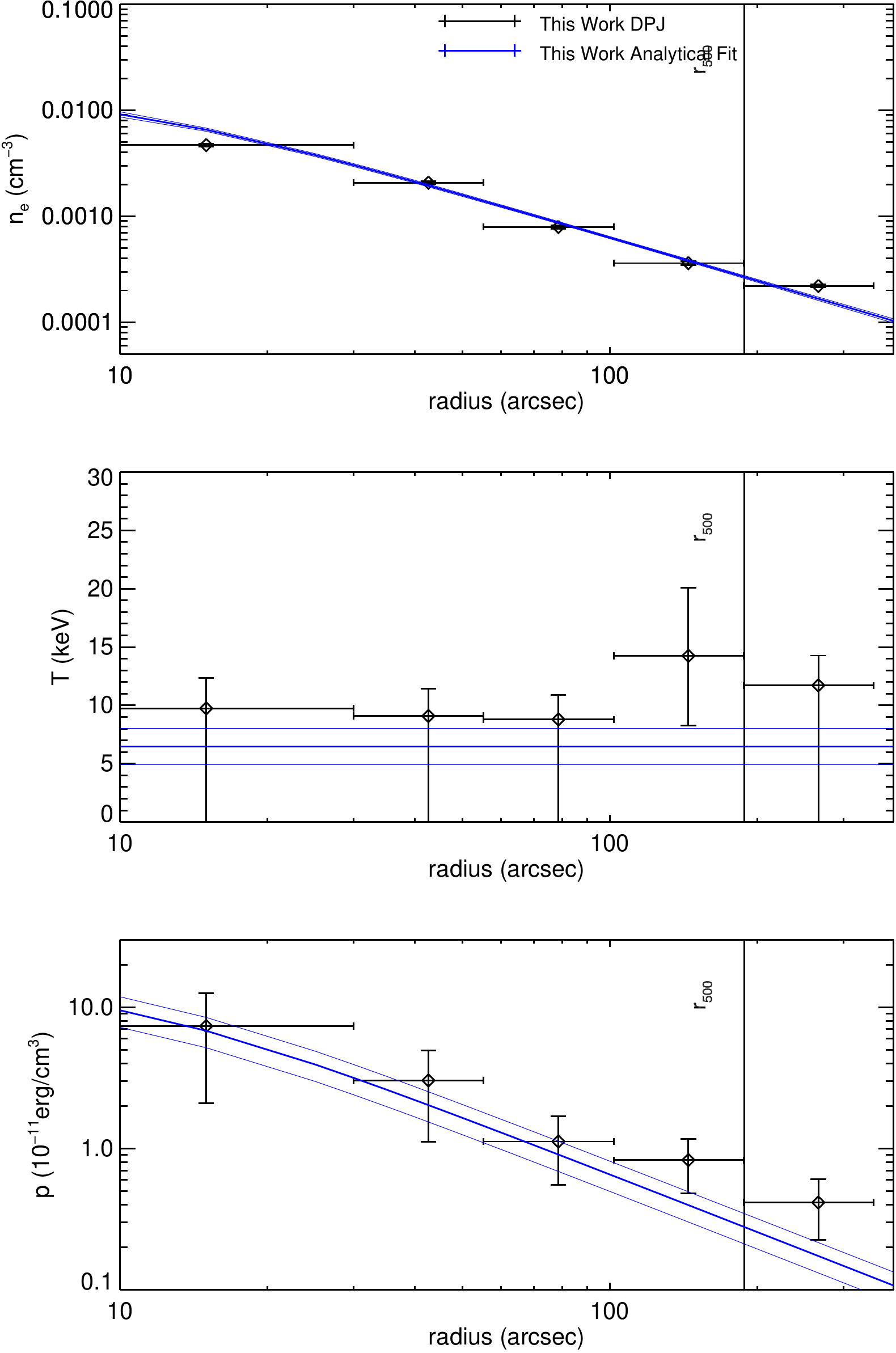}
	\caption{{\bf ZWCL 0024+17.} \citet{2005A&A...429...85Z} performed an \xmm\ study using imaging and spectroscopy, and found a temperature decrease at large radii (80\arcsec--180\arcsec) from the isothermal $\sim$4keV central region, which is fairly consistent with our results.  They also found an elongation in the X-ray hard ratio map on large scales, and substructure at large radii.  \citet{2004ApJ...601..120O} used \chandra\ spectroscopy and found a nearly isothermal profile of $\sim$4.5 keV out to 600 kpc (110\arcsec), again quite similar to our results.  \citet{2000A&A...353..124B} used \rosat\ to look at the cluster's X-ray morphology, and found a very small core radius in the surface brightness.  They also found an elongation in the X-ray emission, however it is consistent with a spherical model.  \citet{1998ApJ...498L.107T} obtained a mass map using strong lensing measurements, and found a relaxed distribution.  \citet{2010ApJ...714.1470U} performed a full-lensing analysis, including Subaru and \hubble/ACS/NIC3 observations, and looked at X-ray data along with simulations.  They suggest that the cluster is in a post-collisional state, with two clusters at the same line of sight, as well as finding the mass profile of the cluster.  \citet{2007ApJ...661..728J} noted that the X-ray surface brightness from \chandra\ is better fit by two isothermal beta models, suggesting that possibly one could be seeing two systems that are along the same line of sight. }
	\label{fig:cl0024p17}
\end{figure}

\begin{figure}
	\includegraphics[width=\linewidth]{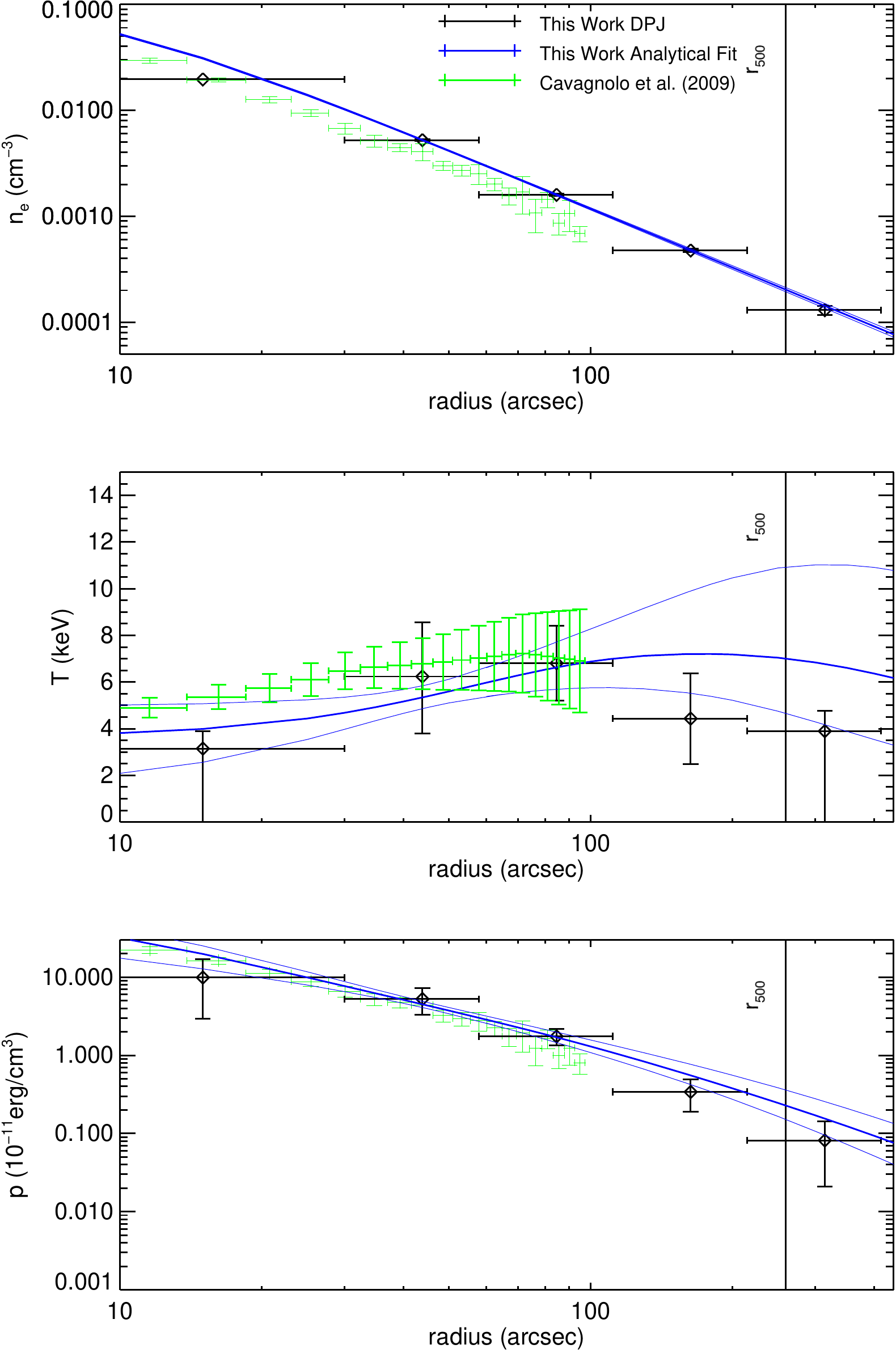}
	\caption{{\bf MACS J1532.9+3021.} Our results for this cluster are in good agreement with the ACCEPT results. \citet{2014ApJ...794..136D} obtained temperature profiles from X-ray spectroscopy (\xmm\ and \chandra) with the temperature profile showing a central value of 3-4 keV before rising to 8 keV at larger radii, consistent with our analysis. The \xmm\ profile decreased down to $\sim$4keV at 900 kpc ($\sim$180\arcsec), while the \chandra\ profile does not show the temperature decrease, instead plateauing at $\sim$9 keV. \citet{2013ApJ...777..163H} performed a \chandra, \xmm, VLA, and \hubble\ analysis.  They observed evidence of AGN feedback, a cold front, and X-ray cavities. They note a difference in the temperature map from the east and the west side of the cluster, and find a central temperature of 4 keV rising to 9 keV at 250 kpc ($\sim$50\arcsec). They also detect slight differences in the temperature profile at small radii ($<$25\arcsec) in different directions, with higher temperatures in the S and W directions. }
	\label{fig:macsj15329}
\end{figure}

\begin{figure}
	\includegraphics[width=\linewidth]{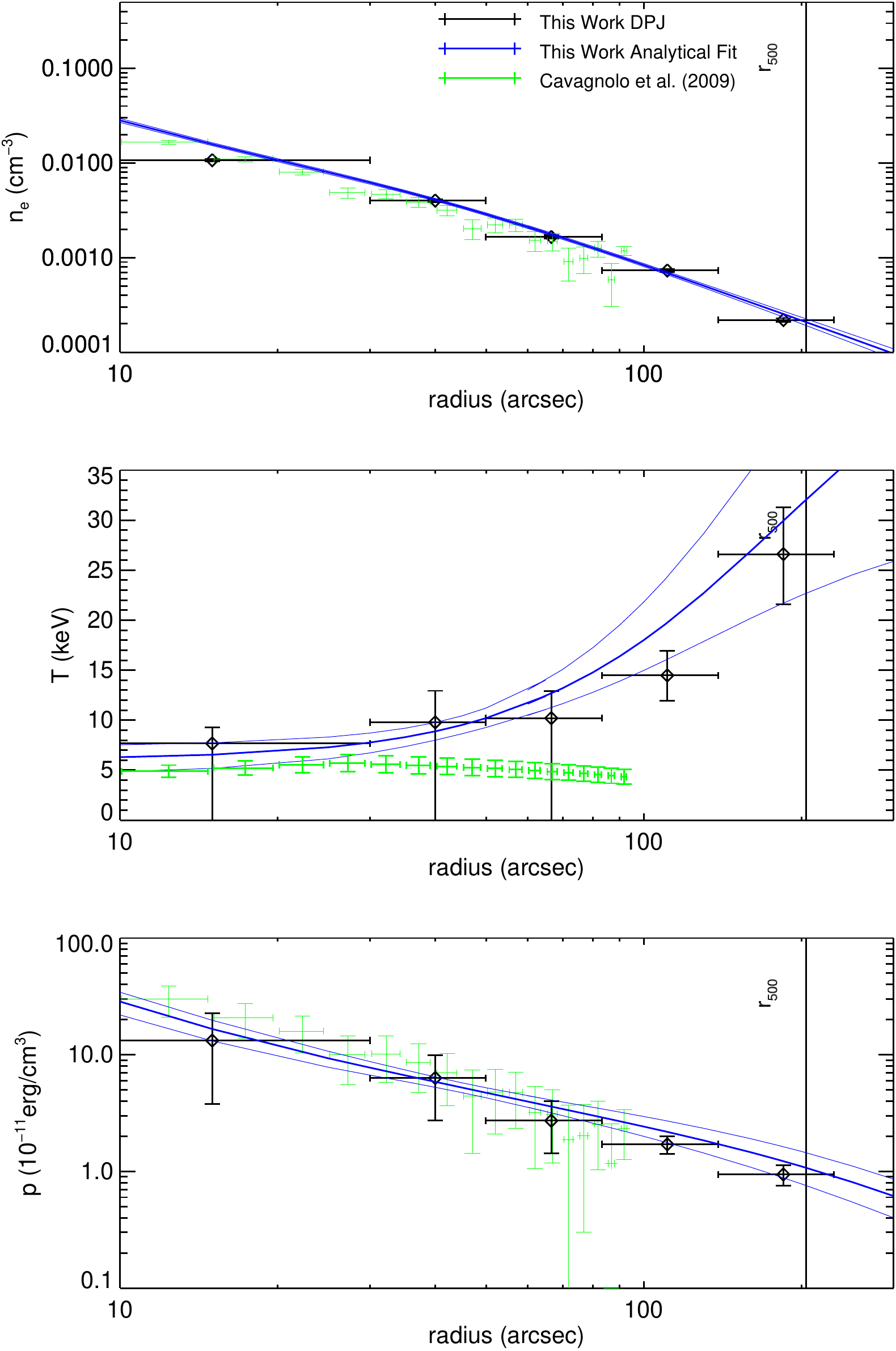}
	\caption{{\bf MACS J0429.6-0253.} Our analysis indicates a temperature increase at large radius that is not seen in the ACCEPT results, and our smooth parametric fits indicate a drop in temperature towards the core at a significance of $\sim 2\sigma$. \citet{2014ApJ...794..136D} found temperature profiles using X-ray spectroscopy from \xmm\ and \chandra, with the temperature profile showing a central temperature of 4 keV and a peak of 9 keV at 200 kpc (40\arcsec), roughly consistent with our results over the same range.}
	\label{fig:macsj04296}
\end{figure}

\begin{figure}
	\includegraphics[width=\linewidth]{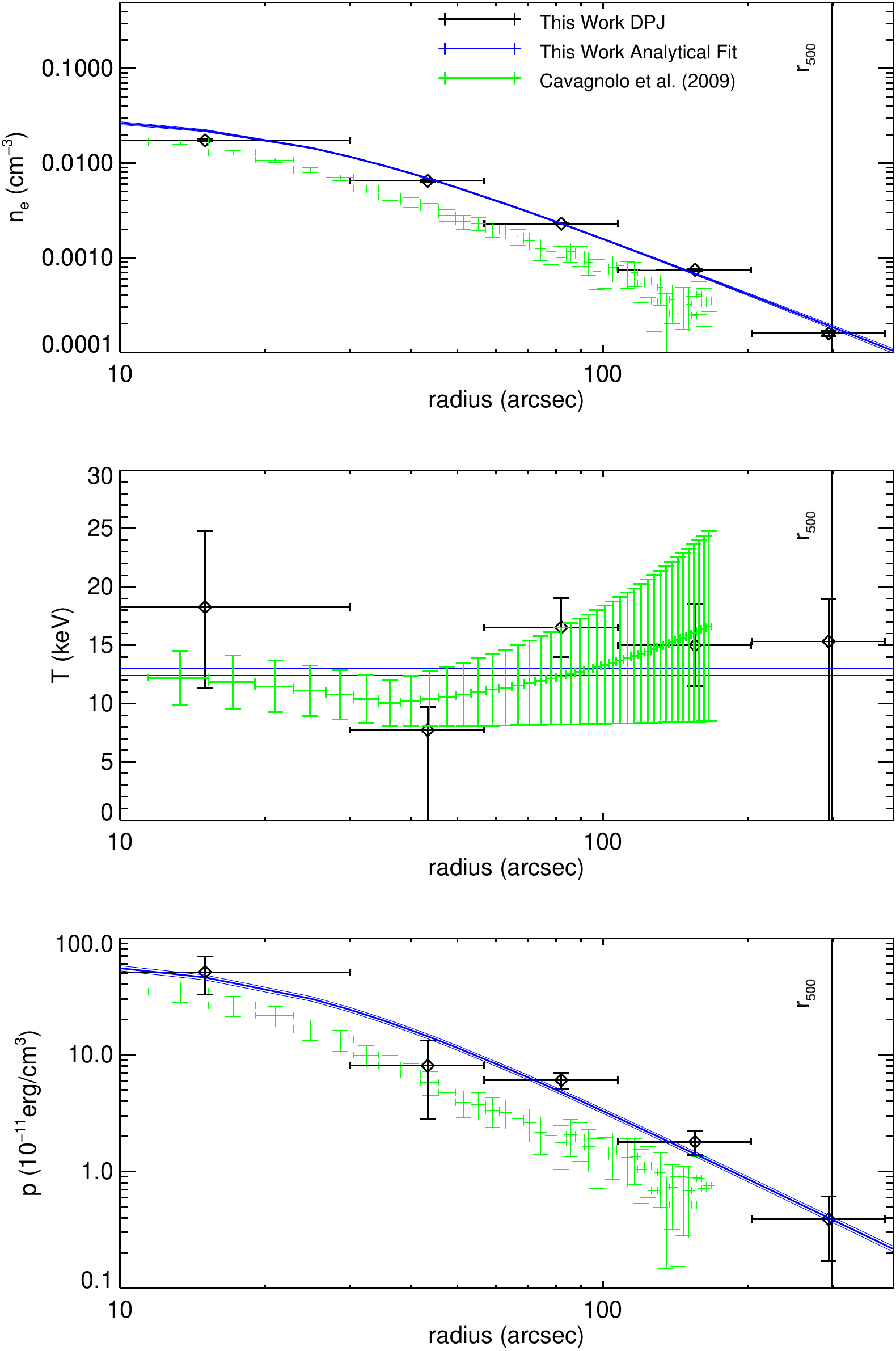}
	\caption{{\bf MACS J2211.7-0349.} Our densities and pressures are systematically and significantly higher than those found in the ACCEPT study.  Due to the lack of independent results for this cluster, it is not possible to determine which analysis may be in error.}
	\label{fig:macsj22117}
\end{figure}
\clearpage 

\begin{figure}
	\includegraphics[width=\linewidth]{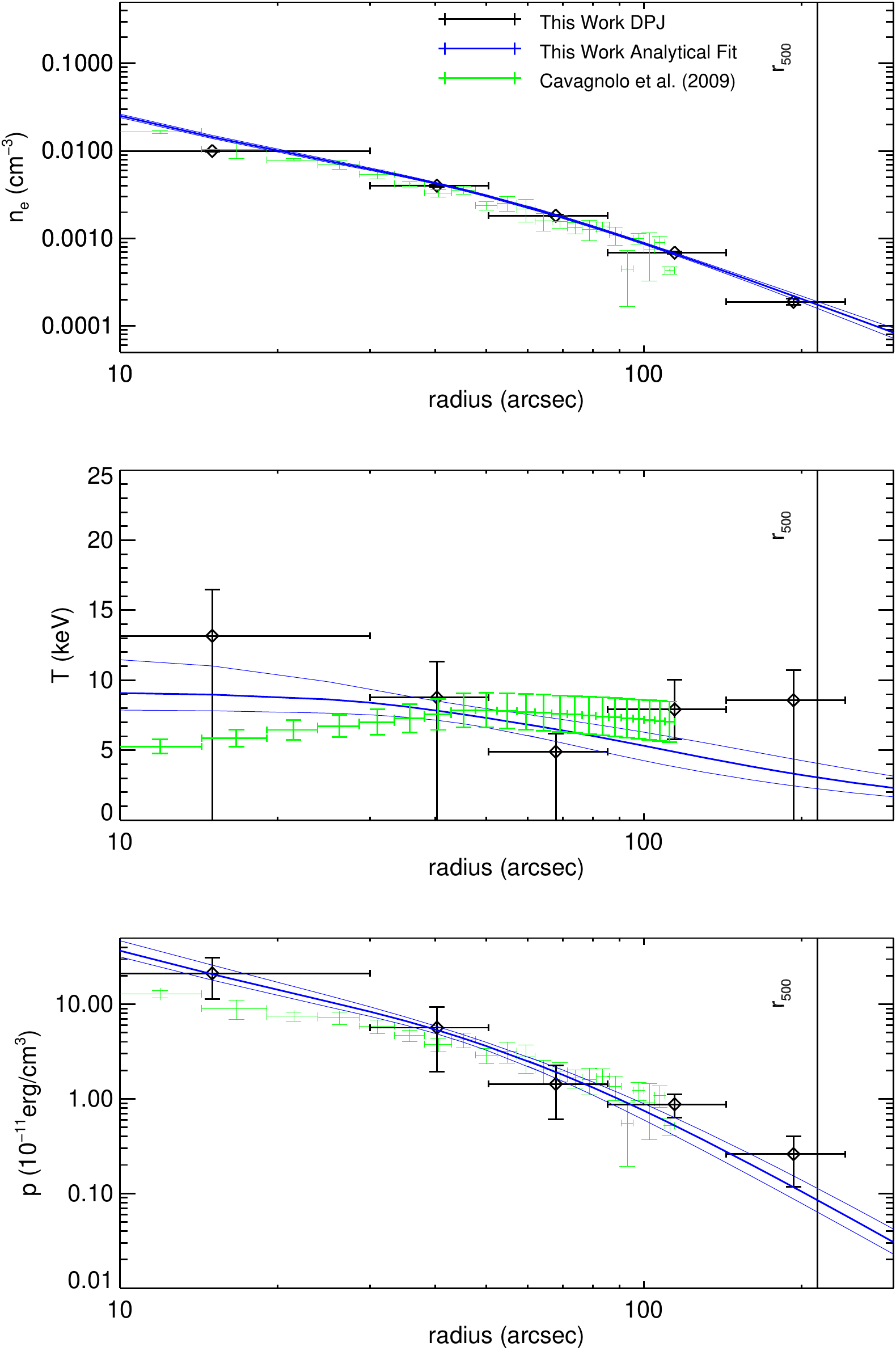}
	\caption{{\bf MACS J1720.3+3536.} Our data indicate a decrease in the temperature in the outskirts at a significance of $6.5\sigma$ and are generally in good agreement with the ACCEPT results.  \citet{2014ApJ...794..136D} obtained temperature profiles using X-ray spectroscopy from \xmm\ and \chandra, and found that the temperature profiles from the two instrument have different shapes. Both have a central temperature of $\sim$3.5 keV. \xmm\ jumps to 10 keV then slowly decreases down to $\sim$2 keV at 900 kpc (170\arcsec), consistent with our results outside of the core region. In contrast, \chandra\ slowly reaches 10 keV but then drops down to temperatures close to zero around 400kpc (70\arcsec).  Essentially, the outer temperature slope is steeper for \chandra, but the inner temperature slope is steeper for \xmm.  \citet{2008ApJS..174..117M} used \chandra\ spectroscopy to obtain an isothermal temperature of $\sim$6.1 keV for radii $< r_{500}$ (7.8 keV if the core region is excised), broadly consistent with our results.}
	\label{fig:macsj17203}
\end{figure}

\begin{figure}
	\includegraphics[width=\linewidth]{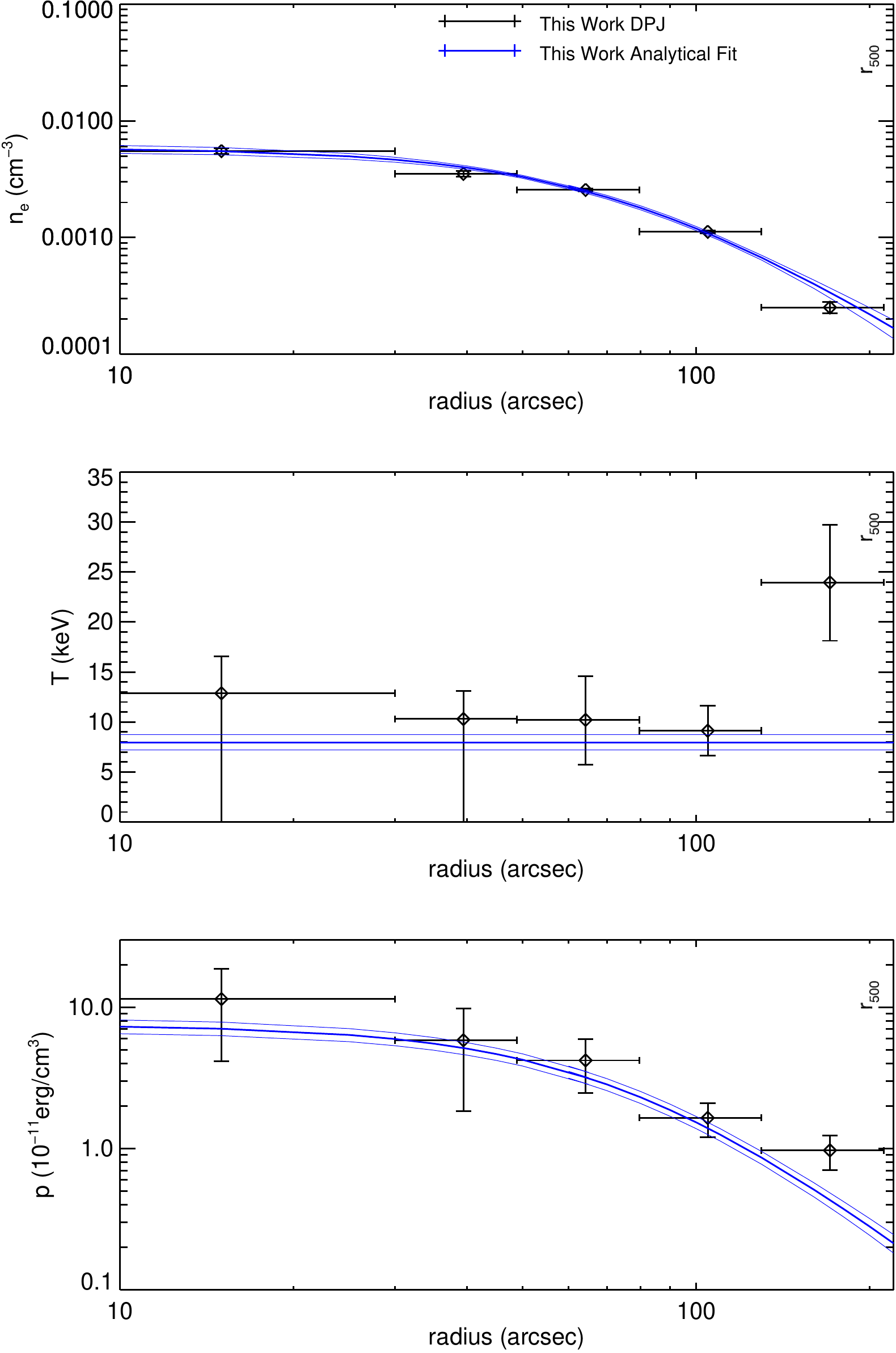}
	\caption{{\bf MACS J0416.1-2403.} Several studies have identified this cluster as a binary merger system.  \citet{2012MNRAS.420.2120M}, using optical and X-ray data, found this system to contain a possible binary head-on collision, just after the first collision. \citet{2015MNRAS.446.4132J} conducted a joint X-ray and optical study, and unveiled a large structure associated with a line-of-sight filament that could not be seen in the X-ray.  A large offset in the radial velocity between two subclusters was found, and their temperatures were determined to be 10 and 13.6 keV, slightly higher than our isothermal smooth parametric fit.  \citet{2015ApJ...812..153O} created temperature, pressure, and entropy maps using a multiwavelength analysis of \chandra, the VLA, the GMRT, and \hubble.  The temperature map is elongated, and has a relatively high mean temperature of 10 keV.  The radio halo is also elongated along the same direction.}
	\label{fig:macsj04161}
\end{figure}

\begin{figure}
	\includegraphics[width=\linewidth]{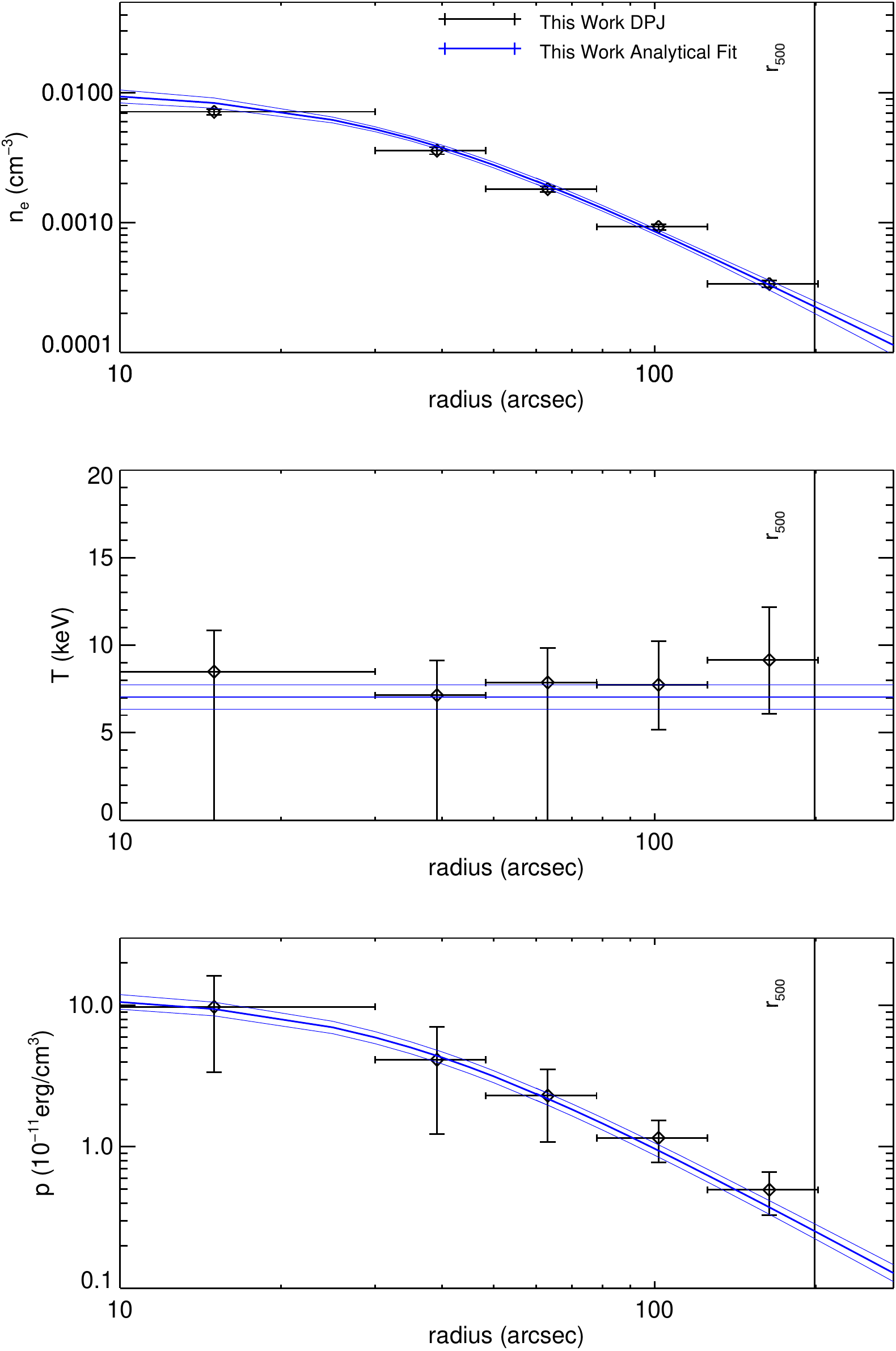}
	\caption{{\bf MACS J0451.9+0006.} \citet{2008ApJS..174..117M} used \chandra\ spectroscopy to obtain an isothermal temperature of $\sim 5.6$~keV at radii $< r_{500}$ (4.8~keV if the central region is excised). This is somewhat lower, but reasonably consistent with, the approximately isothermal profile found in our analysis. \citet{2012MNRAS.420.2120M} found this cluster to have highly irregular morphology, but it was not classified as an extreme or active merger due to its small BCG-Xray peak and BCG-Xray center separations.}
	\label{fig:macsj04519}
\end{figure}

\begin{figure}
	\includegraphics[width=\linewidth]{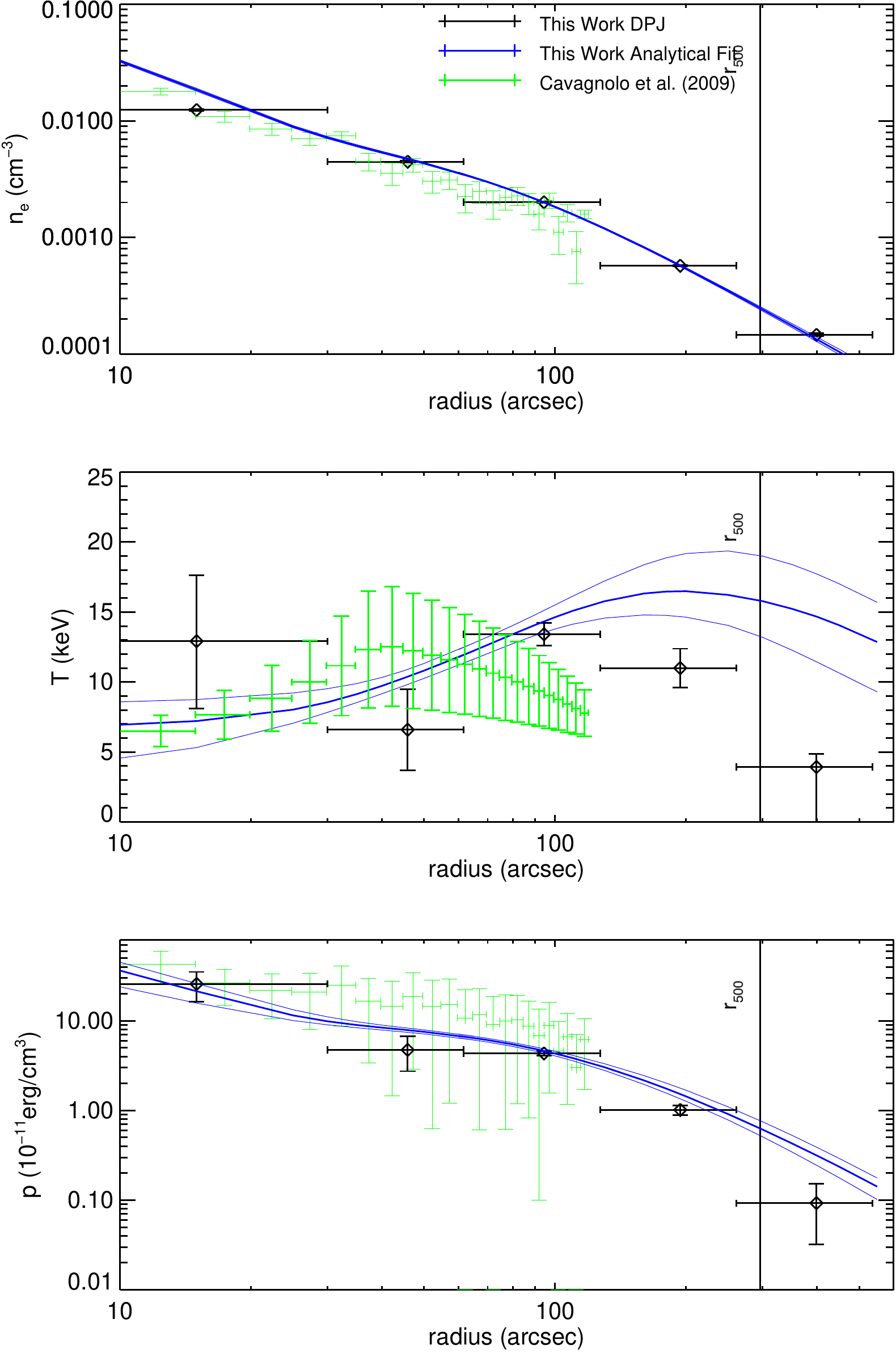}
	\caption{{\bf MACS J0417.5-1154.}From our smooth parametric fits, we detect a cool-core at a significance of $3.3\sigma$, and both our density and temperature profiles are reasonably consistent with the ACCEPT results. \citet{2012MNRAS.420.2120M} identified this cluster as a primary candidate for a binary, head-on collision type merger: the X-ray core aligns with one of two optical cores, but the X-ray emission bleeds into the outskirts and meets with the second of the optical cores. }
	\label{fig:macsj04175}
\end{figure}

\begin{figure}
	\includegraphics[width=\linewidth]{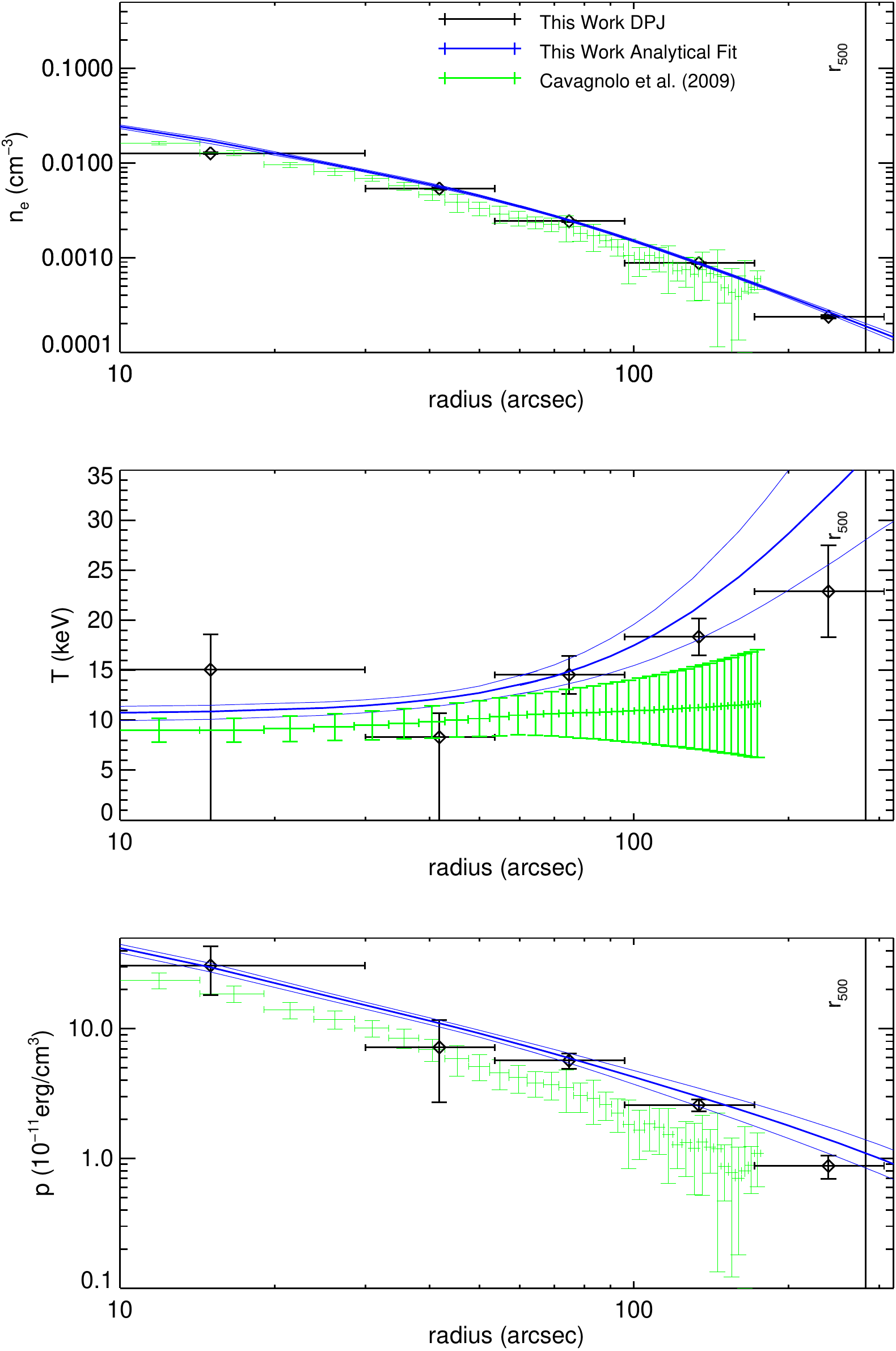}
	\caption{{\bf MACS J1206.2-0847.} Both our density and temperature profiles are in reasonable agreement with the ACCEPT results, and our smooth parametric fits indicate a drop in temperature near the core at a significance of $2.3\sigma$. \citet{2014ApJ...794..136D} obtained temperature profiles using X-ray spectroscopy from \xmm\ and \chandra.  The \chandra\ profile has a large temperature peak in the center (15 keV), dropping to a nearly isothermal profile of 10 keV to the maximum radius probed by the data (1000 kpc or 180\arcsec).   The \xmm\ profile, on the other hand, has a cool core of $\sim$7~keV, which increases with radius to 10 keV before dropping to 5 keV in the outskirts. Our analysis is consistent with these results at intermediate radii, but we find an approximately isothermal profile into the core and a slight temperature increase at large radius. Several studies, using X-ray, optical, and/or lensing data have found this cluster to be relaxed \citep{2009MNRAS.392.1509G,2012MNRAS.420.2120M,2012ApJ...755...56U}.}
	\label{fig:macsj12062}
\end{figure}

\begin{figure}
	\includegraphics[width=\linewidth]{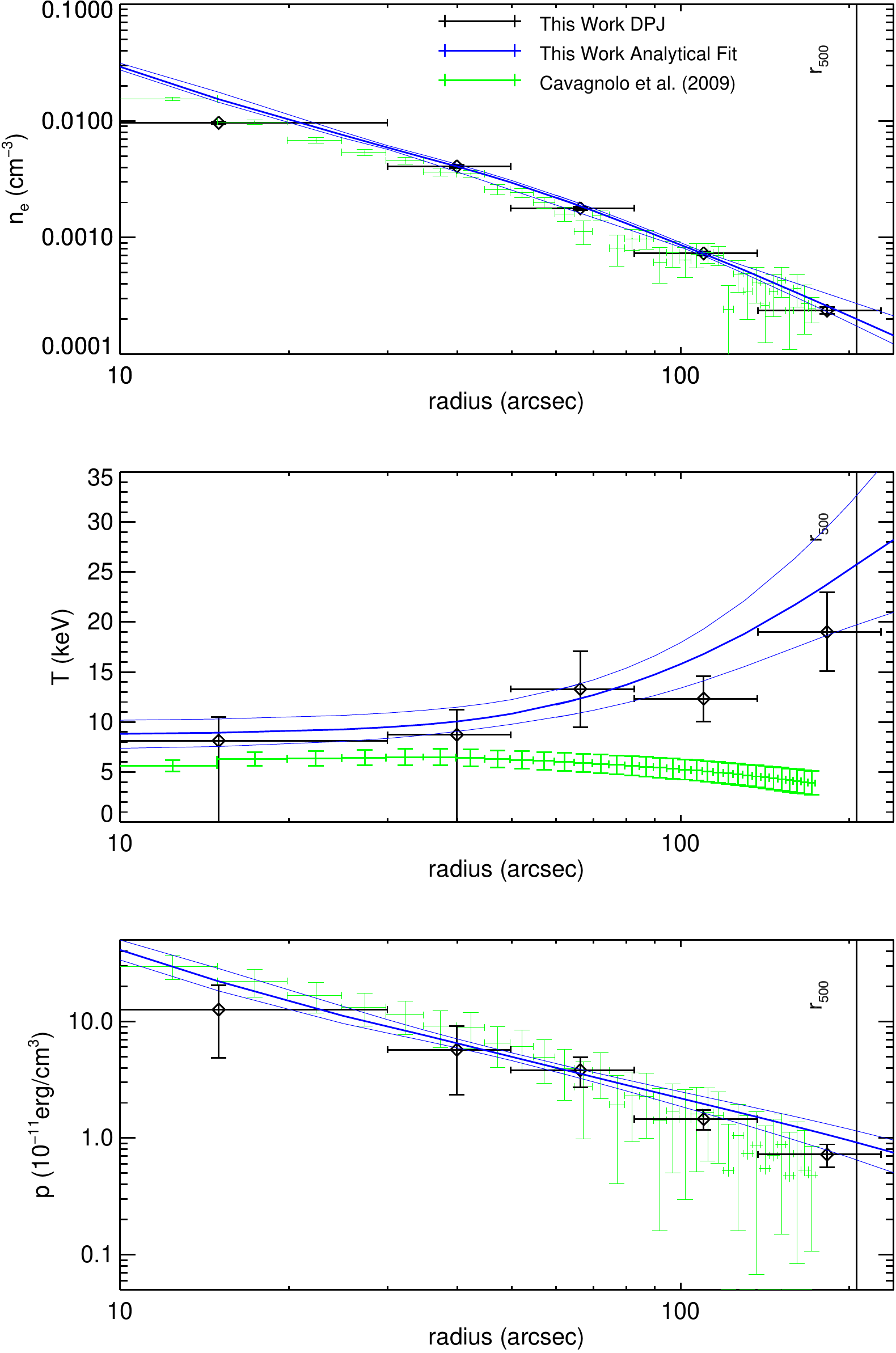}
	\caption{{\bf MACS J0329.6-0211.} Our smooth parametric fit finds a temperature drop in the core at modest significance ($2.1\sigma$). Compared to ACCEPT, our temperature profile is notably higher outside of the core region. \citet{2008ApJS..174..117M} found a temperature of 4.5 keV for $r<r_{500}$, and 4.4 keV for radii within $0.15 < r <1 r_{500}$, in good agreement with the ACCEPT results. \citet{2012MNRAS.420.2120M} found this to be a relaxed cluster according to X-ray and optical alignment.  \citet{2006ApJ...641..752K}, using \chandra\ and \xmm\ data, found this cluster to follow a standard cool-core temperature profile, except for a dip in temperature at intermediate radii (around 100 kpc), which could hint at some substructure.  \citet{2014ApJ...781....9G} found a possible radio minihalo centered in the cluster using VLA data, filling out much of the core to 150 kpc.}
	\label{fig:macsj03296}
\end{figure}

\begin{figure}
	\includegraphics[width=\linewidth]{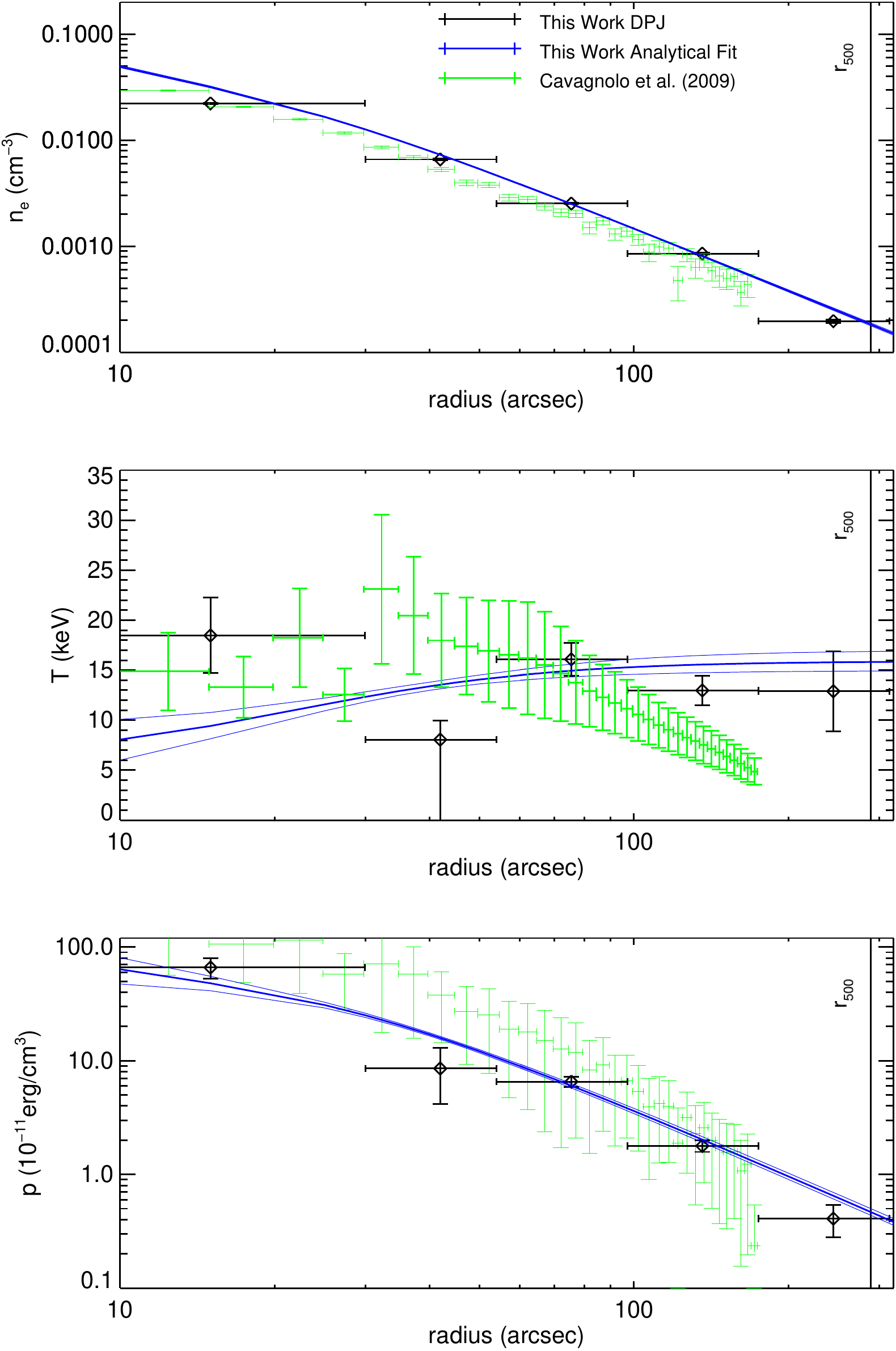}
	\caption{{\bf MACS J1347.5-1144.} We detect a cool-coor in our smooth parametric fit at a significance of $3.1\sigma$, but we do not find strong evidence for a temperature drop in the outskirts. Compared to the ACCEPT results, our temperature is somewhat higher at large radius. \citet{2014ApJ...794..136D} obtained temperature profiles with a central temperature of 5 keV, with a peak of 17 keV that decreases in the outskirts, in good agreement with the ACCEPT results.  \citet{2008ApJS..174..117M} used \chandra\ imaging and spectra and assumed isothermal profiles, finding 12.2 keV for $r<r_{500}$, and 11.7 keV for $0.15<r<1 r_{500}$, roughly consistent with our profiles. \citet{2010MNRAS.403.1787L} studied this cluster using the optical waveband, and found a possible filament between a cluster 7 Mpc away.  This cluster has many SZ-focused studies devoted to it.  \citet{2011A&A...534L..12F} compared radio GMRT data with MUSTANG, \chandra, and \xmm\ X-ray data, and found a correlation between the intracluster radio emission and X-ray and SZ emission.  \citet{2011ApJ...734...10K} made a high resolution SZ map using MUSTANG, and was able to find substructure 20\arcsec from the center in the form of gas that has been heated through shocks caused by mergers. \citet{2011A&A...534L..12F} combined X-ray and SZ data to study the cluster, finding substructure near the core and very hot gas in the cluster up to 20 keV without the use of X-ray spectroscopy.  \citet{2001PASJ...53...57K} mapped the SZ signal at 150 GHz with 13\arcsec resolution using the Nobeyama telescope, and detected the excess SZ emission at 20\arcsec.  \citet{2001ApJ...552...42P} on the other hand, made a similarly resolved map, and could not constrain substructure, although they did observe that the signal was not completely spherical.  Most recently, \citet{2016ApJ...832...26S} was able to place an upper limit on the amplitude of the peculiar velocity of the cluster using 5-band SZ measurements.}
	\label{fig:rxj1347}
\end{figure}

\begin{figure}
	\includegraphics[width=\linewidth]{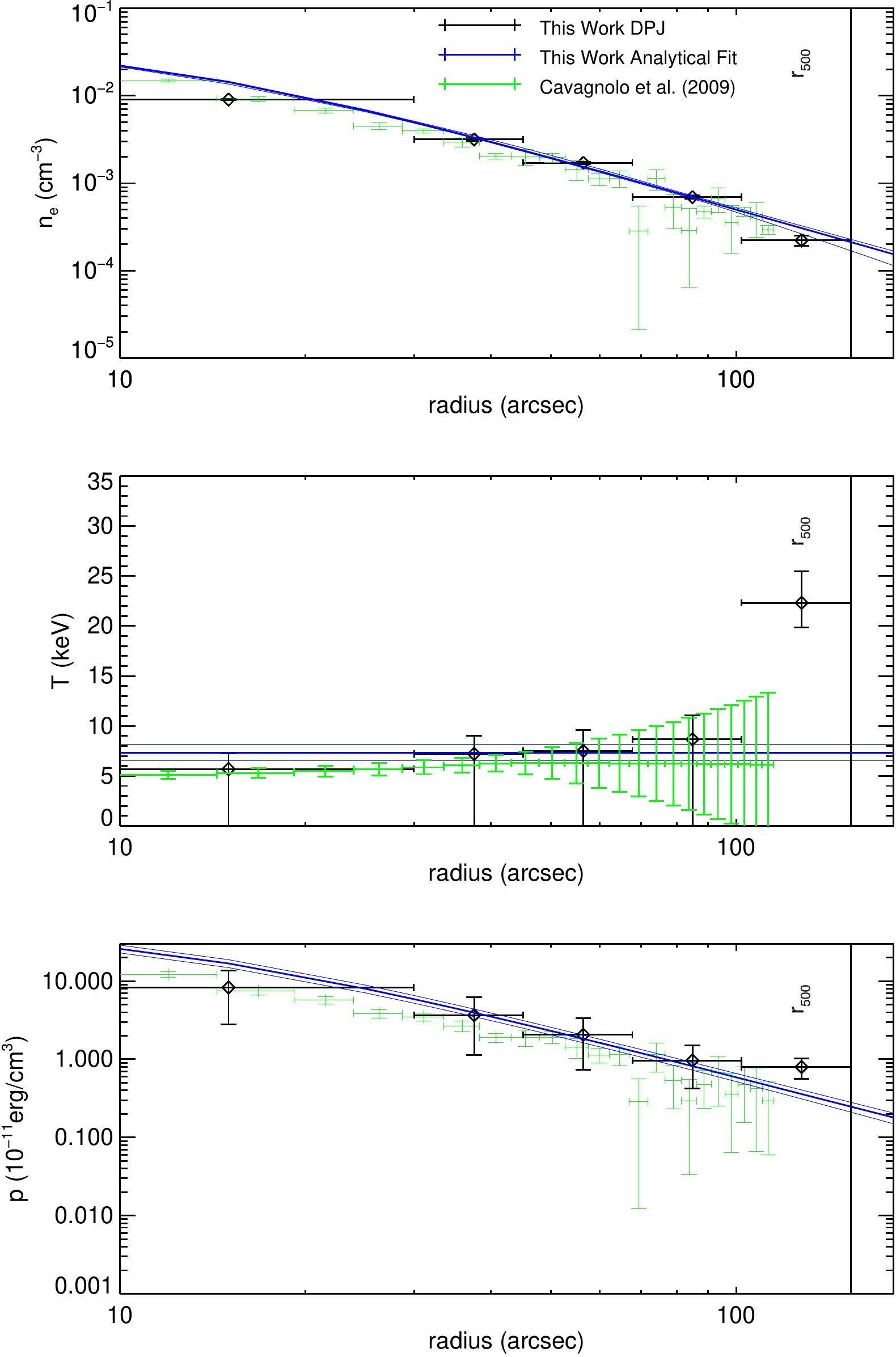}
	\caption{{\bf MACS J1311.0-0310.} Our results for this cluster are in excellent agreement with the ACCEPT results.}
	\label{fig:macsj13110}
\end{figure}

\begin{figure}
	\includegraphics[width=\linewidth]{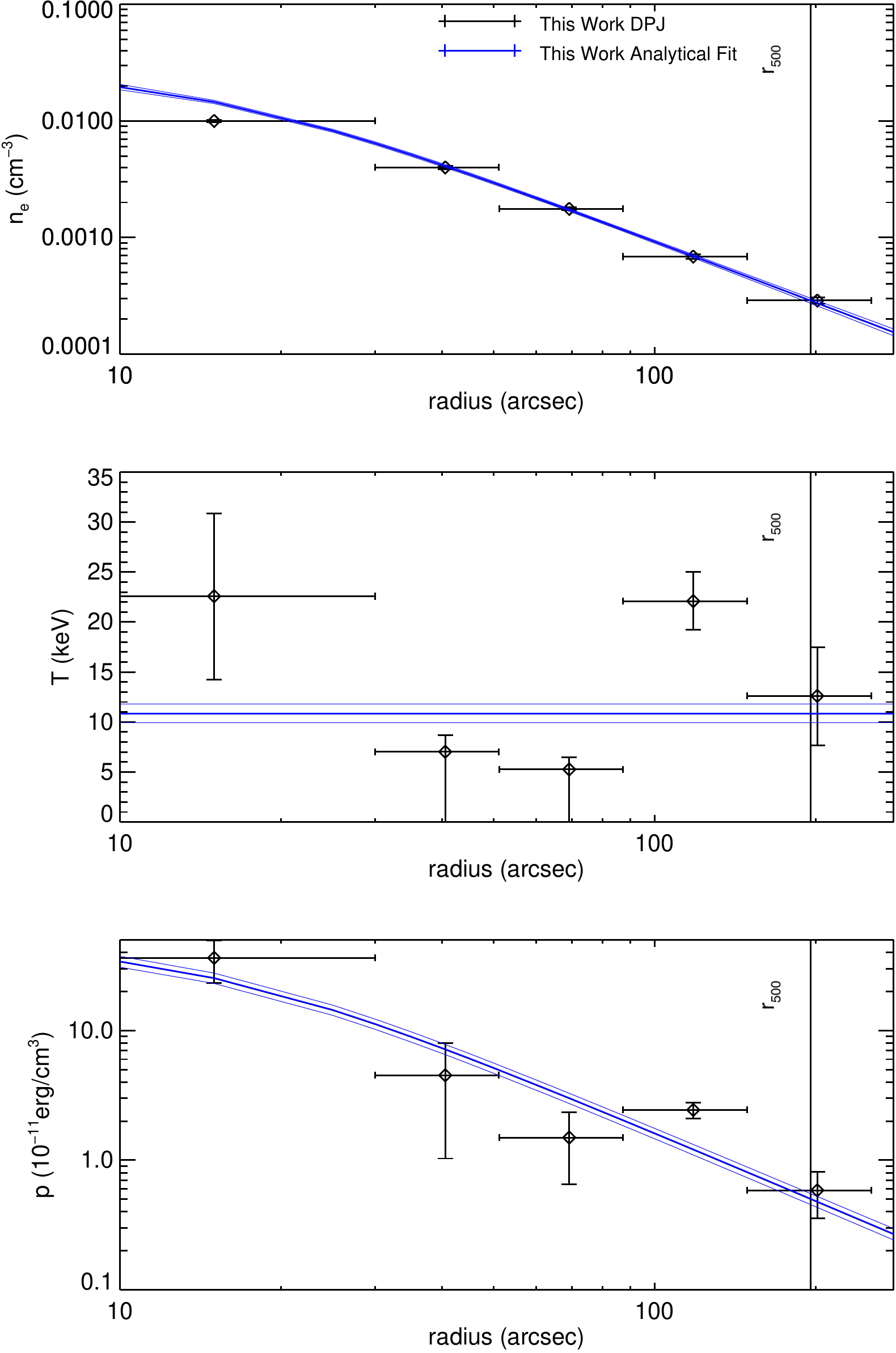}
	\caption{{\bf MACS J0257.1-2325.} This cluster was part of the ACCEPT study, but the density profile is significantly low compared to our and other results \citep{2016A&A...590A.126A}, and so the ACCEPT profile is not plotted. Through the optical, \citet{2008MNRAS.389.1240K} found that, on large scales, this system has filaments directed into the cluster. }
	\label{fig:macsj02571}
\end{figure}

\begin{figure}
	\includegraphics[width=\linewidth]{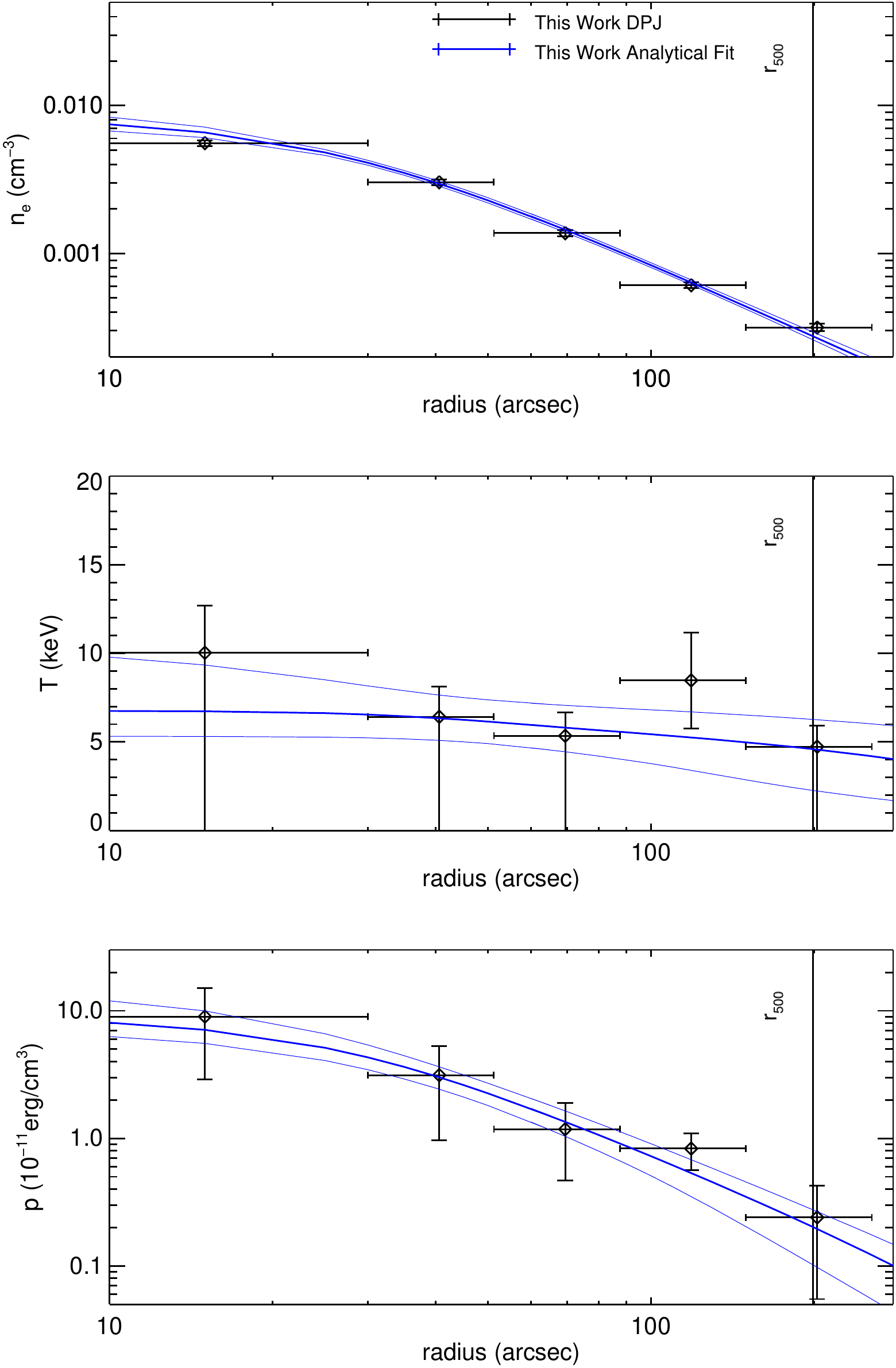}
	\caption{{\bf MACS J0911.2+1746.} \citet{2008MNRAS.389.1240K}, using optical data, found this cluster consists of two subclusters with different masses separated by 1 Mpc.  The X-ray centroid was found to be significantly offset from the galaxy surface density for the smaller subcluster, suggesting that this system has undergone a recent merger.}
	\label{fig:macsj09112}
\end{figure}

\begin{figure}
	\includegraphics[width=\linewidth]{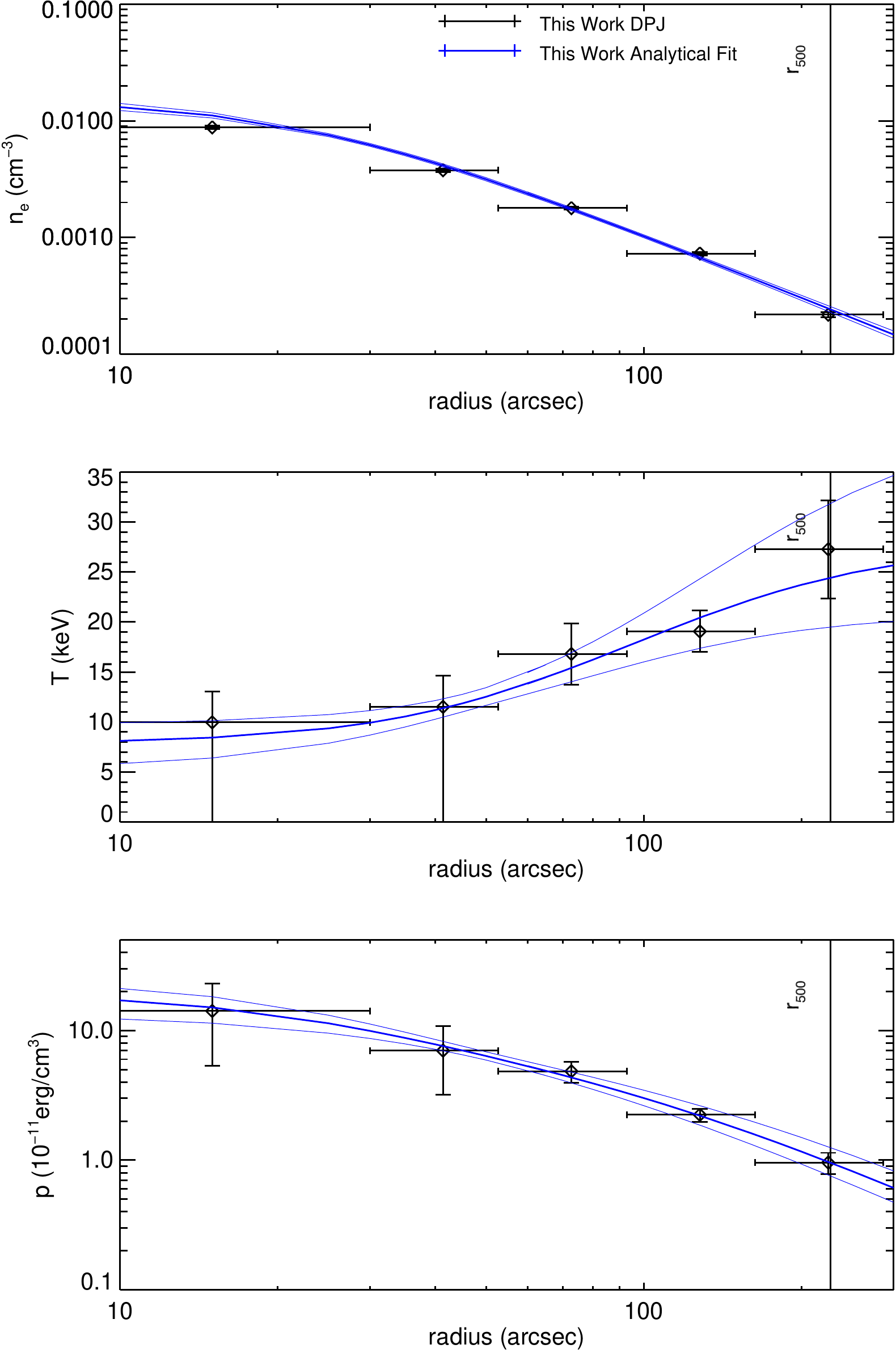}
	\caption{{\bf MACS J2214.9-1359.} This cluster is part of the ACCEPT study, but the density profile is systematically lower than both ours and two other analyses \citep{2006ApJ...647...25B,2016A&A...590A.126A}, and so the ACCEPT profile is not plotted. Our smooth parametric fits indicate a temperature drop in the center at a significance of $2.2\sigma$. \citet{2012MNRAS.420.2120M} found this cluster to have the most relaxed morphological denotation, with a prominent cool core and perfect alignment between the X-ray peak and BCG.}
	\label{fig:macsj22149}
\end{figure}

\begin{figure}
	\includegraphics[width=\linewidth]{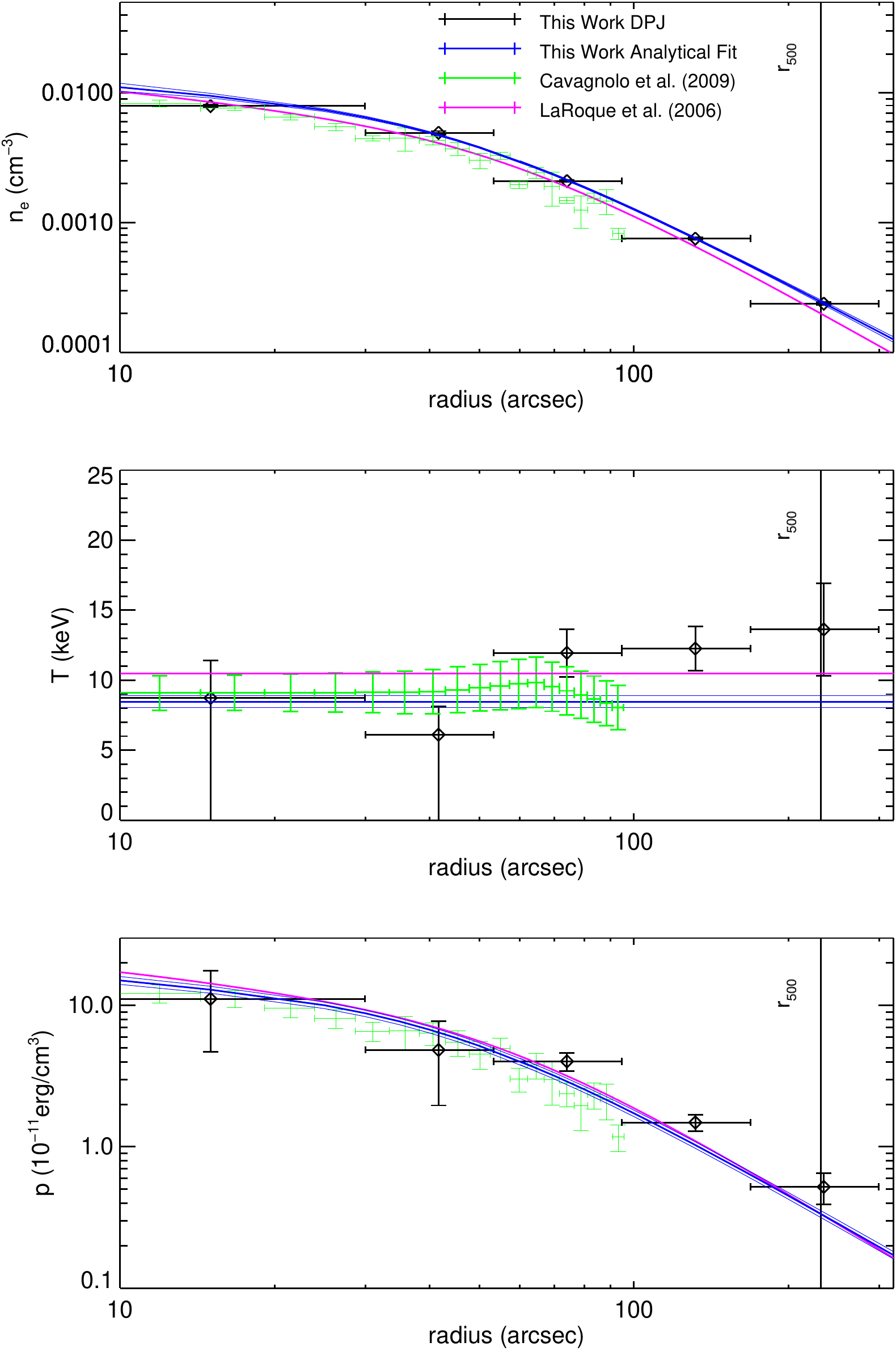}
	\caption{{\bf MACS J0018.5+1626.} The profiles we obtain for this cluster are in excellent agreement with the ACCEPT and \citet{2006ApJ...652..917L} results.  Several studies have found this cluster to be relaxed \citep[e.g.,][]{2005ApJ...633..781K}, however, \citet{2007A&A...476...63S} found evidence for it to be undergoing a merger near the center of the cluster using \xmm\ observations.   They also found a temperature profile out to $r_{200}$, showing a clear decrease in the outskirts, to 4 keV, which does not match the approximately isothermal profile found in our analysis.  \citet{2003MNRAS.340.1261W} found no evidence of a cool core, but they did find evidence for a merger in the center of the cluster due to the non-spherical X-ray shape in that region.  Overall, they find an isothermal temperature of $\sim$9.13 keV, consistent with our results.}
	\label{fig:cl0016}
\end{figure}
\clearpage

\begin{figure}
	\includegraphics[width=\linewidth]{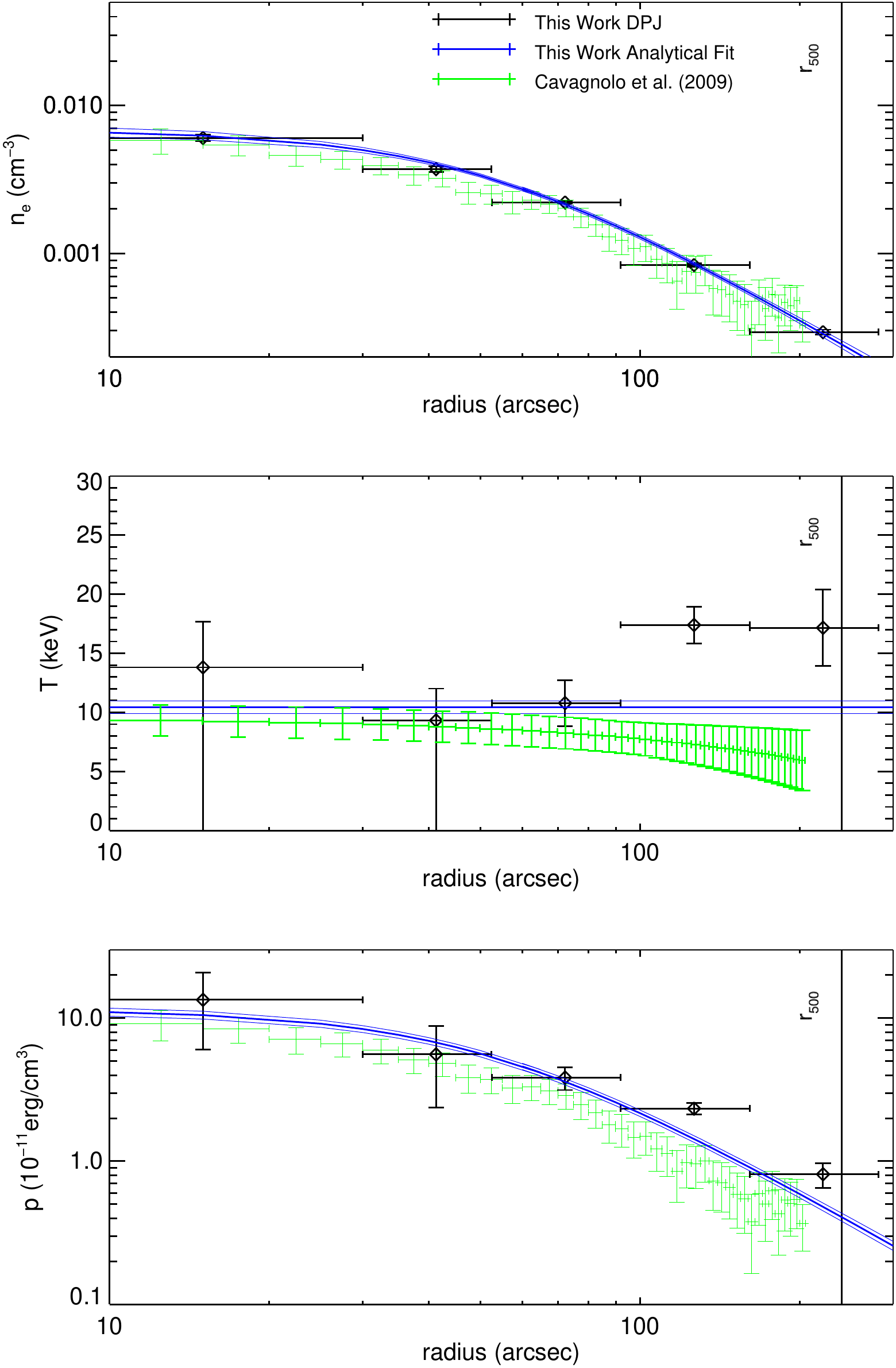}
	\caption{{\bf MACS J1149.5+2223.} Our results for this cluster are in general agreement with the ACCEPT results, although our deprojection indicates a somewhat higher temperature at large radius. This massive cluster has been widely studied through gravitational lensing.  Using \hubble\ data, \citet{2016MNRAS.459.1698M} constrained the mass distribution of this cluster and found several main peaks and clear non-sphericity, suggesting this is a merging cluster.}
	\label{fig:macsj11495}
\end{figure}

\begin{figure}
	\includegraphics[width=\linewidth]{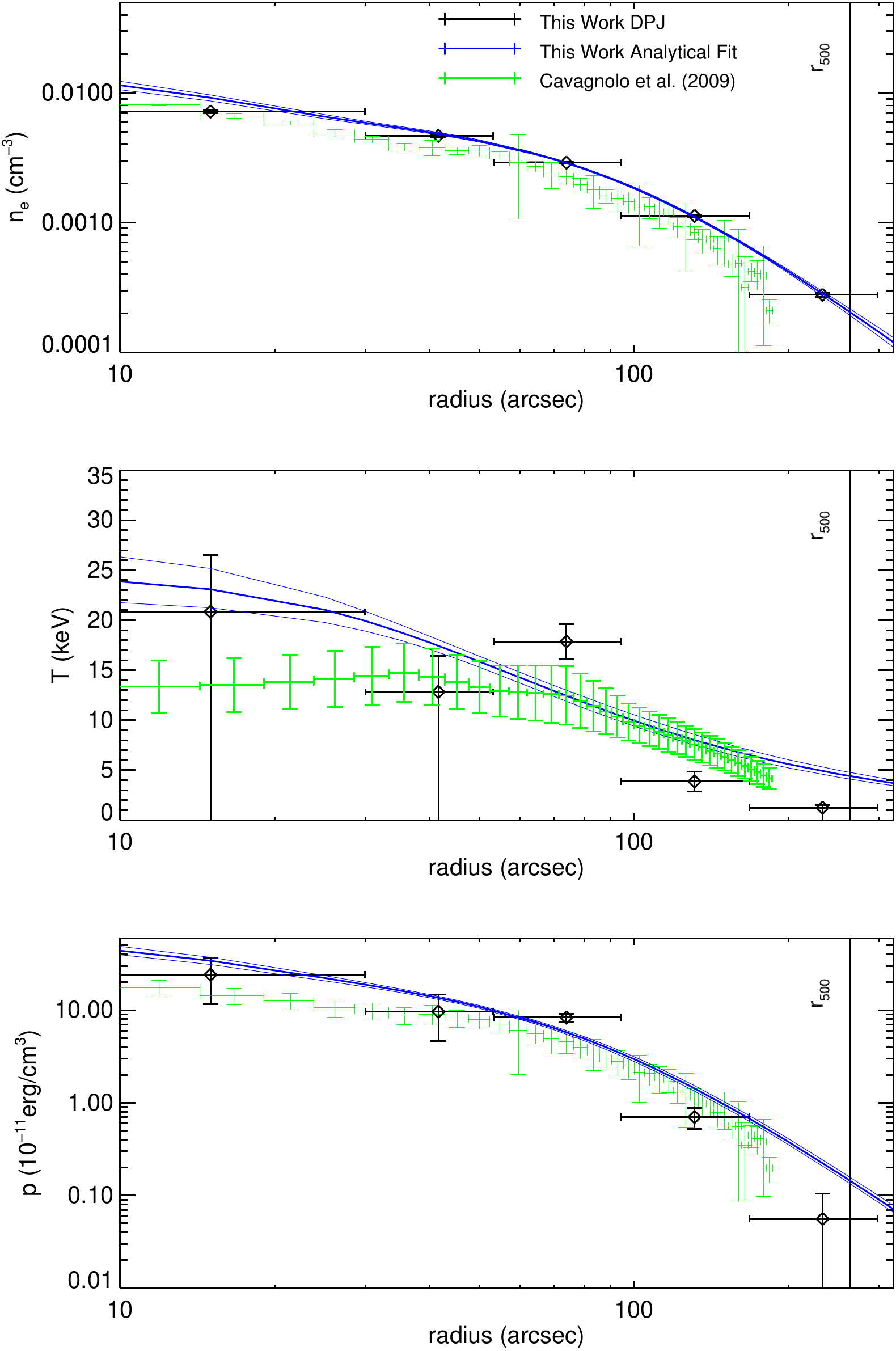}
	\caption{{\bf MACS J0717.5+3745.} Our profiles for this cluster are in excellent agreement with the profiles obtained in the ACCEPT analysis, and our smooth parametric fit indicates a temperature drop at large radius at a significance of $24\sigma$. This cluster has been studied extensively in different wavelengths.  \citet{2009ApJ...693L..56M} found that the cluster consists of four large subclusters based on X-ray imaging and galaxy positions.  Through lensing \citep{2013ApJ...777...43M,2015MNRAS.451.3920D,2016A&A...588A..99L,2016ApJ...821..116U}, four large masses are also clearly found, reflecting that this cluster is a complex merger.  \citet{2012ApJ...761...47M} studied this cluster through the SZ effect, X-ray and lensing observations.  They also found very hot gas near 30 keV in some regions of the cluster.  A large kSZ signal in one of the subclusters, which has not been accounted for in our analysis, could potentially be a source of bias in our results \citep{2013ApJ...778...52S}.}
	\label{fig:macsj07175}
\end{figure}

\begin{figure}
	\includegraphics[width=\linewidth]{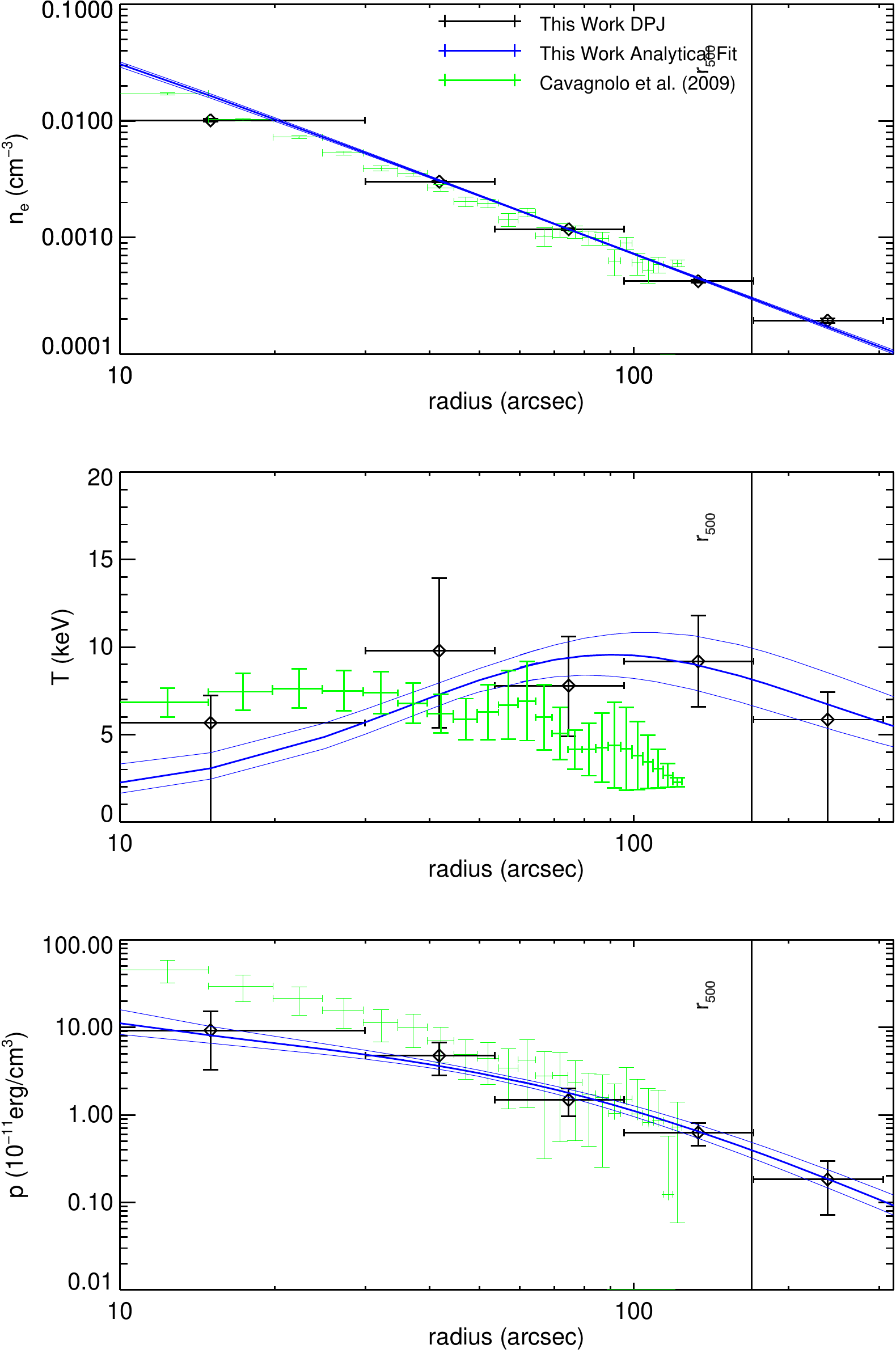}
	\caption{{\bf MACS J1423.8+2404.} Our profile fit detects a cool core at a significance of $3.3\sigma$, and both our density and temperature profiles are consistent with those from the ACCEPT analysis. \citet{2016A&A...586A.122A} conducted a multi-wavelength analysis of the cluster, including X-ray surface brightness and spectroscopy and high resolution SZ data to jointly constrain smooth ICM profiles that are in good agreement with our results.  \citet{2006ApJ...641..752K} used \chandra\ X-ray spectroscopy to constrain the temperature profile, which is again consistent with our results. \citet{2010ApJ...713..491M} used \chandra\ data, and found a cool-core (3 keV core and 7 keV peak at 300 kpc), again in good agreement with our fits. Based on lensing measurements, this cluster has been found to be slightly elongated and relaxed, with little substructure \citep{2010MNRAS.405..777L, 2011MNRAS.410.1939Z}.}
	\label{fig:macsj14238}
\end{figure}

\begin{figure}
	\includegraphics[width=\linewidth]{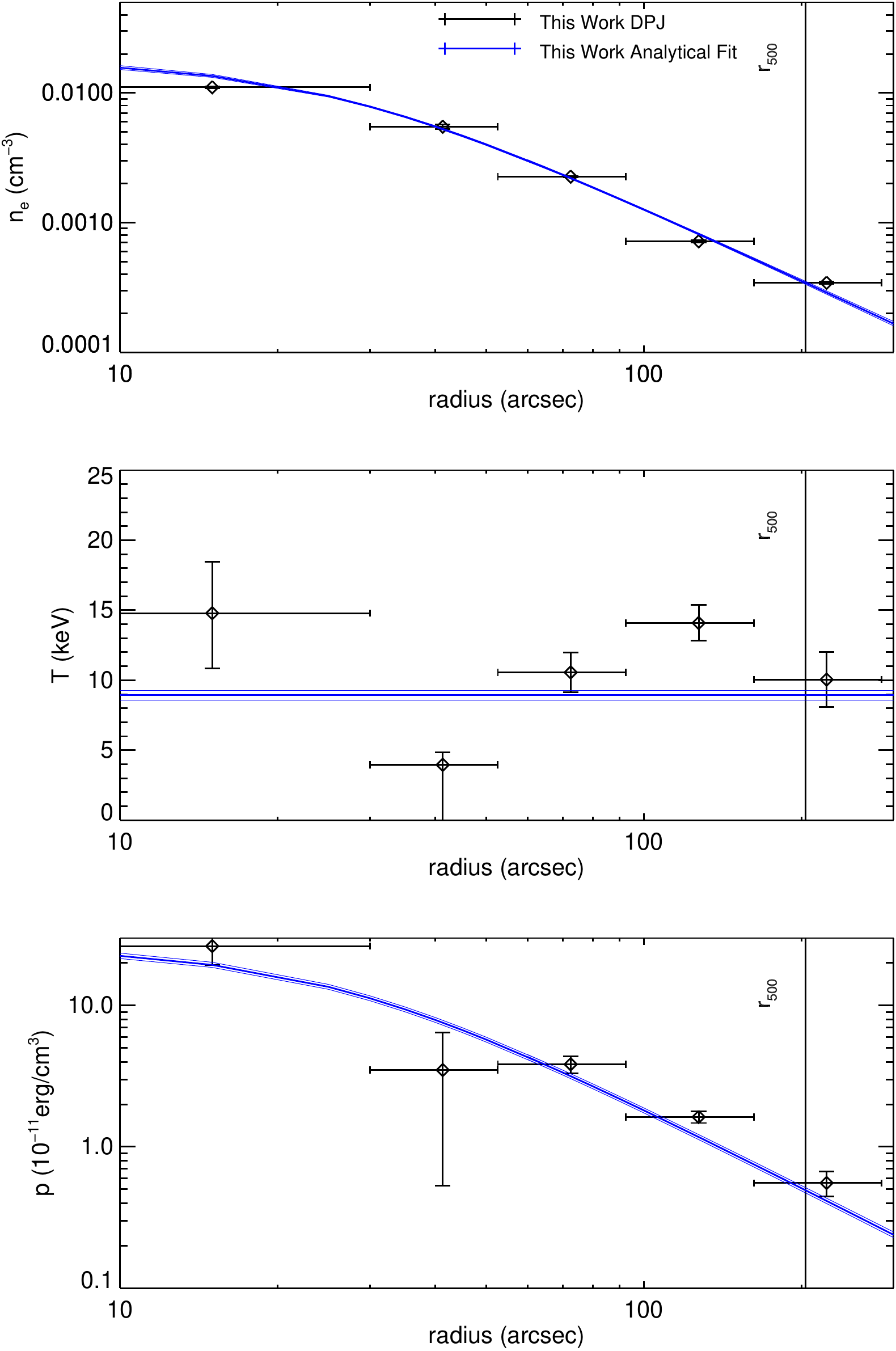}
	\caption{{\bf MACS J0454.1-0300.} The densities from the ACCEPT analysis are systematically lower than ours. However, the ACCEPT results are also inconsistent with two other analyses \citep{2006ApJ...647...25B,2016A&A...590A.126A}, and so the ACCEPT profiles are not plotted. \citet{2003ApJ...598..190D} used \chandra\ X-ray spectroscopy to obtain an isothermal temperature of 10.2 keV, consistent with our results.  They also found an elliptical cluster morphology, and a shift in the BCG location and X-ray centroid.}
	\label{fig:ms04516}
\end{figure}

\begin{figure}
	\includegraphics[width=\linewidth]{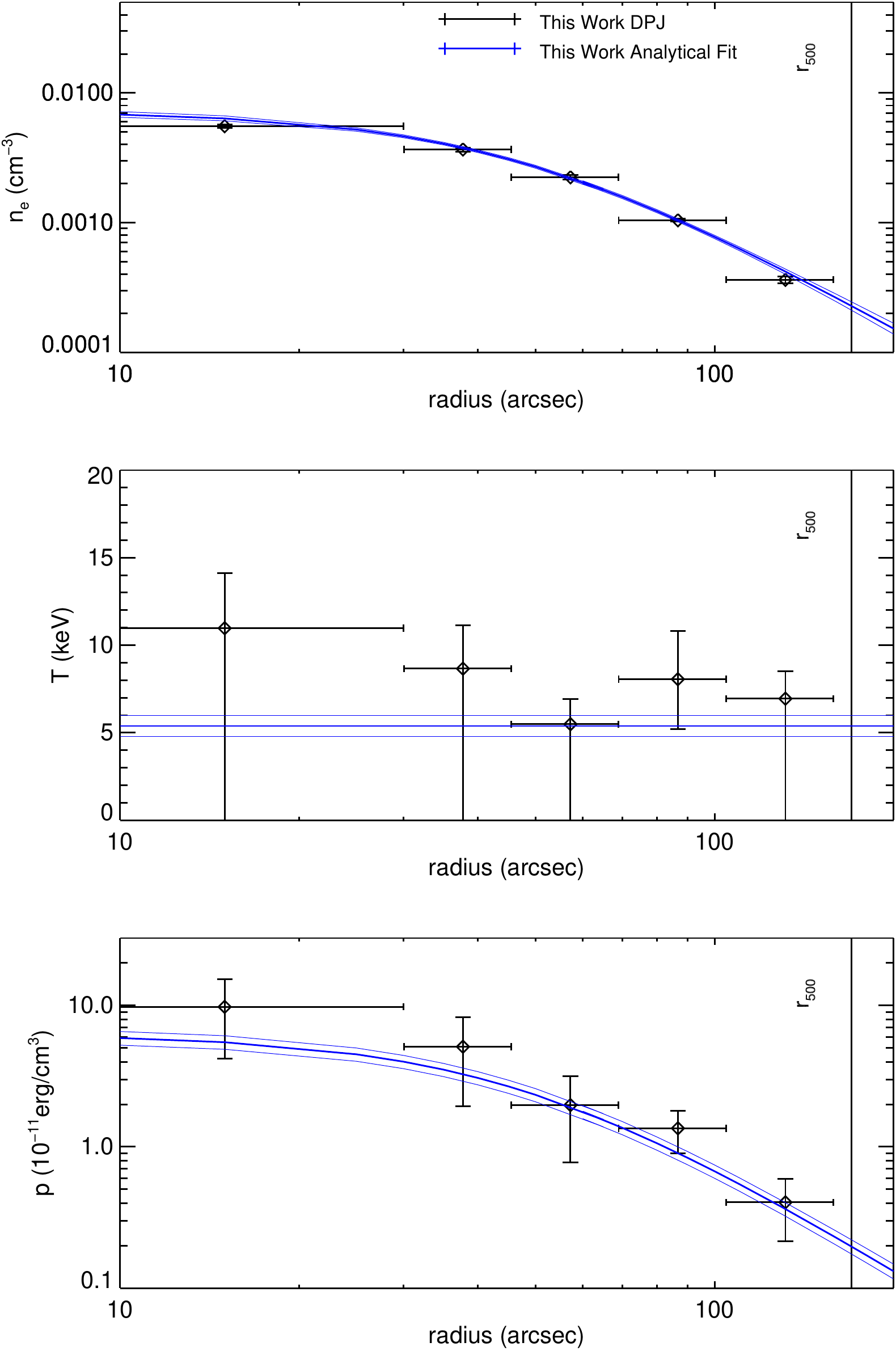}
	\caption{{\bf MACS J0025.4-1222.} Although this cluster is classified as non-disturbed in our analysis, two other studies find that it is clearly a major merger between two clusters of similar masses \citep{2008ApJ...687..959B,2008MNRAS.389.1240K}. Furthermore, \citet{2008ApJ...687..959B} found that the cluster total mass distribution is not consistent with the gas distribution.}
	\label{fig:macsj00254}
\end{figure}

\begin{figure}
	\includegraphics[width=\linewidth]{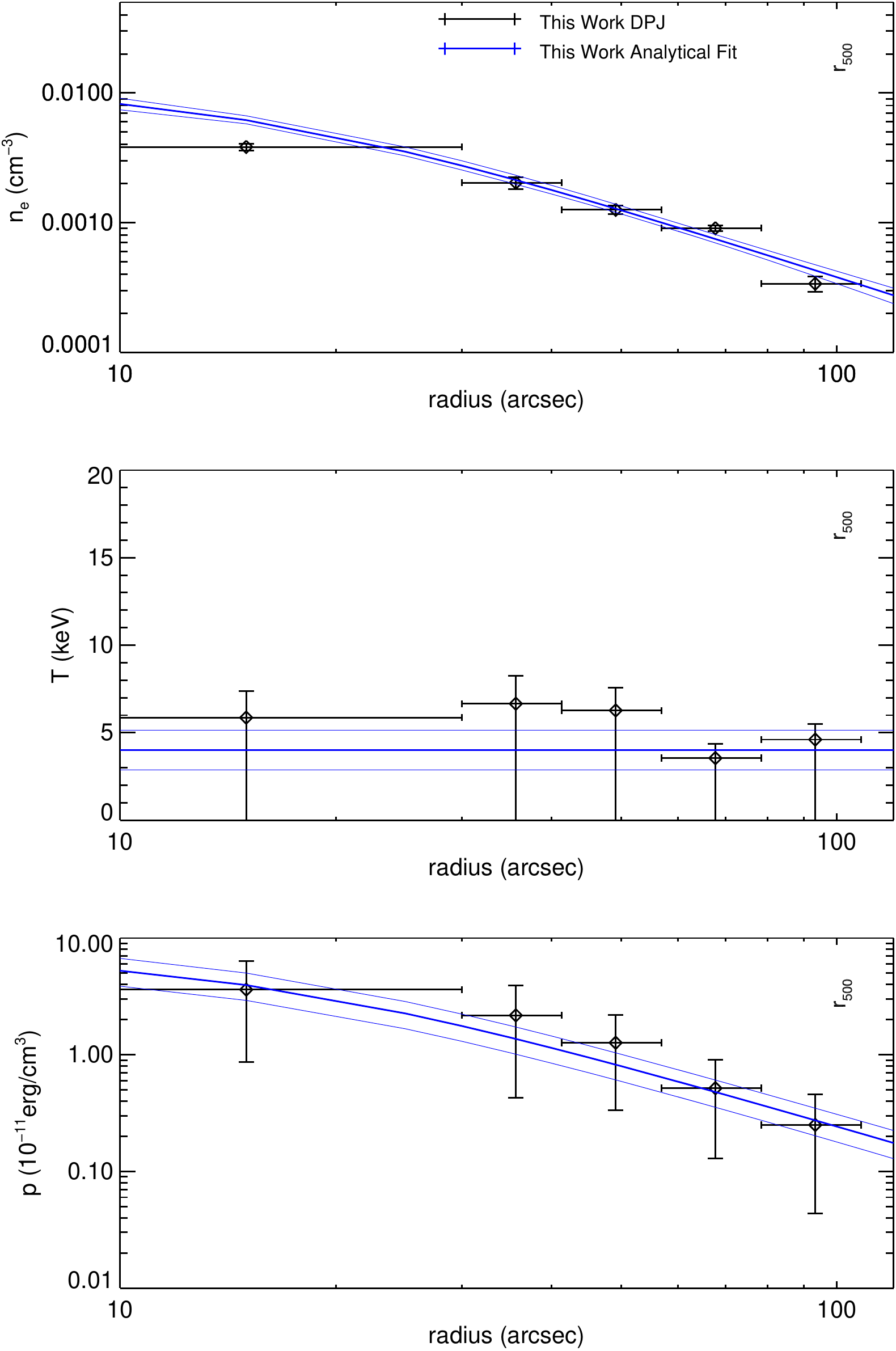}
	\caption{{\bf MS 2053.7-0449.} This is the least massive cluster in our sample, with $M_{500} \approx 3 \times 10^{14} M_{\astrosun}$, and one of the dimmest clusters in the SZ.  The low mass and high redshift also produce a weak X-ray signal. This makes it one of the few clusters in our analysis where the constrains do not extend to $r_{500}$.  \citet{2007ApJ...664..702V} found a bimodal and elongated mass distribution for this cluster using \hubble\ data, indicating a merger.}
	\label{fig:ms2053}
\end{figure}

\begin{figure}
	\includegraphics[width=\linewidth]{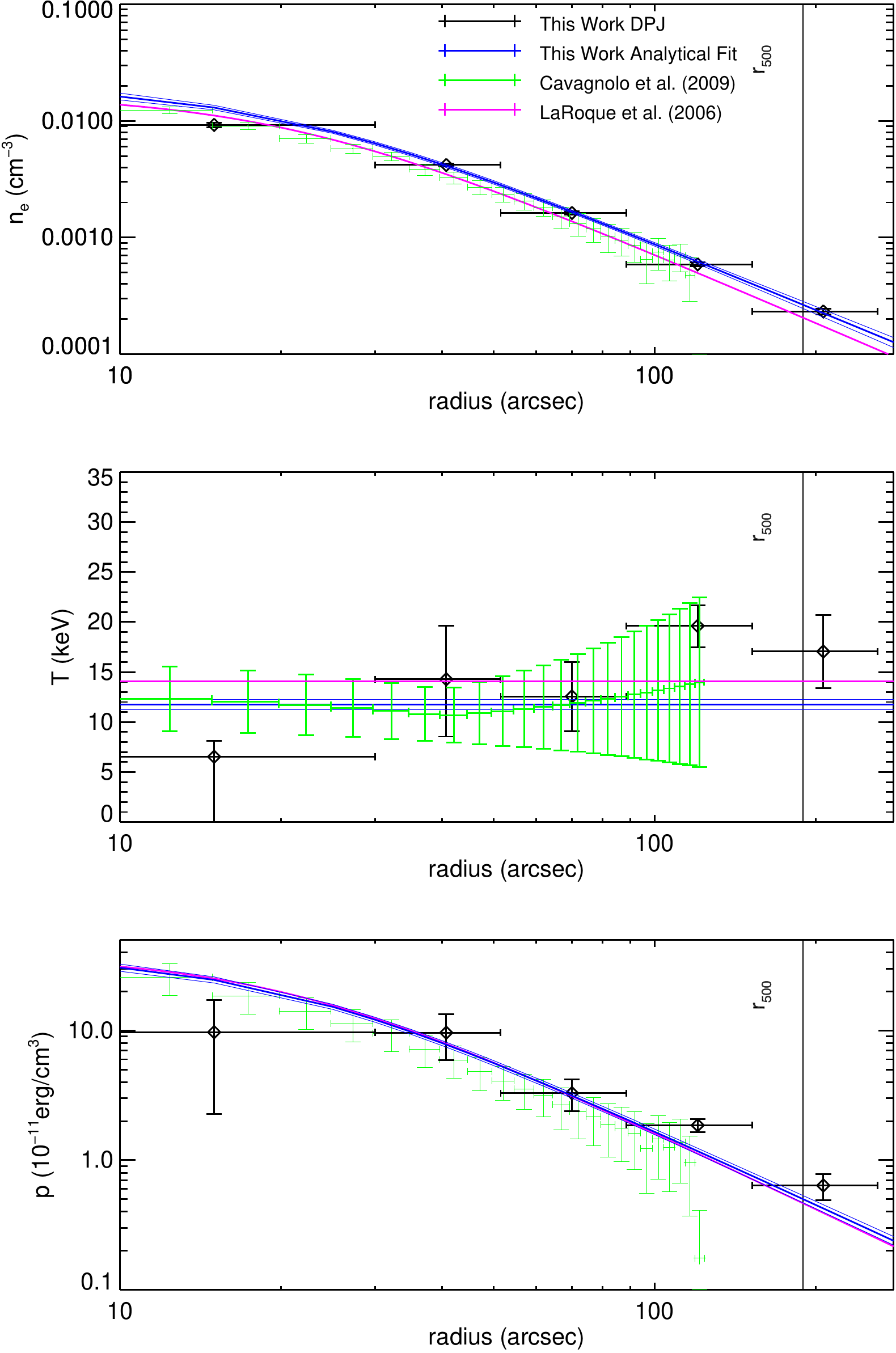}
	\caption{{\bf MACS J0647.7+7015.} Our results for this cluster are in good agreement with the ACCEPT fits. Although our smooth parametric density profile shows a statistically significant offset from the \citet{2006ApJ...652..917L} results, the absolute magnitude of the offset is modest.}
	\label{fig:macsj06477}
	\normalsize
\end{figure}

\begin{figure}
	\includegraphics[width=\linewidth]{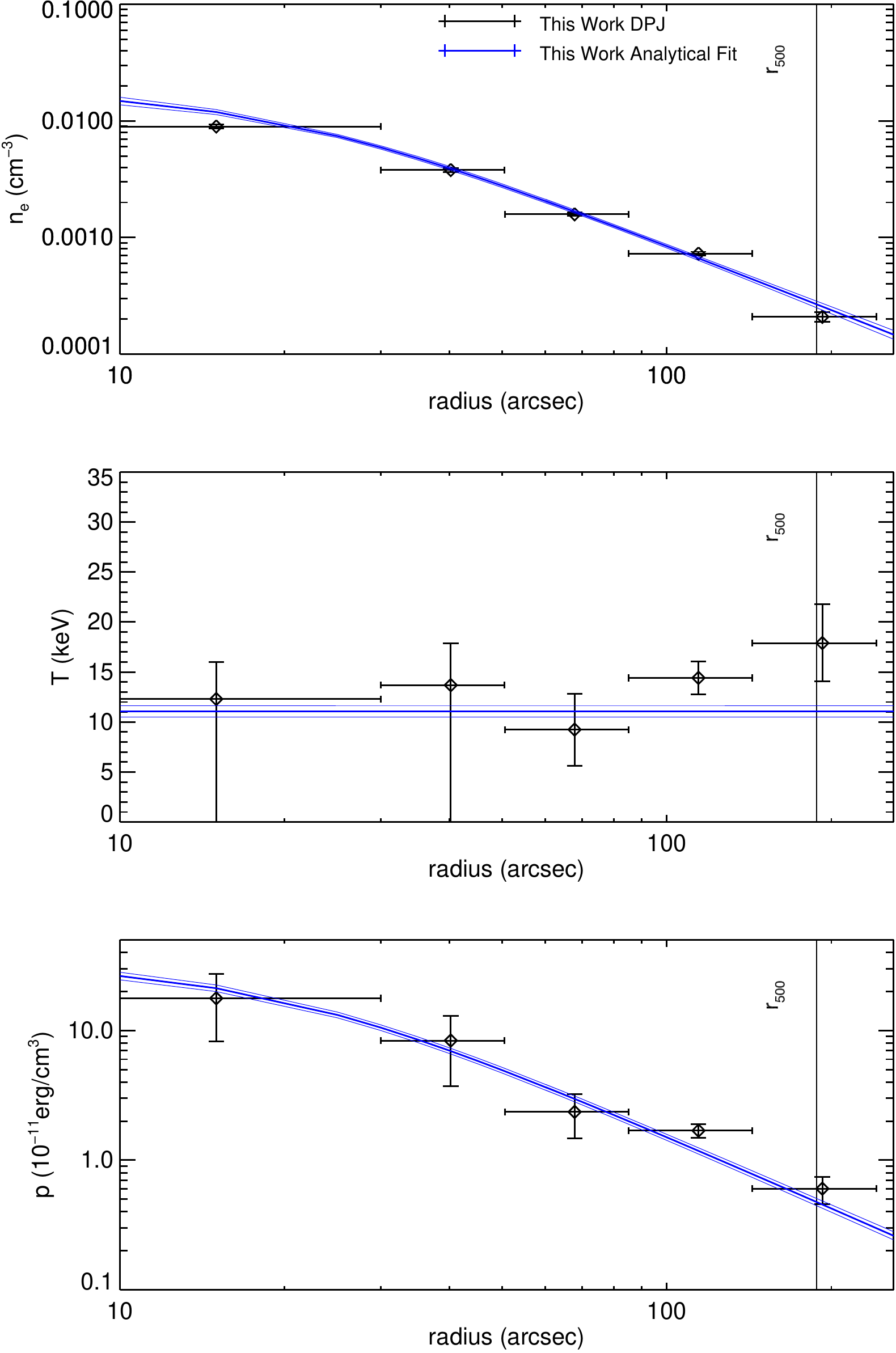}
	\caption{{\bf MACS J2129.4-0741.} \citet{2012MNRAS.420.2120M} conducted a classification of clusters based on morphology, using a combination of X-ray and optical data.  For this cluster they deduced it was a merger because the X-ray centroid location is significantly different than the BCG's.}
	\label{fig:macsj21294}
	\normalsize
\end{figure}

\begin{figure}
	\includegraphics[width=\linewidth]{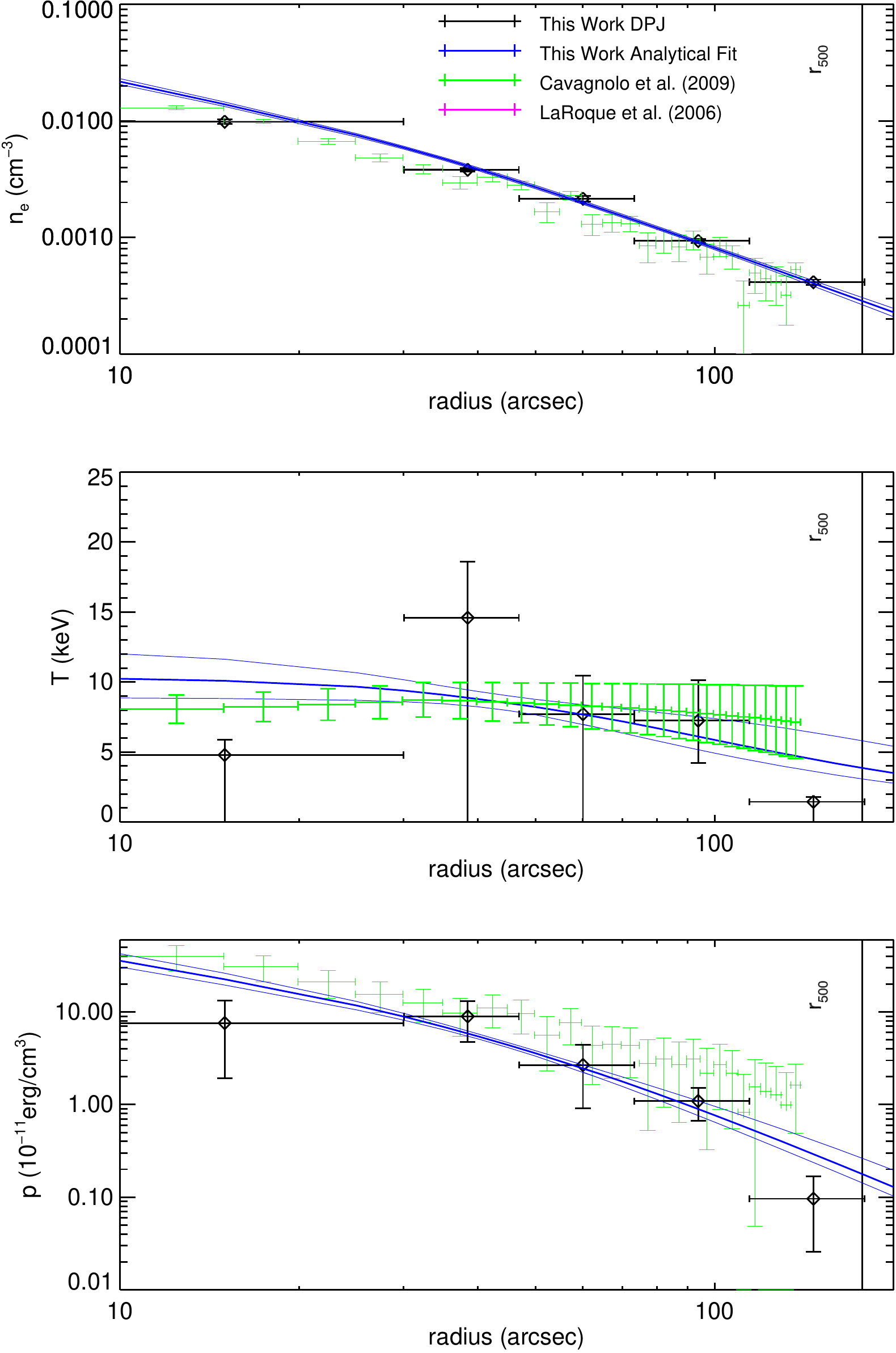}
	\caption{{\bf MACS J0744.8+3927.} Our profile fits for this cluster are in good agreement with the ACCEPT results.  \citet{2003ApJ...583..559L} studied this cluster using OVRO/BIMA and found an average temperature of 17.9 keV, significantly higher than our results.  \citet{2011ApJ...734...10K} studied this cluster through the high resolution SZ images from MUSTANG, and compared it with X-ray data, finding evidence of a merger related shock-front. }
	\label{fig:macsj07448}
\end{figure}

\begin{figure}
	\includegraphics[width=\linewidth]{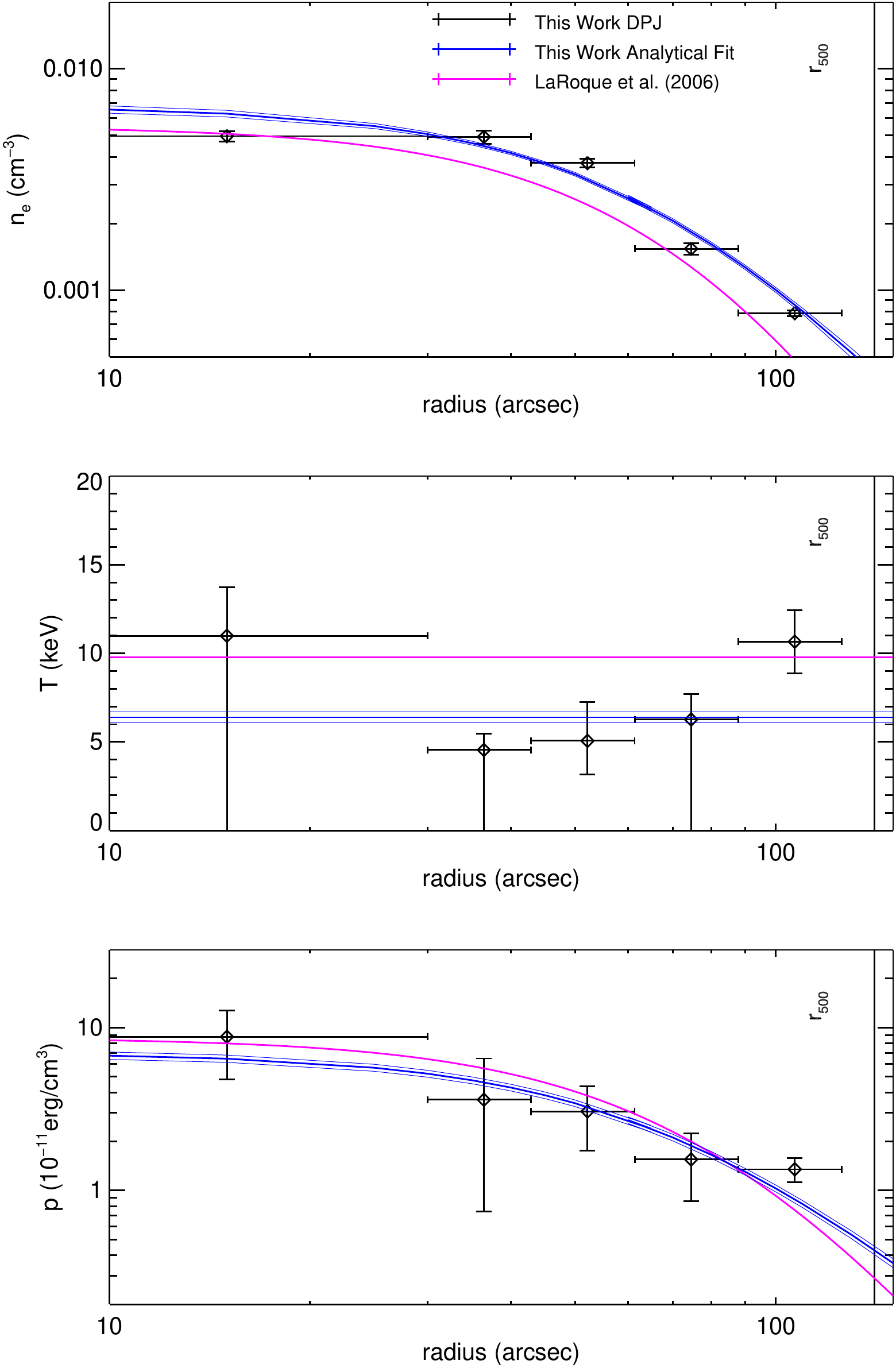}
	\caption{{\bf MS1054.4-0321.} Compared to the smooth parametric fits of \citet{2006ApJ...652..917L}, our smooth parametric fits show a significant offset in both density and temperature.  \citet{2004A&A...419..517G} used \xmm\ to find an average temperature of $\sim7 keV$, in good agreement with our results. \citet{2000ApJ...542...35N} studied this cluster using the X-ray (\rosat), finding substructure and other signs of recent merging processes (e.g., BCG and X-ray peak offset, X-ray centroid and peak offset).}
	\label{fig:ms1054}
\end{figure}

\begin{figure}
	\includegraphics[width=\linewidth]{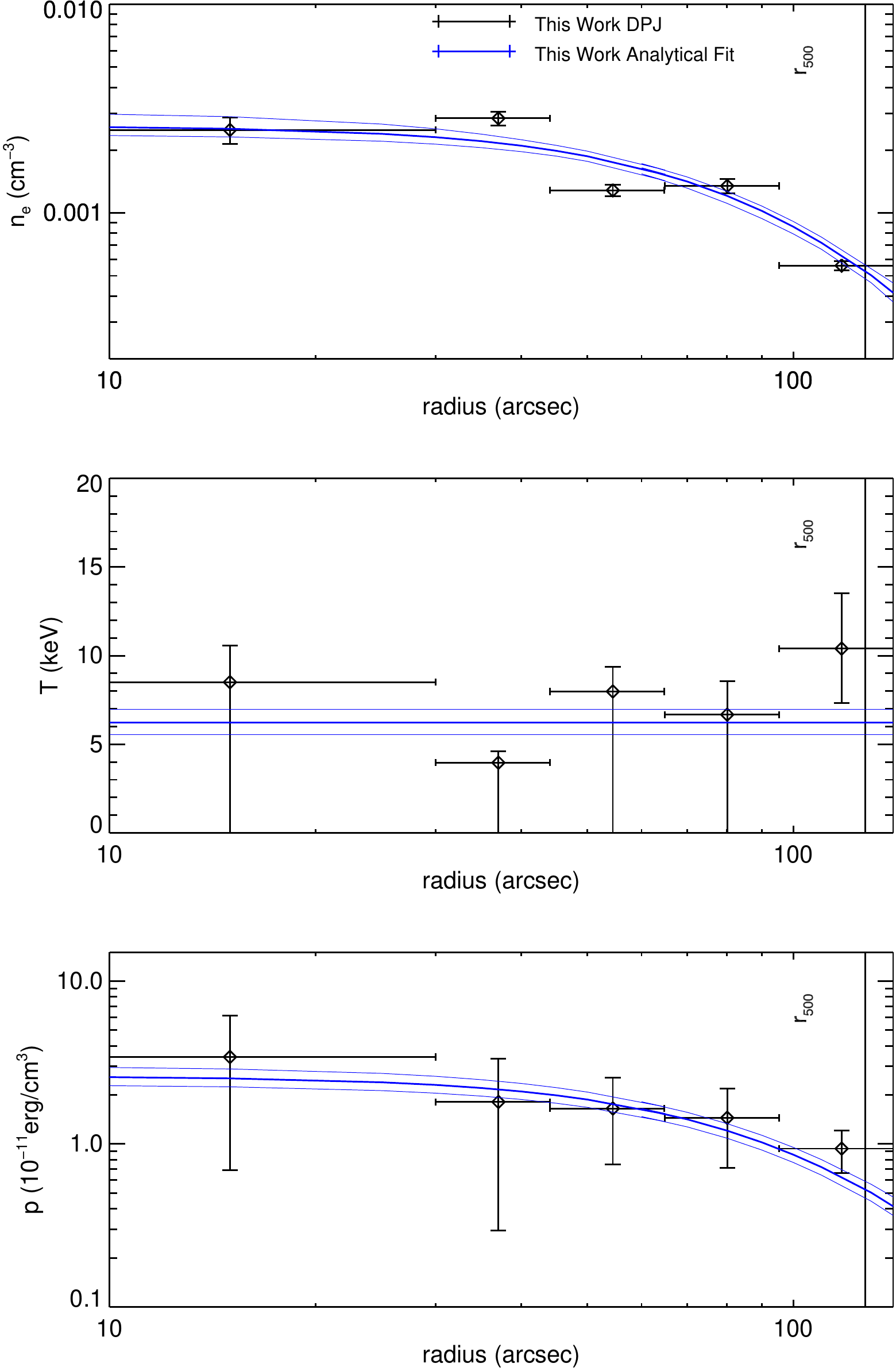}
	\caption{{\bf CL J0152.7.} The irregular shape of our deprojected density profile is possibly due to the disturbed nature of the cluster.  Several studies found evidence of subclusters and of merging based on: X-ray and optical subclusters \citep{2004AJ....127.1263H}, a high velocity dispersion of galaxy cluster members \citep{2005A&A...442...29G}, and an offset between the X-ray and SZ emission peaks \citep{2010ApJ...718L..23M}.  \citet{2006ApJ...640..219M} used \xmm\ to study the X-ray emission distribution in detail, finding many substructures and smaller groups, concluding that the cluster has recently undergone many mergers.}
	\label{fig:clj01527}
\end{figure}

\begin{figure}
	\includegraphics[width=\linewidth]{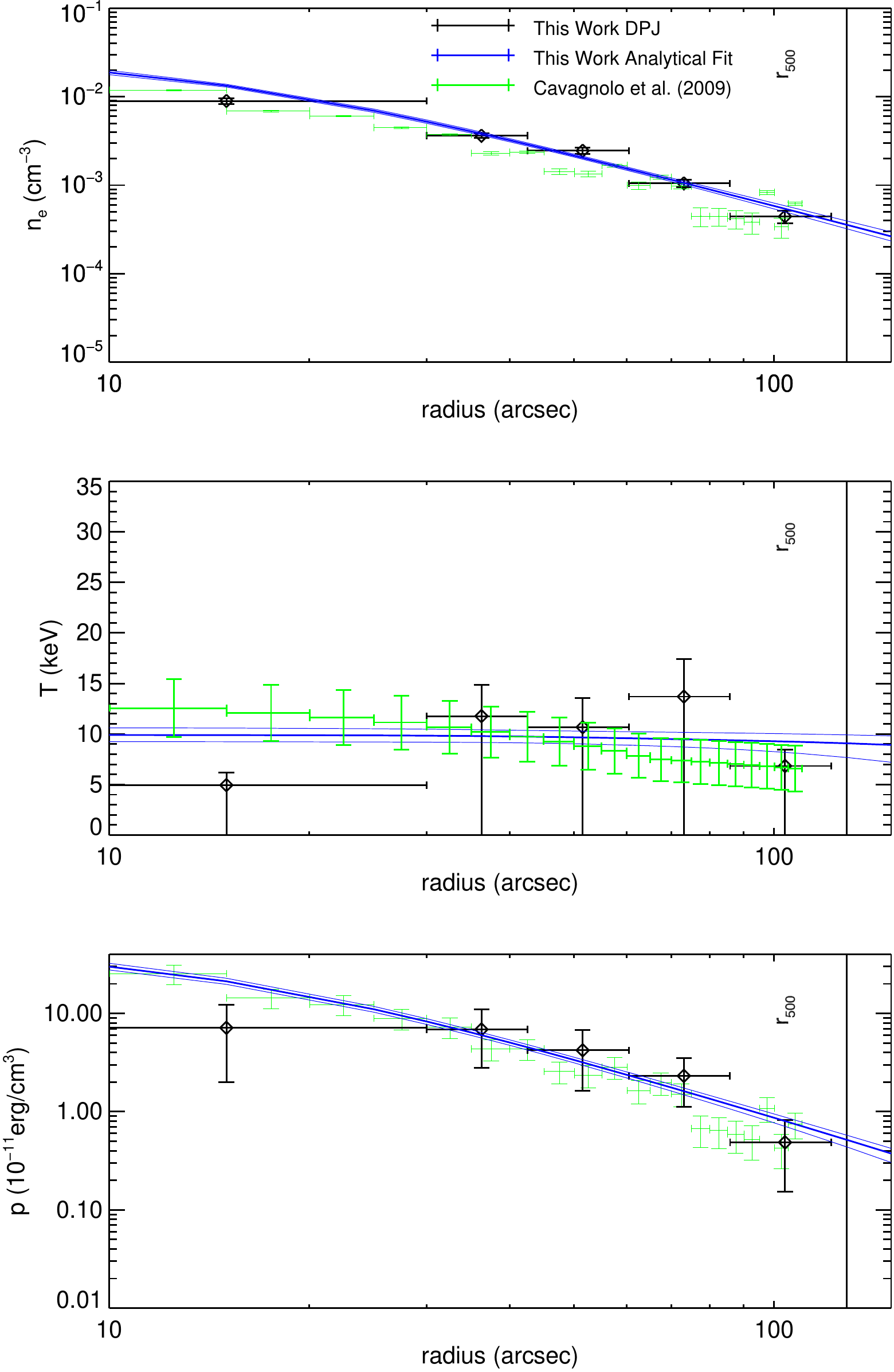}
	\caption{{\bf CL J1226.9+3332.} Our results for this cluster are in relatively good agreement with the ACCEPT fits. Although this cluster is classified as non-disturbed in our analysis, several other works have found evidence of merging through X-ray temperature map asymmetries \citep{2007ApJ...659.1125M}, subclumps through weak lensing \citep{2009ApJ...691.1337J}, and multiple peaks in the SZ through high resolution MUSTANG maps \citep{2011ApJ...734...10K}.}
	\label{fig:clj12269}
\end{figure}


\bsp	
\label{lastpage}
\end{document}